\documentclass[onecolumn]{IEEEtran}
\pdfoutput=1
\usepackage{amsmath,amssymb,amsthm,mathrsfs,bm}
\usepackage{graphicx}
\usepackage{subcaption}
\usepackage{xcolor}
\usepackage{pgfplots}
\usepackage{tikz}
    \usetikzlibrary{patterns}
    \usetikzlibrary{decorations}
    \usetikzlibrary{decorations.pathreplacing}
    \usetikzlibrary{arrows}
\usepackage{multirow}
\usepackage{hyperref}
\usepackage[numbers,sort&compress,square]{natbib}

\newtheorem{theorem}{Theorem}
    \newtheorem{lemma}[theorem]{Lemma}
    
    \newtheorem{claim}[theorem]{Claim}
    \newtheorem*{claim*}{Claim}
    
\theoremstyle{definition}
    \newtheorem*{example*}{Example}
    
    \newtheorem{definition}{Definition}
\theoremstyle{remark}
    \newtheorem*{remark}{Remark}
    
    \newenvironment{secproof}{}{\hfill$\blacksquare$\par\smallskip}

\DeclareMathOperator{\DoF}{\mathsf{DoF}}
\DeclareMathOperator{\LDoF}{\mathsf{DoF}_{L-sum}}

\DeclareMathOperator{\trace}{tr}

\DeclareMathOperator{\rank}{rk}
\DeclareMathOperator{\Proj}{Proj}
\DeclareMathOperator{\im}{im}
\DeclareMathOperator{\spn}{span}
\DeclareMathOperator{\rowspan}{rowspan}
\DeclareMathOperator{\colspan}{colspan}
\DeclareMathOperator{\dimspan}{dim span}
\DeclareMathOperator*{\E}{E}
\DeclareMathOperator*{\maximize}{\mbox{maximize}}
\DeclareMathOperator{\subjectto}{\mbox{subject to}}

\begin{document}

\title{Linear Degrees of Freedom of \\the MIMO X-Channel with Delayed CSIT}
\author{David T.H. Kao and A. Salman Avestimehr%
\thanks{D. T. H. Kao (email: kaod@usc.edu) and A. S. Avestimehr (email: avestimehr@ee.usc.edu) are with the Ming Hsieh Department of Electrical Engineering, University of Southern California, Los Angeles, CA. The research of A. S. Avestimehr and D. T. H. Kao is supported by NSF Grants CAREER 1408639, CCF-1408755, NETS-1419632, EARS-1411244, and ONR award N000141310094. A portion of this work was presented in~\cite{MIMOXC-DCSIT-isit}.}}

\maketitle

\begin{abstract}
We study the degrees of freedom (DoF) of the multiple-input multiple-output X-channel (MIMO XC) with delayed channel state information at the transmitters (delayed CSIT), assuming linear coding strategies at the transmitters. 
We present two results: 1) the linear sum DoF for MIMO XC with general antenna configurations, and 2) the linear DoF region for MIMO XC with symmetric antennas. 
The converse for each result is based on developing a novel rank-ratio inequality that characterizes the maximum ratio between the dimensions of received linear subspaces at the two multiple-antenna receivers.
The achievability of the linear sum DoF is based on a three-phase strategy, in which during the first two phases only the transmitter with fewer antennas exploits delayed CSIT in order to minimize the dimension of its signal at the unintended receiver. During Phase~3, both transmitters use delayed CSIT to send linear combinations of past transmissions such that each receiver receives a superposition of desired message data and known interference, thus simultaneously serving both receivers. 
We also derive other linear DoF outer bounds for the MIMO XC that, in addition to the outer bounds from the sum DoF converse and the proposed transmission strategy, allow us to characterize the linear DoF region for symmetric antenna configurations.
\end{abstract}

\section{Introduction}
The availability of channel state information at transmitters (CSIT) enables considerable capacity gains in wireless networks. However, due to channel variations and feedback delays, acquiring up-to-date channel state information may be practically infeasible. Thus, a realistic assumption in fast fading wireless environments is to assume that CSIT is acquired with some delay.

Interestingly, in~\cite{MaT:2012} it was shown that, in the $K$-user multiple-input single-output broadcast channel (MISO BC), delayed CSIT can still be quite useful. Unlike the MISO BC with no CSIT, where the sum degrees of freedom (DoF) is known to be 1,~\cite{MaT:2012} showed that the the DoF of the MISO BC with delayed CSIT scales almost linearly with the number of users. Motivated by this result, the impact of delayed CSIT has been explored for many other networks.  In particular, in~\cite{MJS:2012,AGK:2013,AA:ARXIV2013,VMA:ARXIV2013,VMA:infocom2014}, it was shown that delayed CSIT can also be useful for interference management in various network configurations; by developing novel transmission strategies that dynamically adapt transmissions to the past receptions at the receivers, distributed transmitters can better manage interference.

In this work, we study the impact of delayed CSIT in the multiple-input multiple-output X-channel (MIMO XC), which is a canonical setting for information-theoretic study of interference management in wireless networks. This channel consists of two transmitters causing interference at two receivers, and each transmitter aims to communicate independent messages to both receivers. In the case of instantaneous CSIT, it is known that the optimal DoF is achieved by \emph{aligning} the interference at each receiver~\cite{MaMK:2008,JS:2008}. With delayed CSIT, it is shown in~\cite{MJS:2012,GMK:isit2011,GAK:isit2012,VV:2012,AGK:2013} that a variation of interference alignment is still feasible, by dynamically aligning interfering transmissions to past receptions (a.k.a retrospective interference alignment).

However, the DoF-optimal transmission strategy for the general MIMO XC with delayed CSIT is still unknown, except for the case where all nodes have a single antenna, for which a novel converse was recently derived in~\cite{LAS:allerton2013,LAS:ARXIV2013} that shows the sum DoF optimality of the scheme proposed in~\cite{GMK:isit2011} under the restriction of linear encoding strategies. The key insight of~\cite{LAS:allerton2013,LAS:ARXIV2013} relied on establishing a \emph{rank-ratio inequality} that shows that, if two distributed single antenna transmitters employ linear strategies, the ratio of the dimensions of received linear subspaces at two arbitrary receivers cannot exceed $\frac{3}{2}$, due to delayed CSIT.

In this paper, we provide two main results. We first establish the sum DoF of MIMO XC with delayed CSIT and for any antenna configuration, assuming linear coding strategies at the transmitters. We then use analytical tools from the linear sum DoF result to establish the DoF region, assuming linear coding strategies, of MIMO XC with symmetric antenna configurations.

Our sum DoF converse generalizes the one given in~\cite{LAS:allerton2013,LAS:ARXIV2013} to the multiple antenna setting, and requires establishing a general rank-ratio inequality for the ratio between dimensions of received signal subspaces at the two receivers. Our rank-ratio inequality results from establishing two bounds. The first is a cooperative bound, which assumes transmitters share message information and thus emulate a MIMO BC with delayed CSIT. The second bound results from focusing on the difference in ranks normalized by number of antennas at each receiver, and analyzing how effectively two transmitters may exploit delayed CSIT to maximize this difference. We arrive at our rank-ratio inequality by taking the minimum of these two bounds, and then apply it in construction of an upper bound on the linear sum DoF.

We then define a class of transmission strategies that achieves the linear sum DoF upper bound, for all antenna configurations. Our strategies consist of three phases. Transmissions during Phase~1 contain only message symbols intended for Receiver~1, and transmissions during Phase~2 contain only symbols intended for Receiver~2. During Phase~3, both transmitters use delayed CSIT to send linear combinations of past transmissions such that each receiver receives a superposition of desired message data and known interference. To maximize the number of Phase~3 transmissions (and thus the sum DoF), our strategy dictates the number of symbols sent from each transmitter during Phases~1 and~2 as a function of the maximum rank-ratios and  exploits delayed CSIT during Phases~1 and~2, but only at the transmitter with fewer antennas.

We identify all antenna configurations where the linear sum DoF of the MIMO XC is strictly less than an analogous MIMO BC. These antenna configurations exhibit a linear sum DoF loss \emph{due to distributed transmitters}. Rather than directly comparing the linear sum DoF expressions for the two networks, we instead classify antenna configurations by focusing on the maximum rank-ratios. Recall that the rank-ratio inequality established in the converse was constructed from two upper bounds, one of which allowed transmitters to cooperate and emulate a MIMO BC with delayed CSIT. When the cooperative bound is tighter than the second bound for both rank-ratios, then we may say no distributed transmitter loss occurs. Conversely, when the second bound is tighter, we have a configuration that exhibits distributed transmitter loss in linear sum DoF.

Finally, we use our results to study the linear DoF region for MIMO XC with delayed CSIT and symmetric antenna configurations. We develop an additional bound on the DoF region and for five specific regimes which span all symmetric antenna configurations, identify all corner points of the region and describe how our general linear sum DoF scheme may be adapted to achieve each corner point.

The paper is organized as follows. In Section~\ref{sec:main} we state the problem formulation and the main results. Additionally we state the core lemma which specifies the maximum rank-ratios. In Section~\ref{sec:rankratio} we prove the main result as well as the core lemma. Proofs of minor steps within the converse may be found in Appendices~\ref{append:BCrankratio} through \ref{app:pAUBc}. In Section~\ref{sec:scheme} we present the transmission scheme to achieve the linear sum DoF of the MIMO XC, as well as highlight the difference between our scheme and previous scheme for symmetric antenna configurations. In Section~\ref{sec:distloss} we identify antenna configurations for the MIMO XC which exhibit a distributed transmitter loss of linear sum DoF (i.e., when linear sum DoF of the MIMO XC is strictly less than the associated MIMO BC). In Section~\ref{sec:dofreg} we prove the linear DoF region result for MIMO XC with symmetric antenna configurations. Concluding remarks may be found in Section~\ref{sec:end}. With respect to notation, unless otherwise stated, we denote random variables using bold type ($\mathbf{x}$), vectors as $\vec{x}$, matrices as capital letters ($X$), and sets as script capital letters ($\mathcal{X}$).
%%%%%%%%%%
%%%%%%%%%%
%%%%%%%%%%
%%%%%%%%%%
%%%%%%%%%%
\section{Problem Statement and Main Result}
\label{sec:main}

\begin{figure}[t]
\centering\vspace{-0.5cm}
    \begin{tikzpicture}[scale=1,font=\scriptsize]
    \node[above] (u11) at (-1,3) {$\vec{\mathbf{u}}_{11}$};
    \node[below] (u21) at (-1,3) {$\vec{\mathbf{u}}_{21}$};
    \node[above] (u12) at (-1,0) {$\vec{\mathbf{u}}_{12}$};
    \node[below] (u22) at (-1,0) {$\vec{\mathbf{u}}_{22}$};

    \node[above] (hu11) at (7,3) {$\widehat{\vec{\mathbf{u}}_{11}}$};
    \node[below] (hu12) at (7,3) {$\widehat{\vec{\mathbf{u}}_{12}}$};
    \node[above] (hu21) at (7,0) {$\widehat{\vec{\mathbf{u}}_{21}}$};
    \node[below] (hu22) at (7,0) {$\widehat{\vec{\mathbf{u}}_{22}}$};

    \node (a) at (0,3) {\begin{tikzpicture}[scale=0.9]
        \draw[thick] (0,0.8) -- (0,-0.8) -- (-0.25,-0.8) -- (-0.25,0.8) -- cycle;
        \draw[thick,-<] (0,-0.7) -- (0.25,-0.7) -- (0.25,-0.45);
        \draw[thick,-<] (0,-0.3) -- (0.25,-0.3) -- (0.25,-0.05);
        \draw[thick,-<] (0,0.1) -- (0.25,0.1) -- (0.25,0.35);
        \draw[thick,-<] (0,0.5) -- (0.25,0.5) -- (0.25,0.75);
    \end{tikzpicture}};
    \node (b) at (0,0) {\begin{tikzpicture}[scale=0.9]
        \draw[thick] (0,0.8) -- (0,-0.4) -- (-0.25,-0.4) -- (-0.25,0.8) -- cycle;
        \draw[thick,-<] (0,-0.3) -- (0.25,-0.3) -- (0.25,-0.05);
        \draw[thick,-<] (0,0.1) -- (0.25,0.1) -- (0.25,0.35);
        \draw[thick,-<] (0,0.5) -- (0.25,0.5) -- (0.25,0.75);
    \end{tikzpicture}};
    \node (c) at (6,3) {\begin{tikzpicture}[scale=0.9]
        \draw[thick] (0,0.8) -- (0,-0.4) -- (0.25,-0.4) -- (0.25,0.8) -- cycle;
        \draw[thick,-<] (0,-0.3) -- (-0.25,-0.3) -- (-0.25,-0.05);
        \draw[thick,-<] (0,0.1) -- (-0.25,0.1) -- (-0.25,0.35);
        \draw[thick,-<] (0,0.5) -- (-0.25,0.5) -- (-0.25,0.75);
    \end{tikzpicture}};
    \node (d) at (6,0) {\begin{tikzpicture}[scale=0.9]
        \draw[thick] (0,0.4) -- (0,-0.4) -- (0.25,-0.4) -- (0.25,0.4) -- cycle;
        \draw[thick,-<] (0,-0.3) -- (-0.25,-0.3) -- (-0.25,-0.05);
        \draw[thick,-<] (0,0.1) -- (-0.25,0.1) -- (-0.25,0.35);
    \end{tikzpicture}};
    \draw[ultra thick,-latex] (a)--node[above]{$\mathbf{G}_{11}$} (c);
    \draw[ultra thick,-latex] (a)--node[above,pos=0.325,sloped]{$\mathbf{G}_{21}$} (d);
    \draw[ultra thick,-latex] (b)--node[above,pos=0.275,sloped]{$\mathbf{G}_{12}$} (c);
    \draw[ultra thick,-latex] (b)--node[above]{$\mathbf{G}_{22}$} (d);
\end{tikzpicture}\vspace{-0.25cm}
\caption{A MIMO X-channel where $M_1=4$, $M_2=3$, $N_1=3$, and $N_2=2$.}\label{fig:XC}
\end{figure}
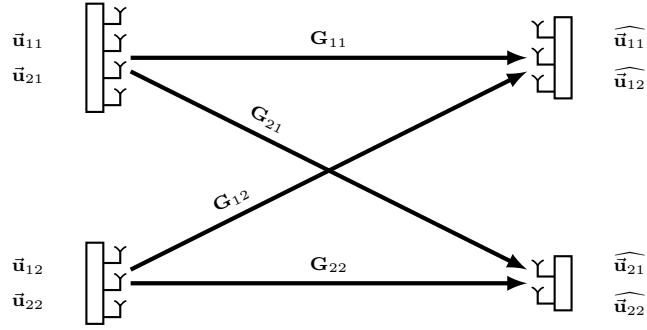

The X-channel is a four-node network containing two transmitters and two receivers, where each transmitter has an independent message for each receiver. This paper focuses on the multiple-input multiple-output X-channel (MIMO XC), where Transmitter~$j$ and Receiver~$i$ have $M_j$ and $N_i$ antennas respectively, with $i,j \in \{1,2\}$.  Without loss of generality, we assume $M_1\geq M_2$. An example of a MIMO XC is shown in Figure~\ref{fig:XC}. The channel output at Receiver~$i$ is 
\begin{align}
    \vec{\mathbf{y}}_i[t] ={}& \sum_{j=1}^2 \mathbf{G}_{ij}[t]\vec{\mathbf{x}}_j[t] + \vec{\mathbf{z}}_i[t],
\end{align}
where $\vec{\mathbf{x}}_j[t]$ and $\vec{\mathbf{y}}_i[t]$ are the (vector) input of the $j$-th transmitter and output of $i$-th receiver respectively at time $t$,  $\vec{\mathbf{z}}_i[t]\sim\mathcal{CN}(0,\mathbb{I}_{N_i})$ is an additive white Gaussian noise vector, and $\mathbf{G}_{ij}[t]$ the fading channel matrix between the $j$-th transmitter and $i$-th receiver. The $N_i\times M_j$ channel matrix $\mathbf{G}_{ij}[t]$ is drawn from a continuous complex distribution, i.i.d. across time. We denote the $m$-th element of $\vec{\mathbf{x}}_i[t]$ as $\mathbf{x}_i^m[t]$ and the $(n,m)$-th element of $\mathbf{G}_{ij}[t]$ as $\mathbf{g}_{ij}^{nm}[t]$. Additionally, we denote 
the set of channel matrices up until time $T$ as $\bm{\mathcal{G}}^T\triangleq\{\mathbf{G}_{ij}[t] :\ i,j\in\{1,2\}, t\in\{1,\ldots,T\}\}$. Since we assume delayed channel state information at the transmitters, at time $t$, transmitters know $\bm{\mathcal{G}}^{t-1}$ whereas receivers know $\bm{\mathcal{G}}^{t}$.\footnote{For i.i.d. channel fades, a delay of one time slot is sufficient to model any finite delay length.}

Each transmitter is subject to an average power constraint evaluated over a block of length $T$, i.e., let $\Sigma_j[t] \triangleq \E[\vec{\mathbf{x}}_j[t]\vec{\mathbf{x}}_j[t]^\dagger]$ with $\dagger$ representing the Hermitian transpose, and $\frac{1}{T}\sum_{t}\trace\Sigma_j[t]\leq P$. 

We restrict ourselves to linear coding strategies as defined in~\cite{BCT:ARXIV2013,LAS:allerton2013,LAS:ARXIV2013}, in which DoF simply represents the dimension of the linear subspace of transmitted signals. More specifically, consider a communication scheme with blocklength $T$, in which Transmitter~$j$ wishes to transmit a vector $\vec{\mathbf{u}}_{ij}\in\mathbb{C}^{m_{ij}^{(T)}}$ of $m_{ij}^{(T)}\in\mathbb{N}$ information symbols to Receiver~$i$. These information symbols are then modulated with precoding matrices$\mathbf{V}_{ij}[t] \in\mathbb{C}^{M_{j}\times m_{ij}^{(T)}}$ at times $t = 1, 2,\ldots , T$. Note that, due to the delayed CSIT, the precoding matrix $\mathbf{V}_{ij}[t]$ can only depend upon the outcome of $\bm{\mathcal{G}}^{t-1}$:
\begin{align}
    {V}_{ij}[t] = f_{ijt}^{(T)}\left(\mathcal{G}^{t-1}\right),\label{eq:encode}
\end{align}
where $f_{ijt}^{(T)}\left(\cdot\right)$ is the precoding function that Transmitter~j uses to choose the precoding matrix ${V}_{ij}[t]$ for symbols for Receiver~$i$ at time $t$.
For notational simplicity, we denote the collection of precoding functions used by Transmitter~$j$ as $f_j^{(T)} = \left\{f_{1jt}^{(T)} , f_{2jt}^{(T)} \right\}_{t=1,\ldots,T}$.

Based on this linear precoding, Transmitter~$j$ will then send 
$\vec{\mathbf{x}}_j[t] = \mathbf{V}_{1j}[t]\vec{\mathbf{u}}_{1j}+\mathbf{V}_{2j}[t]\vec{\mathbf{u}}_{2j}$ at time $t$, and Receiver $i$ receives at time $t$ \begin{align*}
    \vec{\mathbf{y}}_{i}[t] 
        ={}& \mathbf{G}_{i1}[t] \vec{\mathbf{x}}_1[t]
        + \mathbf{G}_{i2}[t]\vec{\mathbf{x}}_2[t]
        + \vec{\mathbf{z}}_i[t]\\
        ={}& \mathbf{G}_{i1}[t]\left(\mathbf{V}_{11}[t]\vec{\mathbf{u}}_{11} + \mathbf{V}_{21}[t]\vec{\mathbf{u}}_{21}\right) \
        + \mathbf{G}_{i2}[t]\left(\mathbf{V}_{12}[t]\vec{\mathbf{u}}_{12} + \mathbf{V}_{22}[t]\vec{\mathbf{u}}_{22}\right)
        + \vec{\mathbf{z}}_i[t].
\end{align*}

We denote by $\mathbf{V}_{ij}^T\in\mathbb{C}^{(TM_2)\times \mathbf{m}_{ij}^{(T)}}$ the overall precoding matrix of Transmitter~$j$ for Receiver~$i$, such that 
\begin{align*}
    \mathbf{V}_{ij}^T \triangleq{}&         
        \begin{bmatrix} 
            \mathbf{V}_{ij} [1] \\ \mathbf{V}_{ij} [2] \\ \vdots \\ \mathbf{V}_{ij} [T]
        \end{bmatrix}.
\end{align*}

Based on the above setting, the received signal at Receiver~$i$ ($i\in\{1, 2\}$) after the $T$ time steps of the communication
will be 
\begin{align}
    \vec{\mathbf{y}}_{i}^T 
        ={}& \mathbf{G}_{i1}^T\left(\mathbf{V}_{11}^T\vec{\mathbf{u}}_{11} + \mathbf{V}_{21}^T\vec{\mathbf{u}}_{21}\right)
            + \mathbf{G}_{i2}^T\left(\mathbf{V}_{12}^T\vec{\mathbf{u}}_{12} + \mathbf{V}_{22}^T\vec{\mathbf{u}}_{22}\right)
            + \vec{\mathbf{z}}_i^T,
\end{align}
where $\mathbf{G}_{ij}^T$ is the $TN_i\times TM_j$ block diagonal matrix
\begin{align}
    \mathbf{G}_{ij}^T \triangleq{}& 
        \begin{bmatrix} 
            \mathbf{G}_{ij}[1] & 0 & \ldots & 0\\
            0 & \mathbf{G}_{ij}[2] & \ldots & 0\\
            \vdots & \vdots & \ddots & \vdots \\
            0 & 0 & \ldots & \mathbf{G}_{ij}[T]
        \end{bmatrix}.
\end{align}

Now consider the decoding of $\vec{\mathbf{u}}_{ij}$ at Receiver~$i$. Let  $i^\prime \triangleq 3-i$ and $j^\prime \triangleq 3-j$. Then the corresponding interference subspace at Receiver~$i$will be
\begin{align*}
    \bm{\mathcal{I}}_{ij} = \colspan\left[\mathbf{G}_{ij}^T\mathbf{V}_{i^\prime j}^T \quad  \mathbf{G}_{ij^\prime}^T\mathbf{V}_{ij^\prime}^T \quad \mathbf{G}_{ij^\prime}^T\mathbf{V}_{i^\prime j^\prime}^T\right],
\end{align*}
where $\colspan(\cdot)$ of a matrix is the space spanned its columns. 
For example, the subspace containing all signals that interfere with $\vec{\mathbf{u}}_{11}$ is $\bm{\mathcal{I}}_{11} = \colspan( \left[\mathbf{G}_{11}^T\mathbf{V}_{21}^T \quad  \mathbf{G}_{12}^T\mathbf{V}_{12}^T \quad \mathbf{G}_{12}^T\mathbf{V}_{22}^T\right])$. 
Let $\bm{\mathcal{I}}_{ij}^c = \mathbb{C}^{TN_i} \setminus \bm{\mathcal{I}}_{ij}$ denote the orthogonal complement of $\bm{\mathcal{I}}_{ij}$. Then, in the regime of asymtotically high transmit powers (i.e., ignoring noise), the decodability of information symbols from Transmitter~$j$ to Receiver~$i$ corresponds to the constraints that the image of $\colspan\left(\mathbf{G}_{ij}^T\mathbf{V}_{i j}^T\right)$ on $\bm{\mathcal{I}}_{ij}^c$ has dimension $m_{ij}^{(T)}$:
\begin{align}
    \dim \left(\Proj_{\bm{\mathcal{I}}_{ij}^c}\colspan\left(\mathbf{G}_{ij}^T\mathbf{V}_{i j}^T\right)\right) 
        ={}& \dim\left(\colspan\left(\mathbf{V}_{ij}^T\right)\right)
        ={} m_{ij}^{(T)}.\label{eq:decode1}
\end{align}

Satisfying (\ref{eq:decode1}) for all $i,j\in \{1,2\}$, is also equivalent to satisfying the following conditions for all $i,j\in \{1,2\}$, as proven in \cite{LAS:ARXIV2013}: 
\begin{align}
    \rank[\mathbf{V}_{ij}^T]
    ={}& m_{ij}^{(T)},\label{eq:decode2a}\\
    \rank[\mathbf{G}_{i1}^T\mathbf{V}_{i1}^T]
        + \rank[\mathbf{G}_{i2}^T\mathbf{V}_{i2}^T]
        + \rank[\mathbf{G}_{i1}^T\mathbf{V}_{i^\prime 1}^T \quad \mathbf{G}_{i2}^T\mathbf{V}_{i^\prime 2}^T]
    ={}& \rank[\mathbf{G}_{i1}^T\mathbf{V}_{i1}^T \quad \mathbf{G}_{i2}^T\mathbf{V}_{i2}^T \quad \mathbf{G}_{i1}^T\mathbf{V}_{i^\prime 1}^T \quad \mathbf{G}_{i2}^T\mathbf{V}_{i^\prime 2}^T].\label{eq:decode2}
\end{align}

%is interference-free, and the projection, $\Proj_{\bm{\mathcal{I}}_{ij}^c}\colspan[\mathbf{G}_{ij}^T\mathbf{V}_{i j}^T]$, provides linear combinations of only the desired symbols. 
%A linear scheme is achievable if, for every $i,j$, all $m_{ij}^{(T)}$ symbols may be linearly decoded, which requires the following decodability conditions:\footnote{This assumes that the impact of noise is negligible which holds at high SNR. Thus, throughout the paper, for ease of exposition we omit noise terms in discussion of linear achievability.}

Based on this setting, we now define the linear sum DoF for the MIMO XC.
\begin{definition}\label{def:linDoF}
The DoF four-tuple $(d_{11},d_{12},d_{11},d_{22})$ is linearly achievable if there exists a sequence of linear encoding strategies with blocklength $T=1,2,\ldots$, such that for every $T$, the message sizes $(m_{11}^{(T)},m_{12}^{(T)},m_{21}^{(T)},m_{22}^{(T)})$,  precoding functions $f_{1}^{(T)},f_{2}^{(T)}$, and corresponding precoding matrices $(\mathbf{V}_{11}^T,\mathbf{V}_{12}^T,\mathbf{V}_{21}^T,\mathbf{V}_{22}^T)$, satisfy the condition given in (\ref{eq:decode1}) with probability 1,
and for all $i,j\in\{1,2\}$,
    \begin{align}
        d_{ij} = \lim_{T\rightarrow\infty}\frac{m_{ij}^{(T)}}{T}.
    \end{align}
We also define the linear DoF region $\mathcal{D}_\mathsf{lin}$ as the closure of the set of all achievable 4-tuples $(d_{11},d_{12},d_{11},d_{22})$. Furthermore, the linear sum DoF ($\LDoF$) is then defined as follows:
\begin{align}
    \LDoF \triangleq{} &\maximize \quad d_{11} + d_{12} + d_{21} + d_{22},\label{eq:DoFdef}\\
        &\subjectto \quad (d_{11},d_{12},d_{11},d_{22})\in\mathcal{D}_\mathsf{lin}.\nonumber
\end{align}
\end{definition}

Before stating results, we define the following parameters. Let $Q_{ij}\triangleq \max\{M_j,N_{i^\prime}\}$, and let the parameter $\Gamma_i$ be defined as
 \begin{align}
    \Gamma_{i} \triangleq{}& 
    \max\left\{
    \min\left\{
    \frac{M_1+M_2}{N_{i^\prime}},
    \frac{N_i+N_{i^\prime}}{N_{i^\prime}},
    \frac{Q_{i1}N_{i^\prime}+Q_{i2}\left(N_{i^\prime} + N_i\right)}{N_{i^\prime}(Q_{i2} + N_{i^\prime})}
    \right\},
    1\right\}.\label{eq:rankratio}
 \end{align}
Using these, we now state the main results of this work:
\begin{theorem}
\label{thm:sumDoF}
For the MIMO X-Channel with delayed CSIT, the linear sum degrees of freedom is
\begin{align}
    \LDoF = 
    \begin{cases}
    \frac{\Gamma_1\Gamma_2(N_1+N_2)-\Gamma_1N_2-\Gamma_2N_1}{\Gamma_1\Gamma_2-1} & \text{ if } M_1+M_2 \geq \max\{N_1,N_2\}\cr
    M_1+M_2 & \text{ otherwise}\cr
    \end{cases}
    ,\label{eq:sumDoF}
\end{align}
where $\Gamma_{1}$ and $\Gamma_{2}$ are as defined in (\ref{eq:rankratio}).
\end{theorem}

The main ingredient of the converse for Theorem~\ref{thm:sumDoF} is a rank-ratio inequality which states that the parameter $\Gamma_i$ given in (\ref{eq:rankratio}) bounds the ratio between dimensions of received signal subspaces at the two receivers. Thus, we formally state the inequality in the following lemma and prove it in Section~\ref{sec:rankratio_b}.
\begin{lemma}[Maximum Rank-Ratio]\hfill\\
\label{lem:rankratioUB}
For any linear coding strategy with precoding functions $f_1^{(T)}$ and $f_2^{(T)}$ with corresponding precoding matrices $\mathbf{V}_{i1}^T$ and $\mathbf{V}_{i2}^T$ as defined in (\ref{eq:encode}) and $\Gamma_i$ defined as in (\ref{eq:rankratio}),
\begin{align}
    \frac{\rank[\mathbf{G}_{i1}^T\mathbf{V}_{i1}^T\quad\mathbf{G}_{i2}^T\mathbf{V}_{i2}^T]}
    {\rank[\mathbf{G}_{{i^\prime}1}^T\mathbf{V}_{i1}^T\quad\mathbf{G}_{{i^\prime}2}^T\mathbf{V}_{i2}^T]} 
    \stackrel{a.s.}{\leq} \Gamma_i.\label{eq:rrub}
\end{align}
\end{lemma}

\begin{remark}
Lemma~\ref{lem:rankratioUB} generalizes the rank-ratio inequality given in Lemma~1 of~\cite{LAS:allerton2013,LAS:ARXIV2013} to all antenna configurations: for the case of single antennas (i.e., $M_1=M_2=N_1=N_2=1$), evaluating (\ref{eq:rankratio}) yields $\Gamma_1 =\Gamma_2 =\frac{3}{2}$, which matches the result of~\cite{LAS:allerton2013,LAS:ARXIV2013}.
\end{remark}

In addition to the linear sum DoF result for general antenna configurations, we also apply the new converse and achievable scheme to study the linear DoF \emph{region} of MIMO XC with delayed CSIT and symmetric antenna configurations. Let $\Gamma$ denote the term defined in (\ref{eq:rankratio})  evaluated for $M_1 = M_2 = M$ and  $N_1 = N_2 = N$, and define $Q\triangleq\max\{M,N\}$. Using this, we state our second result:
\begin{theorem}
For symmetric antenna configurations, the linear degrees of freedom region, $\mathcal{D}_\mathsf{lin}$, is the set of all DoF tuples $(d_{11},d_{12},d_{21},d_{22})$ satisfying, for all $i,j \in\{1,2\}$,
\begin{align}
d_{ij} \geq{}& 0,\label{eq:dofreg1}\\
d_{ij} + \frac{\min\{N,M\}}{\min\{2N,M\}}d_{i^\prime j} \leq{}& \min\{N,M\},\label{eq:dofreg2}\\
d_{i1} + d_{i2} + \frac{1}{\Gamma}(d_{i^\prime 1}+d_{i^\prime 2}) \leq{}& \min\{N,2M\},\label{eq:dofreg3}\\
d_{i1} + d_{i2} + \frac{N}{\min\{Q,2N\}}d_{i^\prime j}
	+ \frac{[N-M]_+}{\min\{\Gamma N-M,M\}}d_{i^\prime j^\prime} \leq{}& \min\{N,2M\}.\label{eq:dofreg4}
\end{align}
\label{thm:DoFreg}
\end{theorem}

The converse for Theorem~\ref{thm:sumDoF} may be found in Section~\ref{sec:rankratio} and the linear strategies that achieve the stated linear sum DoF may be found in Section~\ref{sec:scheme}.
Proofs for Theorem~\ref{thm:DoFreg} (both derivation of outer bounds and presentation of achievability) may be found in Section~\ref{sec:dofreg}.

%%%%%%%%%% %%%%%%%%%% %%%%%%%%%% %%%%%%%%%% %%%%%%%%%% 
%%%%%%%%%% %%%%%%%%%% %%%%%%%%%% %%%%%%%%%% %%%%%%%%%% 
%%%%%%%%%% %%%%%%%%%% %%%%%%%%%% %%%%%%%%%% %%%%%%%%%% 
\section{Converse for Theorem~\ref{thm:sumDoF}}
\label{sec:rankratio}\begin{secproof}
In this section, we prove the linear sum DoF converse. We first state the steps describing how Lemma~\ref{lem:rankratioUB} is used to arrive at a linear sum DoF upper bound. We then prove Lemma~\ref{lem:rankratioUB} in Section~\ref{sec:rankratio_b}.

We prove the main results by first applying Lemma~\ref{lem:rankratioUB} to establish, for a linear encoding strategies with blocklength $T$ as described Definition~\ref{def:linDoF}, two bounds on weighted sums of the message sizes. When normalized by blocklength, these yield linear weighted sum DoF bounds, and by averaging the two bounds, we arrive at the an upper bound on the linear sum DoF.

Recall from Definition~\ref{def:linDoF} that a DoF four-tuple $(d_{11},d_{12},d_{11},d_{22})$ is linearly achievable if there exists a sequence of linear encoding strategies with blocklength $T=1,2,\ldots$, such that for every $T$, the message sizes $(m_{11}^{(T)},m_{12}^{(T)},m_{21}^{(T)},m_{22}^{(T)})$,  precoding functions $f_{1}^{(T)},f_{2}^{(T)}$, and corresponding precoding matrices $(\mathbf{V}_{11}^T,\mathbf{V}_{12}^T,\mathbf{V}_{21}^T,\mathbf{V}_{22}^T)$, satisfy the condition given in (\ref{eq:decode1}) with probability 1,
and for all $i,j\in\{1,2\}$, $d_{ij} = \lim_{T\rightarrow\infty}\frac{m_{ij}^{(T)}}{T}$.  
Now for a fixed $T$ and corresponding linear encoding strategy, consider the following weighted sum:
\begin{align}
m_{i1}^{(T)}&+m_{i2}^{(T)} +\Gamma_i(m_{{i^\prime}1}^{(T)}+m_{{i^\prime}2}^{(T)})\nonumber\\
    \stackrel{(a)}{=}{}&  \rank[\mathbf{V}_{i1}^T] + \rank[\mathbf{V}_{i2}^T]
        + \Gamma_i(\rank[\mathbf{V}_{{i^\prime}1}^T] + \rank[\mathbf{V}_{{i^\prime}2}^T]) \\
    \stackrel{a.s.}{=}{}&  \rank[\mathbf{G}_{i1}^T\mathbf{V}_{i1}^T] + \rank[\mathbf{G}_{i2}^T\mathbf{V}_{i2}^T] 
        + \Gamma_i(\rank[\mathbf{G}_{{i^\prime}1}^T\mathbf{V}_{{i^\prime}1}^T] 
        + \rank[\mathbf{G}_{{i^\prime}2}^T\mathbf{V}_{{i^\prime}2}^T]) \\
    \stackrel{\stackrel{(b)}{a.s.}}{=}{}&  \rank[\mathbf{G}_{i1}^T\mathbf{V}_{i1}^T \quad \mathbf{G}_{i2}^T\mathbf{V}_{i2}^T 
        \quad \mathbf{G}_{i1}^T\mathbf{V}_{{i^\prime}1}^T \quad \mathbf{G}_{i2}^T\mathbf{V}_{{i^\prime}2}^T]
        - \rank[ \mathbf{G}_{i1}^T\mathbf{V}_{{i^\prime}1}^T \quad \mathbf{G}_{i2}^T\mathbf{V}_{{i^\prime}2}^T] \nonumber\\
        &- \Gamma_i\rank[ \mathbf{G}_{{i^\prime}1}^T\mathbf{V}_{i1}^T \quad \mathbf{G}_{{i^\prime}2}^T\mathbf{V}_{i2}^T]
        + \Gamma_i\rank[\mathbf{G}_{{i^\prime}1}^T\mathbf{V}_{i1}^T \quad \mathbf{G}_{{i^\prime}2}^T\mathbf{V}_{i2}^T 
        \quad \mathbf{G}_{{i^\prime}1}^T\mathbf{V}_{{i^\prime}1}^T \quad \mathbf{G}_{{i^\prime}2}^T\mathbf{V}_{{i^\prime}2}^T] \\
    \stackrel{(c)}{\leq}{}& \rank[\mathbf{G}_{i1}^T\mathbf{V}_{i1}^T \quad \mathbf{G}_{i2}^T\mathbf{V}_{i2}^T]
        - \Gamma_i\rank[ \mathbf{G}_{{i^\prime}1}^T\mathbf{V}_{i1}^T \quad \mathbf{G}_{{i^\prime}2}^T\mathbf{V}_{i2}^T]\nonumber\\
        &+\Gamma_i\rank[\mathbf{G}_{{i^\prime}1}^T\mathbf{V}_{i1}^T \quad \mathbf{G}_{{i^\prime}2}^T\mathbf{V}_{i2}^T 
        \quad \mathbf{G}_{{i^\prime}1}^T\mathbf{V}_{{i^\prime}1}^T \quad \mathbf{G}_{{i^\prime}2}^T\mathbf{V}_{{i^\prime}2}^T] \label{eq:convprf1}\\
    \stackrel{(d)}{\leq}{}&  \Gamma_i\rank[\mathbf{G}_{{i^\prime}1}^T\mathbf{V}_{i1}^T \quad \mathbf{G}_{{i^\prime}2}^T\mathbf{V}_{i2}^T 
        \quad \mathbf{G}_{{i^\prime}1}^T\mathbf{V}_{{i^\prime}1}^T \quad \mathbf{G}_{{i^\prime}2}^T\mathbf{V}_{{i^\prime}2}^T] \\
    \stackrel{}{\leq}{}&  T\Gamma_i \min\{N_{i^\prime},M_1+M_2\}.\label{eq:convprf2}
\end{align}
Steps (a) and (b) result of the decodability condition (\ref{eq:decode2}). Step (c) is due to submodularity of rank, and in step (d) we applied Lemma~\ref{lem:rankratioUB} to observe that the sum of the first two terms of (\ref{eq:convprf1}) are negative. 

Normalizing (\ref{eq:convprf2}) by $T$ and evaluating for $i=1,2$, as $T\rightarrow\infty$, we have two linear weighted sum DoF bounds:
\begin{align}
    d_{11}+d_{12} +\Gamma_1(d_{21}+d_{22}) \leq{}& \Gamma_1\min\{N_2,M_1+M_2\},\label{eq:WSM1}\\
    d_{21}+d_{22} +\Gamma_2(d_{11}+d_{12}) \leq{}& \Gamma_2\min\{N_1,M_1+M_2\}.\label{eq:WSM2}
\end{align}
Now we consider two cases. In the first case, if $M_1+M_2\leq N_{i^\prime}$ for either $i=1$ or $i=2$, then evaluating (\ref{eq:rankratio}) and either (\ref{eq:WSM1}) or (\ref{eq:WSM2}) depending on the value of $i$, we find $\Gamma_i=1$ and $d_{11} + d_{12} + d_{21} + d_{22} \leq M_1+M_2$.
In the second case, if $M_1+M_2\geq N_{i^\prime}$ for both $i=1$ and $i=2$, we find 
\begin{align}
    \sum_{i,j\in\{1,2\}} d_{ij} 
        \stackrel{(a)}{=}{}& \frac{\Gamma_2-1}{\Gamma_1\Gamma_2-1}(d_{11} + d_{12} + \Gamma_1d_{21} + \Gamma_1d_{22})
           + \frac{\Gamma_1-1}{\Gamma_1\Gamma_2-1}(\Gamma_2 d_{11} + \Gamma_2 d_{12} + d_{21} + d_{22}) \\
    \stackrel{(b)}{\leq}{}& \frac{(\Gamma_2-1)\Gamma_1 N_2 + (\Gamma_1-1) \Gamma_2 N_1}{\Gamma_1\Gamma_2-1} \\
    ={}& \frac{\Gamma_1\Gamma_2(N_1+N_2)-\Gamma_1N_2-\Gamma_2N_1}{\Gamma_1\Gamma_2-1},
\end{align}
where in step (a) we factored the sum DoF into two parts, and in step (b) we applied the inequalities of both (\ref{eq:WSM1}) and (\ref{eq:WSM2}). This is exactly the bound given in (\ref{eq:sumDoF}).
\end{secproof}

\subsection{Proof of Lemma~\ref{lem:rankratioUB}}
\label{sec:rankratio_b}
\begin{secproof}
To prove Lemma~\ref{lem:rankratioUB}, we derive two upper bounds on the rank-ratio between the outputs at receivers. Evaluating the min of the two bounds yields exactly the definition given in (\ref{eq:rankratio}). The first bound enhances the network by adding cooperation between the two transmitters and applies a rank-ratio inequality for the MIMO BC. The second bound results from focusing on the difference between ranks at each receiver normalized by number of antennas at each receiver. We present a lemma stating how large this difference can be with delayed CSIT, and then use it to arrive at the second rank-ratio bound.

\subsubsection*{Bound~1}
To arrive at the first bound, we first state the following lemma which is a rank-ratio inequality for MIMO BCs, whose proof may be found in Appendix~\ref{append:BCrankratio}. 
\begin{lemma}
\label{lem:BCratio}
Consider a 2-user MIMO BC with delayed CSIT, $M$ antennas at the transmitter, $N_1$ antennas at Receiver~1, and $N_2$ antennas at Receiver~2. For any linear coding strategy $f^T$ and associated $\mathbf{V}_i^T$ defined in a manner similar to (\ref{eq:encode}), we have
\begin{align}
\frac{\rank[\mathbf{G}_{i}^T\mathbf{V}_i^T]}{\rank[\mathbf{G}_{i^\prime}^T\mathbf{V}_i^T]}
    \stackrel{a.s.}{\leq} \frac{\min\{Q_{i},N_i+N_{i^\prime}\}}{N_{i^\prime}}.
\end{align}
\end{lemma}

We now allow transmitters in the MIMO XC with delayed CSIT to share messages. Cooperation emulates a single transmitter with $M_1+M_2$ antennas, and by applying Lemma~\ref{lem:BCratio}, we arrive at the cooperative rank-ratio upper bound:
\begin{align}
\text{Bound~1: }\quad\frac{\rank[\mathbf{G}_{i1}^T\mathbf{V}_{1}^T\quad \mathbf{G}_{i2}^T\mathbf{V}_{2}^T]}
{\rank[\mathbf{G}_{i^\prime 1}^T\mathbf{V}_{1}^T\quad \mathbf{G}_{i^\prime 2}^T\mathbf{V}_{2}^T]} 
    \stackrel{a.s.}{\leq}& 
%    \frac{\min\{\max\{M_1+M_2,N_i^\prime\},N_1+N_2\}}{N_i^\prime} \\
%    ={}& \min\left\{\max\left\{\frac{M_1+M_2}{N_i^\prime},1\right\},\frac{N_1+N_2}{N_i^\prime}\right\} \\
%    ={}& 
    \max\left\{\min\left\{\frac{M_1+M_2}{N_i^\prime},\frac{N_1+N_2}{N_i^\prime}\right\},1\right\}.\label{eq:rankratio_UB1}
\end{align}

\subsubsection*{Bound~2}

To construct the second bound, we focus on the difference of the normalized ranks at each receiver, which will eventually result in a bound on the rank-ratio. We begin by stating the following inequality, which is proven in Appendix~\ref{append:normdiff}.
\begin{lemma}%[Normalized Rank Difference]
\label{lem:diff}
For any linear encoding strategies $f_1^T$ and $f_2^T$ and corresponding $\mathbf{V}_{i1}^T$ and $\mathbf{V}_{i2}^T$ as defined in (\ref{eq:encode}), we have
\begin{align}
    \frac{\rank[\mathbf{G}_{i1}^T\mathbf{V}_{i1}^T\quad\mathbf{G}_{i2}^T\mathbf{V}_{i2}^T]}{N_i}
        &- \frac{\rank[\mathbf{G}_{i^\prime 1}^T\mathbf{V}_{i1}^T\quad\mathbf{G}_{i^\prime 2}^T\mathbf{V}_{i2}^T]}{N_{i^\prime}} 
        \nonumber\\
    \stackrel{a.s.}{\leq}{} \frac{1}{N_i} &\left(\rank[\mathbf{G}_{i^\prime 1}^T\mathbf{V}_{i1}^T\quad\mathbf{G}_{i^\prime 2}^T\mathbf{V}_{i2}^T]
        - \rank[\mathbf{G}_{i^\prime 1}^T\mathbf{V}_{i1}^T]
        + \rank[\mathbf{G}_{i^\prime 1}^T\mathbf{V}_{i1}^T\quad\mathbf{G}_{i^\prime 2}^T\mathbf{V}_{i2}^T]
        - \rank[\mathbf{G}_{i^\prime 2}^T\mathbf{V}_{i2}^T]\right).\label{eq:preratio1}
\end{align}
\end{lemma}

%The following relation, which holds due to submodularity of the rank operation, will also be used.
%\begin{align}
%     \rank[\mathbf{G}_{i1}^T\mathbf{V}_{1}^T\quad\mathbf{G}_{i2}^T\mathbf{V}_{2}^T]
%     \leq \rank[\mathbf{G}_{i1}^T\mathbf{V}_{1}^T]
%         + \rank[\mathbf{G}_{i2}^T\mathbf{V}_{2}^T].\label{eq:preratio2}
%\end{align}

We now observe the following:
\begin{align}
    \rank[\mathbf{G}_{{i^\prime}1}^T\mathbf{V}_{i1}^T] + \rank[\mathbf{G}_{{i^\prime}2}^T\mathbf{V}_{i2}^T] 
        \stackrel{\stackrel{(a)}{a.s.}}{\geq}{}& \frac{N_{i^\prime}}{Q_{i 1}}\rank[\mathbf{G}_{i1}^T\mathbf{V}_{i1}^T] 
            + \frac{N_{i^\prime}}{Q_{i2}}\rank[\mathbf{G}_{i2}^T\mathbf{V}_{i2}^T] \\
        \stackrel{(b)}{\geq}{}& 
            \left(\frac{N_{i^\prime}}{Q_{i1}}-\frac{N_{i^\prime}}{Q_{i2}}\right)\rank[\mathbf{G}_{i1}^T\mathbf{V}_{i1}^T]
            + \frac{N_{i^\prime}}{Q_{i2}}\rank[\mathbf{G}_{i1}^T\mathbf{V}_{i1}^T\quad \mathbf{G}_{i2}^T\mathbf{V}_{i2}^T] \\
        ={}& \frac{N_{i^\prime}}{Q_{i2}}
            \rank[\mathbf{G}_{i1}^T\mathbf{V}_{i1}^T\quad \mathbf{G}_{i2}^T\mathbf{V}_{i2}^T] 
            - \left(\frac{N_{i^\prime}}{Q_{i2}}-\frac{N_{i^\prime}}{Q_{i1}}\right)\rank[\mathbf{G}_{i1}^T\mathbf{V}_{i1}^T] \\
        \stackrel{(c)}{\geq}{}& \frac{N_{i^\prime}}{Q_{i2}}
            \rank[\mathbf{G}_{i1}^T\mathbf{V}_{i1}^T\quad \mathbf{G}_{i2}^T\mathbf{V}_{i2}^T] 
            - \left(\frac{Q_{i1}}{Q_{i2}}-1\right)\rank[\mathbf{G}_{i^\prime 1}^T\mathbf{V}_{i1}^T] \\
        \stackrel{(d)}{\geq}{}& 
            \frac{N_{i^\prime}}{Q_{i2}}\rank[\mathbf{G}_{i1}^T\mathbf{V}_{i1}^T\quad \mathbf{G}_{i2}^T\mathbf{V}_{i2}^T] 
            - \left(\frac{Q_{i1}}{Q_{i2}}-1\right)\rank[\mathbf{G}_{{i^\prime}1}^T\mathbf{V}_{i1}^T\quad \mathbf{G}_{{i^\prime}2}^T\mathbf{V}_{i2}^T].\label{eq:ncbound_p1}
\end{align}
In step (a), we applied Lemma~\ref{lem:BCratio} to both terms. Step (b) results from submodularity of the rank function. In step (c), we applied Lemma~\ref{lem:BCratio} to the second term, and finally, we noted the monotonicity of the rank operation in step (d). The inequality (\ref{eq:ncbound_p1}) may be equivalently stated as
\begin{align}
    \frac{N_{i^\prime}}{Q_{i2}}\rank[\mathbf{G}_{i1}^T\mathbf{V}_{i1}^T\quad \mathbf{G}_{i2}^T\mathbf{V}_{i2}^T] 
    \stackrel{a.s.}{\leq}{}&
        \rank[\mathbf{G}_{{i^\prime}1}^T\mathbf{V}_{i1}^T] + \rank[\mathbf{G}_{{i^\prime}2}^T\mathbf{V}_{i2}^T] 
        + \left(\frac{Q_{i1}}{Q_{i2}}-1\right)\rank[\mathbf{G}_{{i^\prime}1}^T\mathbf{V}_{i1}^T
            \quad \mathbf{G}_{{i^\prime}2}^T\mathbf{V}_{i2}^T],
\label{eq:ncbound_p1a}
\end{align}
which scaled by $\frac{1}{N_i}$ and summing with (\ref{eq:preratio1}) from Lemma~\ref{lem:diff} yields
\begin{align}
    \frac{1}{N_i}\left(1+\frac{N_{i^\prime}}{Q_{i2}}\right)&\rank[\mathbf{G}_{i1}^n\mathbf{V}_{i1}^n\quad\mathbf{G}_{i2}^n\mathbf{V}_{i2}^n]
            -\frac{1}{N_{i^\prime}}\rank[\mathbf{G}_{{i^\prime}1}^n\mathbf{V}_{i1}^n\quad\mathbf{G}_{{i^\prime}2}^n\mathbf{V}_{i2}^n]\nonumber\\
        \stackrel{a.s.}{\leq}{}& \frac{2}{N_i}\rank[\mathbf{G}_{{i^\prime}1}^n\mathbf{V}_{i1}^n\quad\mathbf{G}_{{i^\prime}2}^n\mathbf{V}_{i2}^n] 
            +\frac{1}{N_i}\left(\frac{Q_{i1}}{Q_{i2}}-1\right)\rank[\mathbf{G}_{{i^\prime}1}^n\mathbf{V}_{i1}^n\quad \mathbf{G}_{{i^\prime}2}^n\mathbf{V}_{i2}^n]\\ 
        ={}& \frac{1 + \frac{Q_{i1}}{Q_{i2}}}{N_i}\rank[\mathbf{G}_{{i^\prime}1}^n\mathbf{V}_{i1}^n\quad\mathbf{G}_{{i^\prime}2}^n\mathbf{V}_{i2}^n],
\end{align}
or equivalently
\begin{align}
    \text{Bound~2: }\quad\frac{ \rank[\mathbf{G}_{i1}^n\mathbf{V}_{i1}^n\quad\mathbf{G}_{i2}^n\mathbf{V}_{i2}^n] }
            { \rank[\mathbf{G}_{{i^\prime}1}^n\mathbf{V}_{i1}^n\quad\mathbf{G}_{{i^\prime}2}^n\mathbf{V}_{i2}^n] }
        \stackrel{a.s}{\leq}{}& 
%        \frac{1 + \frac{Q_{i1}}{Q_{i2}}+\frac{N_i}{N_{i^\prime}}}{1+\frac{N_{i^\prime}}{Q_{i2}}} \\
%        ={}& 
        \frac{Q_{i1}N_{i^\prime}+Q_{i2}\left(N_{i^\prime} + N_i\right)}{N_{i^\prime}(Q_{i2}+N_{i^\prime})}.\label{eq:rankratio_UB2}
\end{align}

To complete the proof of Lemma~\ref{lem:rankratioUB}, we take the minimum of (\ref{eq:rankratio_UB1}) and (\ref{eq:rankratio_UB2}), which gives (\ref{eq:rankratio}).
\end{secproof}

%%%%%%%%%% %%%%%%%%%% %%%%%%%%%% %%%%%%%%%% %%%%%%%%%% 
%%%%%%%%%% %%%%%%%%%% %%%%%%%%%% %%%%%%%%%% %%%%%%%%%% 
%%%%%%%%%% %%%%%%%%%% %%%%%%%%%% %%%%%%%%%% %%%%%%%%%% 
%%%%%%%%%% %%%%%%%%%% %%%%%%%%%% %%%%%%%%%% %%%%%%%%%% 
\section{Transmission Strategy}
\label{sec:scheme}

We now present a linear encoding strategy that, for any antenna configuration, achieves the sum DoF stated in Theorem~\ref{thm:sumDoF}. We first describe in Section~\ref{sec:Gdesc} the general structure of the three-phase strategy, how structure is related to $\Gamma_1$ and $\Gamma_2$ defined in (\ref{eq:rankratio}), and how delayed CSIT is used to determine the input of Transmitter~2 during Phases~1 and 2 and inputs of both transmitters during Phase~3. While describing the strategy, we repeatedly reference a simple example network (one where $M_1=M_2=3$ and $N_1=N_2=2$) to illustrate key points. In Section~\ref{sec:achsumdof} we compute the achieved sum DoF and show that our strategy achieves (\ref{eq:sumDoF}) for all antenna configurations. The special case of symmetric antenna configurations is discussed at the end, in Section~\ref{sec:sym}, with comparison to the prior art.

\subsection{General Description}
\label{sec:Gdesc}
Our transmission strategy consists of 3 phases. During Phase~1 (Phase~2), symbols desired by Receiver~1 (Receiver~2) are sent by both transmitters. In Phase~3, we use a retrospective interference alignment approach and each transmitter sends linear combinations of past transmissions such that each receiver receives a sum of a desired message signal and an undesired, but already known, interference signal.

%Phases~1 and 2 are further divided into transmission rounds. Symbols used to generate transmissions within each round are independent of other rounds. We let $\kappa_1$ and $\kappa_2$ denote the number of rounds in Phase~1 and 2 respectively. Later, we describe how $\kappa_1$ and $\kappa_2$ are chosen to maximize achieved sum DoF.
%Before describing the structure of transmissions in each phase, we point out two key features of our strategy:
%\begin{enumerate}
%\item For each round in Phase~1 (Phase~2), the number of channel uses per round and the numbers of symbols sent by each transmitter are functions of $\Gamma_1$ ($\Gamma_2$). 
%\item Phase~3 applies retrospective interference alignment to simultaneously deliver desired message information to both receivers. Therefore, we exploit delayed CSIT in Phases~1 and 2 in order to \emph{maximize the number of Phase~3 transmissions}. Although both transmitters are given delayed CSIT, only Transmitter~2 uses it during Phases~1 and 2. Transmitter~1 does not use its delayed CSIT until Phase~3. 
%\end{enumerate}

\subsubsection*{\bf Phases~1 and 2}

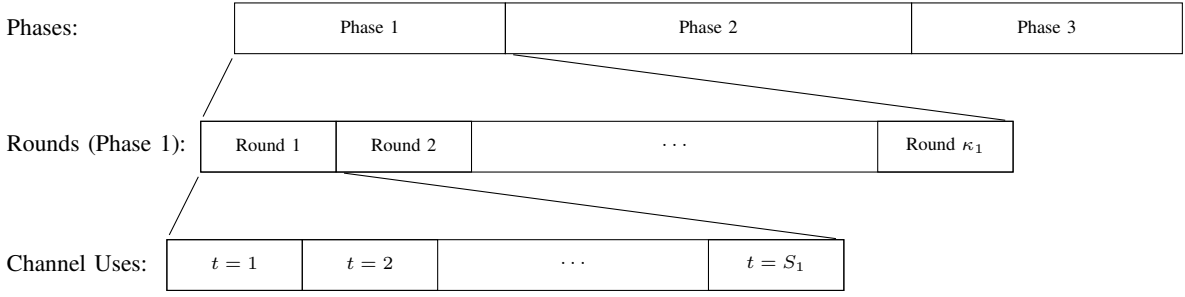
\begin{figure}[ht]
\centering
\begin{tikzpicture}[font=\scriptsize,scale=0.9]
    \draw (0,0) rectangle (4,0.75);
    \draw (2,0.4) node {Phase~1};
    \draw (4,0) rectangle (10,0.75);
    \draw (7,0.4) node {Phase~2};
    \draw (10,0) rectangle (14,0.75);
    \draw (12,0.4) node {Phase~3};
    
    \draw[shorten <=2pt,shorten >=2pt] (0,0) -- (-0.5,-1);
    \draw[shorten <=3pt,shorten >=3pt] (4,0) -- (11.5,-1);
    \draw (-0.5,-1.75) rectangle (11.5,-1);
    \draw (-0.5,-1.75) rectangle (1.5,-1);
    \draw (0.5,-1.35) node{Round 1};
    \draw (1.5,-1.75) rectangle (3.5,-1);
    \draw (2.5,-1.35) node{Round 2};
        \draw (6.5,-1.35) node{$\ldots$};
    \draw (9.5,-1.75) rectangle (11.5,-1);
    \draw (10.5,-1.35) node{Round $\kappa_1$};
    
    \draw[shorten <=2pt,shorten >=2pt] (-0.5,-1.75) -- (-1,-2.75);
    \draw[shorten <=3pt,shorten >=3pt] (1.5,-1.75) -- (9,-2.75);
    \draw (-1,-3.5) rectangle (9,-2.75);
    \draw (-1,-3.5) rectangle (1,-2.75);
    \draw (0,-3.1) node{$t=1$};
    \draw (1,-3.5) rectangle (3,-2.75);
    \draw (2,-3.1) node{$t=2$};
        \draw (5,-3.1) node{$\ldots$};
    \draw (7,-3.5) rectangle (9,-2.75);
    \draw (8,-3.1) node{$t=S_1$};
    
    \draw (-3.5,-3.1) node[right]{\small Channel Uses:};
    \draw (-3.5,-1.35) node[right]{\small Rounds (Phase~1):};
    \draw (-3.5,0.4) node[right]{\small Phases:};
\end{tikzpicture}
\caption{Temporal structure of proposed linear encoding strategy: The overall schemes is made up of three transmission phases. Phases~1 and 2 are in turn made up of $\kappa_1$ and $\kappa_2$ rounds, where rounds in Phase~1 are made up of $S_1$ channel uses and rounds in Phase~2 are made up of $S_2$ channel uses.}\label{fig:timedecomp}
\end{figure}

We first describe the overall temporal structure of the transmission strategy, as illustrated in Figure~\ref{fig:timedecomp}. Of the three transmission phases, Phase~$i$ ($i=1,2$) is further divided into $\kappa_i$ rounds. Symbols transmitted in any single round of Phase~$i$ are desired by Receiver~$i$, and are independent of those used in other rounds. Each round of Phase~$i$ is further made up of of $S_i$ channel uses. %The number channel uses per round ($S_i$) and how symbols are transmitted 

We now proceed to describe the structure of a single round of transmission in Phase~$i$ for antenna configurations where $\Gamma_i N_{i^\prime} > M_1$.
The alternative antenna configurations (i.e., those where $\Gamma_iN_{i^\prime}\leq M_1$) uses a simpler scheme  and its description may be found in Appendix~\ref{app:othercase}. We point out that in the simpler scheme, Transmitter~2 remains silent during all phases of transmission, and thus delayed CSIT is not used in Phases~1 and 2.

When $\Gamma_i N_{i^\prime} > M_1$, let $\xi_i\in\mathbb{N}$ be the smallest positive integer such that 
\begin{align}
    S_i ={}& \frac{M_2\xi_i}{\min\{(\Gamma_i N_{i^\prime}-M_1),M_2\}},\label{eq:chanperrnd}
\end{align}
is an integer. Notice that $\xi_i\leq S_i$. 
Let $\vec{\mathbf{u}}_{ij}[k]$ be a vector containing the symbols sent from Transmitter~$j$ during only Round $k$. Our strategy dictates that,  the total number of symbols sent per round by Transmitter~1 is $|\vec{\mathbf{u}}_{i1}[k]| = M_1S_i$, and the total number sent per round by Transmitter~2 is $|\vec{\mathbf{u}}_{i2}[k]|M_2\xi_i$. 

During each of the $S_i$ channel uses in the round, Transmitter~1 broadcasts a new symbol on each antenna, for a total of $M_1$ symbols sent per channel use. During only the first $\xi_i$ channel uses, Transmitter~2 also broadcasts a new symbol on each antenna, for a total of $M_2$ symbols sent per channel use. 

So far, no CSIT has been used in the round. We now describe precisely how delayed CSIT is used, and stress that it is only used by Transmitter~2. After the initial $\xi_i$ channel uses of Round~$k$, the image of Transmitter~2's symbols, $\vec{\mathbf{u}}_{i2}[k]$, at Receiver~$i^\prime$ is the linear transformation, $\bm{\Phi}_i[k]\vec{\mathbf{u}}_{i2}[k]$, where $\bm{\Phi}_i[k]$ is defined as
\begin{align}
    \bm{\Phi}_i[k] = \begin{bmatrix}
            \mathbf{G}_{i^\prime 2}[t_0[k]]   \begin{bmatrix} I_{M_2\times M_2} & 0_{M_2\times M_2(\xi_i-1)}\end{bmatrix}\\
            \mathbf{G}_{i^\prime 2}[t_0[k]+1] \begin{bmatrix} 0_{M_2\times M_2} & I_{M_2\times M_2} & 0_{M_2\times M_2(\xi_i-2)}\end{bmatrix}\\
            \vdots\\
            \mathbf{G}_{i^\prime 2}[t_0[k]+\xi_i-1]\begin{bmatrix} 0_{M_2\times M_2(\xi_i-1)} & I_{M_2\times M_2}\end{bmatrix}
        \end{bmatrix},
\end{align}
with $t_0[k]$ denoting the first time index for the $k$-th round of transmission. The matrix $\bm{\Phi}_i[k]$ is a $N_{i^\prime}\xi_i\times M_2\xi_i$ matrix, which is almost surely full rank (i.e. $\rank[\bm{\Phi}_i[k]]\stackrel{a.s.}{=} \xi_i\min\{M_2,N_{i^\prime}\}$).

Because Transmitter~2 has delayed CSIT, it knows $\bm{\Phi}_i[k]$ for any time $t\geq t_0[k]+\xi_i$. 
In the remaining $S_i-\xi_i$ channel uses of the round, Transmitter~2 sends linearly precoded combinations of the previously transmitted symbols on each antenna. 
Precoding vectors are created from elements of the set $\{\vec{\bm{\phi}}_{i,1}[k],\vec{\bm{\phi}}_{i,2}[k],\ldots,\vec{\bm{\phi}}_{i,\xi_i N_{i^\prime}}[k]\}$, where $\vec{\bm{\phi}}_{i,\ell}[k]$ denotes the $\ell$-th row of $\bm{\Phi}_i[k]$. 
Each element is scaled in order to satisfy the transmit power constraint and applied sequentially (e.g., the first antenna at Transmitter~2 at time $t_0[k]+\xi_i$ transmits $\vec{\bm{\phi}}_{i,1}[k]\vec{\mathbf{u}}_{i2}[k]$, the second antenna at time $t_0[k]+\xi_i$ transmits $\vec{\bm{\phi}}_{i,2}[k]\vec{\mathbf{u}}_{i2}[k]$, etc.). 
When the elements of $\{\vec{\bm{\phi}}_{i,1}[k],\vec{\bm{\phi}}_{i,2}[k],\ldots,\vec{\bm{\phi}}_{i,\xi_i N_{i^\prime}}[k]\}$ have been exhausted, they are repeated, and this repetition continues until the end of the round. 

We now briefly justify our method of using delayed CSIT during Phases~1 and 2. Our method minimizes the dimension of the image of Transmitter~2 symbols at Receiver~$i^\prime$. Consider the vectors $\vec{\bm{\phi}}_{i,\ell}[k]$, which when scaled are used as precoding vectors at Transmitter~2 for the later portion of each round. 
By definition, we know that $\{\vec{\bm{\phi}}_{i,1}[k],\ldots,\vec{\bm{\phi}}_{i,{\xi_i N_{i^\prime}}}[k]\}$ is a spanning set for the rowspan of $\bm{\Phi}_i[k]$. 
When $M_2\leq N_{i^\prime}$, the set $\{\vec{\bm{\phi}}_{i,1}[k],\ldots,\vec{\bm{\phi}}_{i,\xi_i N_{i^\prime}}[k]\}$ is also almost surely a basis (i.e., when $M_2\leq N_{i^\prime}$ the rows of $\bm{\Phi}_i[k]$ are almost surely linearly independent). Using basis elements as precoding vectors for each antenna ensures that, \emph{regardless of channel coefficients}, the dimension of the image of $\vec{\mathbf{u}}_{i2}[k]$ received at Receiver~$i^\prime$, from $t_0[k]+\xi_i$ onward, does not increase the dimension beyond what was received from $t_0[k]$ to $t_0[k]+\xi_i-1$. The following example illustrates this key feature:

\begin{example*}[Phase~1]
Consider the MIMO XC depicted in Figure~\ref{fig:examplescheme}, where $M_1=M_2=3$ and $N_1=N_2=2$. From evaluating (\ref{eq:rankratio}) and (\ref{eq:chanperrnd}) we have $\Gamma_1 = \Gamma_2 = \frac{9}{5}$, $S_1= S_2 = 5$ and $\xi_1 = \xi_2=1$. Therefore, a round of Phase~1 lasts 5 channel uses, during which Transmitter~1 sends $S_1M_1=15$ symbols and Transmitter~2 sends $\xi_1M_2=3$. 

For the first channel use, both transmitters broadcast 3 independent symbols (Figure~\ref{fig:exscheme1}). During the 2nd--5th channel uses Transmitter~1 continues to broadcast 3 new symbols each channel use. Transmitter~2 instead applies delayed CSIT to retransmit the two rows of $\bm{\Phi}_i[k]\vec{\mathbf{u}}_{i2}[k]$ as seen in Figure~\ref{fig:exscheme2}. 

We denote with $a_{i2}^{\ell m}[t]$ the effective channel gain that results from sending the same signal through multiple antennas (e.g., in Figure~\ref{fig:exscheme2}, $a_{i2}^{\ell 1}[t_0[k]+1] = g_{i2}^{\ell 1}[t_0[k]+1] + g_{i2}^{\ell 3}[t_0[k]+1]$), which allows us to illustrate in Figure~\ref{fig:exscheme2} that, regardless of the channel, the image of Transmitter~2's symbols at Receiver~2 may be expressed as a function of only two variables, $\vec{\bm{\phi}}_{1,1}[k]\vec{\mathbf{u}}_{12}[k]$ and $\vec{\bm{\phi}}_{1,2}[k]\vec{\mathbf{u}}_{12}[k]$, instead of the three symbols, $\mathbf{u}_{12}^1[k]$, $\mathbf{u}_{12}^2[k]$, and $\mathbf{u}_{12}^3[k]$.

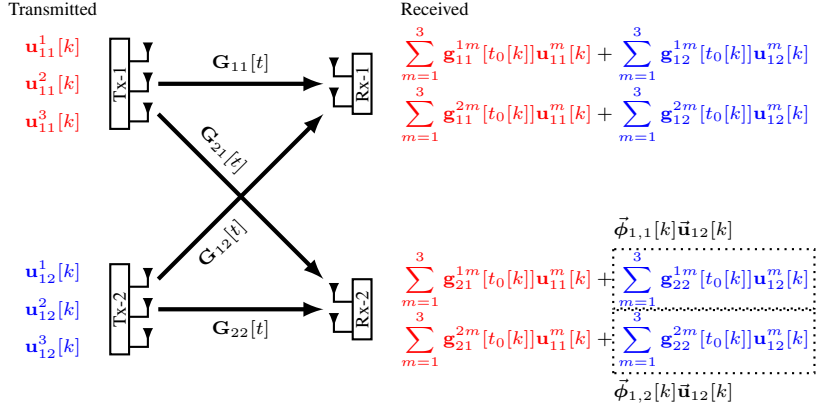
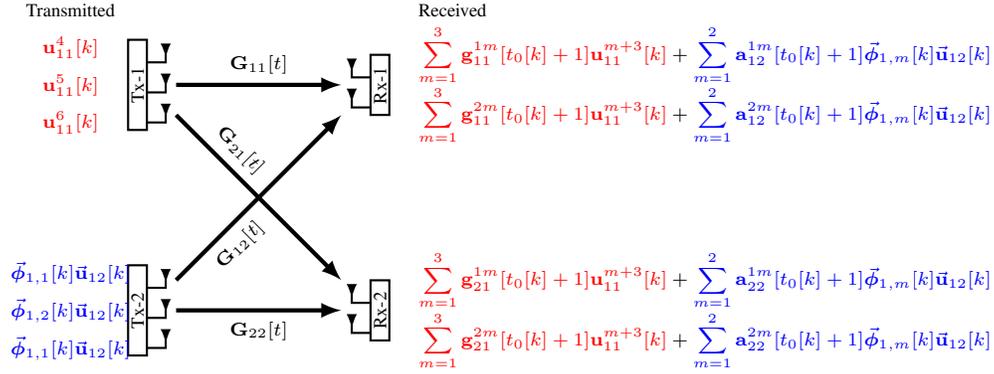
\begin{figure}[ht]
\centering
\begin{subfigure}[b]{\textwidth}\centering
    \begin{tikzpicture}[scale=1,font=\scriptsize]
    \node (a) at (0,3) {\begin{tikzpicture}[yscale=1, xscale=1]
        \draw (-0.125,0.2) node[rotate=90] {Tx-1};
        \draw[thick] (0,0.8) -- (0,-0.4) -- (-0.25,-0.4) -- (-0.25,0.8) -- cycle;
        \draw[thick,-latex reversed] (0,-0.3) -- (0.25,-0.3) -- (0.25,-0.05);
        \draw[thick,-latex reversed] (0,0.1) -- (0.25,0.1) -- (0.25,0.35);
        \draw[thick,-latex reversed] (0,0.5) -- (0.25,0.5) -- (0.25,0.75);
    \end{tikzpicture}};
    \node (b) at (0,0) {\begin{tikzpicture}[yscale=1, xscale=1]
        \draw (-0.125,0.2) node[rotate=90] {Tx-2};
        \draw[thick] (0,0.8) -- (0,-0.4) -- (-0.25,-0.4) -- (-0.25,0.8) -- cycle;
        \draw[thick,-latex reversed] (0,-0.3) -- (0.25,-0.3) -- (0.25,-0.05);
        \draw[thick,-latex reversed] (0,0.1) -- (0.25,0.1) -- (0.25,0.35);
        \draw[thick,-latex reversed] (0,0.5) -- (0.25,0.5) -- (0.25,0.75);
    \end{tikzpicture}};
    \node (c) at (3,3) {\begin{tikzpicture}[yscale=1, xscale=1]
        \draw (0.125,0) node[rotate=90] {Rx-1};
        \draw[thick] (0,0.4) -- (0,-0.4) -- (0.25,-0.4) -- (0.25,0.4) -- cycle;
        \draw[thick,-latex reversed] (0,-0.3) -- (-0.25,-0.3) -- (-0.25,-0.05);
        \draw[thick,-latex reversed] (0,0.1) -- (-0.25,0.1) -- (-0.25,0.35);
    \end{tikzpicture}};
    \node (d) at (3,0) {\begin{tikzpicture}[yscale=1, xscale=1]
        \draw (0.125,0) node[rotate=90] {Rx-2};
        \draw[thick] (0,0.4) -- (0,-0.4) -- (0.25,-0.4) -- (0.25,0.4) -- cycle;
        \draw[thick,-latex reversed] (0,-0.3) -- (-0.25,-0.3) -- (-0.25,-0.05);
        \draw[thick,-latex reversed] (0,0.1) -- (-0.25,0.1) -- (-0.25,0.35);
    \end{tikzpicture}};
    \draw[ultra thick,-latex] (a)--node[above]{$\mathbf{G}_{11}[t]$} (c);
    \draw[ultra thick,-latex] (a)--node[above,pos=0.3,sloped]{$\mathbf{G}_{21}[t]$} (d);
    \draw[ultra thick,-latex] (b)--node[below,pos=0.3,sloped]{$\mathbf{G}_{12}[t]$} (c);
    \draw[ultra thick,-latex] (b)--node[below]{$\mathbf{G}_{22}[t]$} (d);
\draw[dashed]  (-1,4) node[] {Transmitted};
\node[red] (A) at (-1,3.5) {$\mathbf{u}_{11}^1[k]$};
\node[red] (B) at (-1,3) {$\mathbf{u}_{11}^2[k]$};
\node[red] (C) at (-1,2.5) {$\mathbf{u}_{11}^3[k]$};
\node[blue] (X) at (-1,0.5) {$\mathbf{u}_{12}^1[k]$};
\node[blue] (Y) at (-1,0) {$\mathbf{u}_{12}^2[k]$};
\node[blue] (Z) at (-1,-0.5) {$\mathbf{u}_{12}^3[k]$};

\draw[dashed]  (3.5,4) node[right] {Received};
\node[right] (R111_1) at (3.5,3.4) 
    {$\displaystyle{\color{red}\sum_{m=1}^3 \mathbf{g}_{11}^{1 m}[t_0[k]]\mathbf{u}_{11}^m[k]} + {\color{blue}\sum_{m=1}^3 \mathbf{g}_{12}^{1 m}[t_0[k]]\mathbf{u}_{12}^m[k]}$};
\node[right] (R112_1) at (3.5,2.6) 
    {$\displaystyle{\color{red}\sum_{m=1}^3 \mathbf{g}_{11}^{2 m}[t_0[k]]\mathbf{u}_{11}^m[k]} + {\color{blue}\sum_{m=1}^3 \mathbf{g}_{12}^{2 m}[t_0[k]]\mathbf{u}_{12}^m[k]}$};
\node[right] (R111_1) at (3.5,0.4) 
    {$\displaystyle{\color{red}\sum_{m=1}^3 \mathbf{g}_{21}^{1 m}[t_0[k]]\mathbf{u}_{11}^m[k]} + {\color{blue}\sum_{m=1}^3 \mathbf{g}_{22}^{1 m}[t_0[k]]\mathbf{u}_{12}^m[k]}$};
\node[right] (R112_1) at (3.5,-0.4) 
    {$\displaystyle{\color{red}\sum_{m=1}^3 \mathbf{g}_{21}^{2 m}[t_0[k]]\mathbf{u}_{11}^m[k]} + {\color{blue}\sum_{m=1}^3 \mathbf{g}_{22}^{2 m}[t_0[k]]\mathbf{u}_{12}^m[k]}$};

    \draw[dotted,thick] (6.45,0.8) rectangle (9.125,0.01);
    \draw (7.25,0.8) node[above] {$\vec{\bm{\phi}}_{1,1}[k]\vec{\mathbf{u}}_{12}[k]$};
    \draw[dotted,thick] (6.45,-0.85) rectangle (9.125,-0.01);
    \draw (7.25,-0.8) node[below] {$\vec{\bm{\phi}}_{1,2}[k]\vec{\mathbf{u}}_{12}[k]$};
    \draw[white](-1.8,3) -- (-1.8,0);
    \draw[white](10.2,3) -- (10.2,0);
\end{tikzpicture}\caption{First channel use of Phase~1, Round~$k$ (i.e., $t=t_0[k]$)}\label{fig:exscheme1}
\end{subfigure}
\begin{subfigure}[b]{\textwidth}\centering
    \begin{tikzpicture}[scale=1,font=\scriptsize]
    \node (a) at (0,3) {\begin{tikzpicture}[yscale=1, xscale=1]
        \draw (-0.125,0.2) node[rotate=90] {Tx-1};
        \draw[thick] (0,0.8) -- (0,-0.4) -- (-0.25,-0.4) -- (-0.25,0.8) -- cycle;
        \draw[thick,-latex reversed] (0,-0.3) -- (0.25,-0.3) -- (0.25,-0.05);
        \draw[thick,-latex reversed] (0,0.1) -- (0.25,0.1) -- (0.25,0.35);
        \draw[thick,-latex reversed] (0,0.5) -- (0.25,0.5) -- (0.25,0.75);
    \end{tikzpicture}};
    \node (b) at (0,0) {\begin{tikzpicture}[yscale=1, xscale=1]
        \draw (-0.125,0.2) node[rotate=90] {Tx-2};
        \draw[thick] (0,0.8) -- (0,-0.4) -- (-0.25,-0.4) -- (-0.25,0.8) -- cycle;
        \draw[thick,-latex reversed] (0,-0.3) -- (0.25,-0.3) -- (0.25,-0.05);
        \draw[thick,-latex reversed] (0,0.1) -- (0.25,0.1) -- (0.25,0.35);
        \draw[thick,-latex reversed] (0,0.5) -- (0.25,0.5) -- (0.25,0.75);
    \end{tikzpicture}};
    \node (c) at (3,3) {\begin{tikzpicture}[yscale=1, xscale=1]
        \draw (0.125,0) node[rotate=90] {Rx-1};
        \draw[thick] (0,0.4) -- (0,-0.4) -- (0.25,-0.4) -- (0.25,0.4) -- cycle;
        \draw[thick,-latex reversed] (0,-0.3) -- (-0.25,-0.3) -- (-0.25,-0.05);
        \draw[thick,-latex reversed] (0,0.1) -- (-0.25,0.1) -- (-0.25,0.35);
    \end{tikzpicture}};
    \node (d) at (3,0) {\begin{tikzpicture}[yscale=1, xscale=1]
        \draw (0.125,0) node[rotate=90] {Rx-2};
        \draw[thick] (0,0.4) -- (0,-0.4) -- (0.25,-0.4) -- (0.25,0.4) -- cycle;
        \draw[thick,-latex reversed] (0,-0.3) -- (-0.25,-0.3) -- (-0.25,-0.05);
        \draw[thick,-latex reversed] (0,0.1) -- (-0.25,0.1) -- (-0.25,0.35);
    \end{tikzpicture}};
    \draw[ultra thick,-latex] (a)--node[above]{$\mathbf{G}_{11}[t]$} (c);
    \draw[ultra thick,-latex] (a)--node[above,pos=0.3,sloped]{$\mathbf{G}_{21}[t]$} (d);
    \draw[ultra thick,-latex] (b)--node[below,pos=0.3,sloped]{$\mathbf{G}_{12}[t]$} (c);
    \draw[ultra thick,-latex] (b)--node[below]{$\mathbf{G}_{22}[t]$} (d);
\draw[dashed]  (-1,4) node[] {Transmitted};
\node[red] (A) at (-1,3.5) {$\mathbf{u}_{11}^4[k]$};
\node[red] (B) at (-1,3) {$\mathbf{u}_{11}^5[k]$};
\node[red] (C) at (-1,2.5) {$\mathbf{u}_{11}^6[k]$};
\node[blue] (X) at (-1,0.5) {$\vec{\bm{\phi}}_{1,1}[k]\vec{\mathbf{u}}_{12}[k]$};
\node[blue] (Y) at (-1,0) {$\vec{\bm{\phi}}_{1,2}[k]\vec{\mathbf{u}}_{12}[k]$};
\node[blue] (Z) at (-1,-0.5) {$\vec{\bm{\phi}}_{1,1}[k]\vec{\mathbf{u}}_{12}[k]$};

\draw[dashed]  (3.5,4) node[right] {Received};
\node[right] (R111_1) at (3.5,3.4) 
    {$\displaystyle{\color{red}\sum_{m=1}^3 \mathbf{g}_{11}^{1 m}[t_0[k]+1]\mathbf{u}_{11}^{m+3}[k]} 
    + {\color{blue}\sum_{m=1}^2 \mathbf{a}_{12}^{1 m}[t_0[k]+1]\vec{\bm{\phi}}_{1,m}[k]\vec{\mathbf{u}}_{12}[k]}$};
\node[right] (R112_1) at (3.5,2.6) 
    {$\displaystyle{\color{red}\sum_{m=1}^3 \mathbf{g}_{11}^{2 m}[t_0[k]+1]\mathbf{u}_{11}^{m+3}[k]}
    + {\color{blue}\sum_{m=1}^2 \mathbf{a}_{12}^{2 m}[t_0[k]+1]\vec{\bm{\phi}}_{1,m}[k]\vec{\mathbf{u}}_{12}[k]}$};
\node[right] (R111_1) at (3.5,0.4) 
    {$\displaystyle{\color{red}\sum_{m=1}^3 \mathbf{g}_{21}^{1 m}[t_0[k]+1]\mathbf{u}_{11}^{m+3}[k]} 
    + {\color{blue}\sum_{m=1}^2 \mathbf{a}_{22}^{1 m}[t_0[k]+1]\vec{\bm{\phi}}_{1,m}[k]\vec{\mathbf{u}}_{12}[k]}$};
\node[right] (R112_1) at (3.5,-0.4) 
    {$\displaystyle{\color{red}\sum_{m=1}^3 \mathbf{g}_{21}^{2 m}[t_0[k]+1]\mathbf{u}_{11}^{m+3}[k]}
    + {\color{blue}\sum_{m=1}^2 \mathbf{a}_{22}^{2 m}[t_0[k]+1]\vec{\bm{\phi}}_{1,m}[k]\vec{\mathbf{u}}_{12}[k]}$};
    \draw[white](-1.8,3) -- (-1.8,0);
    \draw[white](10.2,3) -- (10.2,0);
\end{tikzpicture}\caption{Second channel use of Phase~1, Round~$k$ (i.e., $t=t_0[k]+1$)}\label{fig:exscheme2}
\end{subfigure}
\caption{First two channel uses from Round~$k$ of Phase~1 when $M_1=M_2=3$ and $N_1=N_2=2$. For simplicity, noise and scaling to satisfy power constraints have been omitted.  Only the first two out of five channel uses are shown. Red and blue are used to signify information originating from Transmitter~1 and 2, respectively. Notice that, although 3 symbols are broadcast from Transmitter~2, their image at Receiver~2 has dimension of 2 instead of 3: all blue terms at Receiver~2 are linear functions of the two variables $\vec{\bm{\phi}}_{1,1}[k]\vec{\mathbf{u}}_{12}[k]$ and $\vec{\bm{\phi}}_{1,2}[k]\vec{\mathbf{u}}_{12}[k]$.}\label{fig:examplescheme}
\end{figure}
\end{example*}

\subsubsection*{\bf Phase~3}

During Phase~3, all transmissions are linear equations of previously transmitted message symbols. To motivate the how these linear equations are chosen, we again consider the example MIMO XC presented in Figure~\ref{fig:examplescheme}. 

\begin{example*}[Phase~3]
Recall from the previous example that, for the MIMO XC in Figure~2, $S_i=5$ and $\xi_i=1$. 
Therefore, during a single round ($k$) of Phase~1, Receiver~1 received a total of $S_1\times N_1 = 10$ equations, which are almost surely linearly independent due to the continuously distributed channels. 
Each of these 10 equations are linear combinations of $S_1M_1 = 15$ variables from Transmitter~1 (elements of $\vec{\mathbf{u}}_{11}[k]$) and $\xi_1 M_2=3$ variables from Transmitter~2 (elements of $\vec{\mathbf{u}}_{12}[k]$). 
Consequently, Receiver~1 still requires $8$ equations of these symbols, linearly independent of what it already received, in order to decode.

In comparison, during the same round of Phase~1, Receiver~2 also received 10 almost surely linearly independent equations, but each of its 10 equations are linear combinations of $S_1M_1 = 15$ variables from Transmitter~1 (elements of $\vec{\mathbf{u}}_{11}[k]$) and $\xi_1\min\{M_2,N_2\}=2$ variables from Transmitter~2 ($\vec{\bm{\phi}}_{1}^1[k]\vec{\mathbf{u}}_{12}[k]$ and $\vec{\bm{\phi}}_{1}^2[k]\vec{\mathbf{u}}_{12}[k]$). Notice that if Receiver~2 eliminates the variables $\vec{\bm{\phi}}_{1}^1[k]\vec{\mathbf{u}}_{12}[k]$ and $\vec{\bm{\phi}}_{1}^2[k]\vec{\mathbf{u}}_{12}[k]$ from its equations, it is left with $10-2=8$ linear equations in terms of only $\vec{\mathbf{u}}_{11}[k]$.\footnote{This may be accomplished by using two equations to solve for $\vec{\bm{\phi}}_{1}^1[k]\vec{\mathbf{u}}_{12}[k]$ and $\vec{\bm{\phi}}_{1}^2[k]\vec{\mathbf{u}}_{12}[k]$ in terms of elements of $\vec{\mathbf{u}}_{11}[k]$, and substituting them back into the remaining 8 equations.} 
If we provide these 8 equations to Receiver~1, then it could almost surely decode its 18 desired symbols for the round. 

Similarly, due to the symmetric antenna configuration, after each round of Phase~2 there exist 8 equations Receiver~1 can create from past receptions that would allow Receiver~2 to decode its desired message symbols. In both cases, these equations are only in terms of symbols from Transmitter~1, thus Transmitter~1 can identify these equations using delayed CSIT and broadcast sums of one equation from Phase~1 and one from Phase~2. At each receiver, because the part that is not desired message information is already known, it may be subtracted, and every transmission benefits both receivers. 

The effect in Phase~3 of exploiting delayed CSIT during Phases~1 and 2 should now be clear. For the network in Figure~\ref{fig:examplescheme}, if we had simply repeatedly retransmitted the symbols in $\vec{\mathbf{u}}_{12}[k]$ during the duration of the round in Phase~1, the image of $\vec{\mathbf{u}}_{12}[k]$ at Receiver~2 would have dimension 3, and the maximum number of linearly independent equations Receiver~2 could reconstruct in terms of $\vec{\mathbf{u}}_{11}[k]$ would almost surely be $(S_1N_2-3) = 7$. However, because we used delayed CSIT, the image of Transmitter~2 symbols, $\vec{\mathbf{u}}_{12}[k]$, at Receiver~2 had had a dimension of 2 instead of 3. This allows Receiver~2 to identify one extra linearly independent equation of $\vec{\mathbf{u}}_{11}[k]$ per round, which results in $\lambda_i=8$ instead of 7 (i.e., an additional opportunity for a Phase~3 transmission was created without increasing the number of Phase~1 channel uses).
\end{example*}

This example captures the intuition behind all Phase~3 transmissions: delayed CSIT is used to identify equations in terms of each transmitter's symbols that are desired by one receiver and already known to another and these equations are used to create transmissions that simultaneously benefit both receivers. In general, at the end of each round ($k$) of Phase~$i$, each transmitter ($j$) uses delayed CSIT to identify linearly independent equations of its own symbols, $\vec{\mathbf{u}}_{ij}[k]$, that satisfy:
\begin{enumerate}
    \item Receiver~$i^\prime$ can (linearly) compute the equation from what it received during Round~$k$ of Phase~$i$. 
    \item Receiver~$i$ cannot compute the equation from what it received. 
\end{enumerate}
Each transmitter buffers the equations it identifies. The number of such equations is dependent on the antenna configuration, so to simplify explanation of Phase~3, we denote as $\lambda_i$ the total number of such equations per round of Phase~$i$, buffered by either Transmitter~1 or Transmitter~2. Consequently, the total number of equations buffered during Phase~$i$ is $\kappa_i\lambda_i$. 

Buffered equations from Phases~1 and 2 are then transmitted during Phase~3 as follows. 
During each channel use, the Transmitter~$j$ selects at most $M_j$ equations buffered during Phase~1 and at most $M_j$ equations buffered during Phase~2. 
The total number of equations from Phase~1 selected by Transmitter~1 and Transmitter~2 is $\min\{M_1+M_2,N_1\}$. 
Similarly, the total number of equations from Phase~2 selected by Transmitter~1 and Transmitter~2 is $\min\{M_1+M_2,N_2\}$. 
Each transmitter sends the sum of one selected equation from Phase~1 and one selected equation from Phase~2 (scaled to satisfy power constraints) using a different antenna, with no repeats. 
If the number of selected equations for each phase is unequal, then some antennas broadcast a single equation. 

Receiver~$i$ uses the side information it overheard during Phase~$i^\prime$ ($i\in\{1,2\}$) and knowledge of the channel states to cancel linear combinations of equations buffered during Phase~$i^\prime$. Consequently, as long as buffers are not empty, during each channel use of Phase~3, Receiver $i$ gains $\min\{M_1+M_2,N_i\}$ new independent linear equations that describe its desired symbols.

Finally, we clarify the manner in which the number of rounds for Phases~1 and 2 (i.e., $\kappa_1$ and $\kappa_2$) are chosen.
Notice that the efficiency of Phase~3 is maximized when, for every channel use of Phase~3, the buffer from Phase~1 has exactly $\min\{M_1+M_2,N_1\}$ equations to send and the buffer from Phase~2 has exactly $\min\{M_1+M_2,N_2\}$ equations to send. Therefore, $\kappa_1$ and $\kappa_2$ should satisfy
\begin{align}
    \frac{\kappa_1\lambda_1}{\min\{M_1+M_2,N_1\}} = \frac{\kappa_2\lambda_2}{\min\{M_1+M_2,N_2\}} \in \mathbb{Z}_+.\label{eq:balance}
\end{align}
If $\lambda_i$ is zero, we set $\kappa_i=1$. Otherwise, we set $\kappa_1$ and $\kappa_2$ equal to the smallest non-negative integers that satisfy (\ref{eq:balance}). The resulting number of channel uses used in Phase~3 is therefore $\frac{\kappa_1\lambda_1}{\min\{M_1+M_2,N_1\}}$.

%%%%%%%%%%
%%%%%%%%%%
%%%%%%%%%%
%%%%%%%%%%
%%%%%%%%%%
\subsection{Proof of Achievability of (\ref{eq:sumDoF})}
\label{sec:achsumdof}

To prove that our transmission strategy achieves the sum DoF stated in (\ref{eq:sumDoF}), we must accomplish two things for all antenna configurations:
\begin{enumerate}
 \item Verify that all symbols desired by Receiver~$i$ are decodable at Receiver~$i$.
 \item Evaluate the achieved sum DoF, and show that it is equal to (\ref{eq:sumDoF}).
\end{enumerate}

To verify that all symbols intended for Receiver~$i$ are decodable at Receiver~$i$, we claim the following:
\begin{claim}\label{cl:decoderound}
Receiver~$i$ can almost surely decode all of the symbols from each round if
\begin{align}
    \Lambda_i \leq S_i\min\{M_1+M_2,N_i\} + \lambda_i,\label{eq:decode_1}
\end{align}
where $\Lambda_i$ is the number of symbols per round of Phase~$i$:
\begin{align}
    \Lambda_i = \begin{cases}
        \min\{M_1,N_1+N_2\}& \text{ if }\Gamma_i \leq \frac{M_1}{N_{i^\prime}} \cr
        M_1 S_i + M_2\xi_i &\text{ if }\Gamma_i > \frac{M_1}{N_{i^\prime}}\cr
        \end{cases}.\label{eq:decode_2}
\end{align}
\end{claim}
\begin{IEEEproof}
We will show that condition (\ref{eq:decode_1}) is equivalent (for our strategy) to the more general decoding conditions given in (\ref{eq:decode2a}) and (\ref{eq:decode2}), but note that a simple rationale for the claim follows from an equation counting argument. Notice that the probability of rank deficient channels from continuous distribution is zero, and consequently Receiver~$i$ almost surely receives $S_i\min\{M_1+M_2,N_i\}$ linearly independent equations per round of Phase~$i$. The $\lambda_i$ buffered equations are then delivered during Phase~3, and when the number of equations communicated for each round is greater than or equal to the number of variables per round ($\Lambda_i$), the symbols may be decoded.

Notice that  (\ref{eq:decode2a}) is satisfied by the definition of the strategy, since during at least one time instance, every symbol is sent on its own antenna (specifically during rounds of Phase~1 and 2). To see that (\ref{eq:decode2}) is satisfied, we point out that the method of selecting equations to transmit during Phase~3 results in $\rank[\mathbf{G}_{i1}^T\mathbf{V}_{i^\prime 1}^T \quad \mathbf{G}_{i2}^T\mathbf{V}_{i^\prime 2}^T]=\kappa_{i^\prime}S_{i^\prime}\min\{M_1+M_2,N_i\}$ (i.e., only Phase~2 contributes to increasing the rank of undesired transmissions). Using this we see that  almost surely satisfying (\ref{eq:decode_1}) requires  almost surely satisfying
\begin{align}
    \rank[\mathbf{G}_{i1}^T\mathbf{V}_{i 1}^T] + \rank[\mathbf{G}_{i2}^T\mathbf{V}_{i 2}^T] ={}& \rank[\mathbf{V}_{i 1}^T] + \rank[\mathbf{V}_{i 2}^T]\\
        ={}& |m_{i1}^{(T)}| +  |m_{i2}^{(T)}| \\
        \stackrel{}{=}{}& \kappa_i\Lambda_i\\ 
        \leq{}& \kappa_{i}S_{i}\min\{M_1+M_2,N_i\} + \kappa_i\lambda_i\\
        \stackrel{(a)}{=}{}& \left(T - \kappa_{i^\prime}S_{i^\prime}\right)\min\{M_1+M_2,N_i\}\\
        \stackrel{(b)}{=}{}& \rank[\mathbf{G}_{i1}^T\mathbf{V}_{i1}^T \quad \mathbf{G}_{i2}^T\mathbf{V}_{i2}^T \quad \mathbf{G}_{i1}^T\mathbf{V}_{i^\prime 1}^T 
            \quad \mathbf{G}_{i2}^T\mathbf{V}_{i^\prime 2}^T] - 
            \rank[\mathbf{G}_{i1}^T\mathbf{V}_{i^\prime 1}^T \quad \mathbf{G}_{i2}^T\mathbf{V}_{i^\prime 2}^T].
\end{align}
In (a) we observed that by definition of the strategy, $\kappa_{i}S_{i}$ is the total number of Phase~1 channel uses and $\frac{\kappa_i\lambda_i}{\min\{M_1+M_2,N_i\}}$ is the number of Phase~3 channel uses. In (b), we observed that our strategy uses all available spatial dimensions during Phases~1 and 2. To show equivalence to (\ref{eq:decode2}), we need only confirm
\begin{align}
\rank[\mathbf{G}_{i1}^T\mathbf{V}_{i1}^T \quad \mathbf{G}_{i2}^T\mathbf{V}_{i2}^T \quad \mathbf{G}_{i1}^T\mathbf{V}_{i^\prime 1}^T 
            \quad \mathbf{G}_{i2}^T\mathbf{V}_{i^\prime 2}^T] - 
            \rank[\mathbf{G}_{i1}^T\mathbf{V}_{i^\prime 1}^T \quad \mathbf{G}_{i2}^T\mathbf{V}_{i^\prime 2}^T] \leq 
            \rank[\mathbf{G}_{i1}^T\mathbf{V}_{i 1}^T] + \rank[\mathbf{G}_{i2}^T\mathbf{V}_{i 2}^T]
\end{align}
which holds due to submodularity of rank.
\end{IEEEproof}

To evaluate the sum DoF, we divide the total number of symbols sent by the total number of channel uses. Assuming that the number of rounds per phase satisfies (\ref{eq:balance}), we find the achieved sum DoF ($\underline{\DoF}$) is given by
\begin{align}
    \underline{\DoF} = \frac{\kappa_1\Lambda_1+\kappa_2\Lambda_2}{\kappa_1S_1 + \kappa_2S_2 + \frac{\kappa_1\lambda_1}{\min\{M_1+M_2,N_1\}}}.\label{eq:achsumdof1}
\end{align}

Proving (\ref{eq:decode_1}) is satisfied and evaluating (\ref{eq:achsumdof1}) for all antenna configurations requires evaluating $\Gamma_i$, $S_i$, $\xi_i$ and computing $\lambda_i$ for all antenna configurations. We do so now by separating the analysis for three cases that capture all antenna configurations: 

\subsubsection{Case 1 ($\max\{N_1,N_2\}\geq M_1+M_2$)}

If $N_i\geq M_1+M_2$, then even with no CSIT the cut set bound is achievable: independent symbols may be broadcast on all transmit antennas of both transmitters, and Receiver~$i$ uses CSIT to separate independent streams. The achieved DoF is $M_1+M_2$.

Though this argument is straightforward, we now illustrate that it is also consistent with our general strategy. Assume that $N_i\geq M_1+M_2$ and that $N_{i^\prime}$ is any nonnegative integer. Evaluating terms in the $\min$ operation of (\ref{eq:rankratio}) we observe that 
\begin{align}
    \frac{N_{i^\prime}+N_i}{N_{i^\prime}} 
        \geq{}& \frac{N_{i^\prime}+M_1+M_2}{N_{i^\prime}} \\
        \geq{}& \frac{M_1 + M_2}{N_{i^\prime}},
\end{align}
and
\begin{align}
    \frac{Q_{i1}N_{i^\prime}+Q_{i2}\left(N_{i^\prime} + N_i\right)}{N_{i^\prime}(Q_{i2} + N_{i^\prime})}
        ={}& \frac{(Q_{i1}+Q_{i2})N_{i^\prime} +Q_{i2}N_i}{N_{i^\prime}(Q_{i2} + N_{i^\prime})} \\
        \geq{}& \frac{(M_{1}+M_{2})N_{i^\prime} +Q_{i2}(M_1+M_2)}{N_{i^\prime}(Q_{i2} + N_{i^\prime})} \\
        ={}& \frac{M_1 + M_2}{N_{i^\prime}},
\end{align}
which implies that 
$\Gamma_{i}=\max\{\frac{M_1+M_2}{N_{i^\prime}},1\}$. 
Inserting this rank-ratio into (\ref{eq:chanperrnd}), we see $S_i=\xi_i=1$, and from (\ref{eq:decode_2}) we observe $\Lambda_i = M_1+M_2$.
Because channel matrices from each transmitter to Receiver~$i$ are almost surely full rank, Receiver~$i$ can apply a linear transformation to recover transmitted symbols during each time instance. Therefore, $\lambda_i=0$ and $\kappa_i=1$, which is consistent with the decodability condition given in (\ref{eq:decode_1}).

For Receiver~$i^\prime$ regardless of the value of $N_{i^\prime}$, we observe $\Gamma_{i^\prime}=1$, $S_{i^\prime}=\xi_{i^\prime}=1$, and $\Lambda_{i^\prime} = M_1+M_2$. If $N_{i^\prime}\geq M_1+M_2$, then $\lambda_{i^\prime}=0$ and $\kappa_{i^\prime}=1$, and the achieved DoF is, from (\ref{eq:achsumdof1}), $M_1+M_2$.
If on the other hand, $N_{i^\prime}<M_1+M_2$, then $\lambda_{i^\prime}=M_1+M_2-N_{i^\prime}$ and $\kappa_{i^\prime}=0$ (in order to satisfy (\ref{eq:balance})). The resulting achieved DoF computed in (\ref{eq:achsumdof1}) remains $M_1+M_2$, as expected.

\subsubsection{Case 2 ($M_1 \geq N_1+N_2$)}

The case of $M_1\geq N_1+N_2$ is particularly interesting, because in such antenna configurations Transmitter~2 is not needed to achieve the linear sum DoF. We first claim the following, which implies that when $M_1 \geq N_1+N_2$, the simpler transmission strategy of Appendix~\ref{app:othercase} which ignores Transmitter~2 is used for for Phases 1 and 2.
\begin{claim}\label{cl:case2}
The following statements are equivalent:
\begin{itemize}
\item $M_1\geq N_1+N_2$.
\item $\Gamma_1N_2 \leq M_1$.
\item $\Gamma_2N_1 \leq M_1$.
\end{itemize}
\end{claim}
\begin{IEEEproof}
First, note that when $M_1\geq N_1+N_2$
\begin{align}
    \frac{M_1+M_2}{N_{i^\prime}} 
        \geq{}& \frac{N_1+N_2+M_2}{N_{i^\prime}} \\
        \geq{}& \frac{N_1 + N_2}{N_{i^\prime}},
\end{align}
and
\begin{align}
    \frac{Q_{i1}N_{i^\prime}+Q_{i2}\left(N_{i^\prime} + N_i\right)}{N_{i^\prime}(Q_{i2} + N_{i^\prime})}
        \geq{}& \frac{M_{1}N_{i^\prime}+Q_{i2}\left(N_{i^\prime} + N_i\right)}{N_{i^\prime}(Q_{i2} + N_{i^\prime})} \\
        \geq{}& \frac{(N_{1}+N_{2})N_{i^\prime} +Q_{i2}(N_1+N_2)}{N_{i^\prime}(Q_{i2} + N_{i^\prime})} \\
        ={}& \frac{N_1 + N_2}{N_{i^\prime}},
\end{align}
and therefore $\Gamma_i = \frac{N_1+N_2}{N_{i^\prime}}$, and $\Gamma_iN_{i^\prime}\leq M_1$.

To confirm the opposite implication, we first note that because by definition, (\ref{eq:rankratio}), $\Gamma_i\geq 1$,
\begin{align}
\Gamma_i N_{i^\prime}\leq M_1 \quad\Rightarrow{}& \quad N_{i^\prime}\leq M_1 \quad \Leftrightarrow\quad Q_{i1}=M_1.
\end{align}
Using this, we see
\begin{align}
\Gamma_i N_{i^\prime}-M_1\leq 0\\
    \stackrel{(a)}{\Leftrightarrow}{}& \max\left\{
    \min\left\{
    M_2,
    N_1+N_2-M_1,
    \frac{Q_{i2}(N_1+N_2-M_1)}{Q_{i2}+N_{i^\prime}}
    \right\},
    N_{i^\prime}-M_1\right\}\leq 0\\
    \stackrel{}{\Leftrightarrow}{}& 
    \min\left\{
    M_2,
    \frac{Q_{i2}(N_1+N_2-M_1)}{Q_{i2}+N_{i^\prime}}
    \right\}\leq 0\\
    \stackrel{}{\Leftrightarrow}{}& M_1\geq N_1+N_2.
\end{align}
where (a) is by definition of the maximum rank-ratio given in (\ref{eq:rankratio}). 
\end{IEEEproof}

Recall that, by construction of the strategy, Transmitter~2 remains silent during Phase~$i$ if $\Gamma_i N_{i^\prime}\leq M_1$. Therefore we observe from Claim~\ref{cl:case2} that when $M_1\geq N_1+N_2$, Transmitter~2 is silent in \emph{both Phases~1 and 2}, and consequently has nothing to send during Phase~3; our strategy does not require Transmitter~2 to achieve the linear sum DoF.

The resulting transmission strategy remains a three-phase strategy, with $S_i=1$ channel uses and $\Lambda_i = N_1+N_2$ symbols transmitted, as defined in our strategy. Since all of the transmissions originate at Transmitter~1, and because channel matrices are almost surely full rank, everything received by Receiver~2 is almost surely linearly independent of what is received by Receiver~1. Therefore, $\lambda_i=N_{i^\prime}$ equations are buffered during each round in Phase~$i$. Notice that these terms satisfy the decoding condition (\ref{eq:decode_1}).

From (\ref{eq:balance}), the number of rounds per phase, $\kappa_1$ and $\kappa_2$, are integers chosen such that
\begin{align}
    \kappa_1
        ={}& \frac{N_1}{\lambda_1}\frac{\lambda_2}{N_2}\kappa_2\\
        ={}& \left(\frac{N_1}{N_2}\right)^2\kappa_2,
\end{align}
which when substituted into (\ref{eq:achsumdof1}) yields the desired achieved sum DoF:
\begin{align}
    \underline{\DoF} 
        ={}& \frac{\kappa_1\Lambda_1+\kappa_2\Lambda_2}
            {\kappa_1S_1 + \kappa_2S_2 + \frac{\kappa_1\lambda_1}{\min\{M_1+M_2,N_1\}}}\\
        ={}& \frac{\left(\frac{N_1}{N_2}\right)^2(N_1+N_2) + (N_1+N_2)}
            {\left(\frac{N_1}{N_2}\right)^2 + 1 + \left(\frac{N_1}{N_2}\right)^2\frac{N_2}{N_1}}\\
%         ={}& \frac{\frac{N_1}{N_2} +\frac{N_2}{N_1}}
%             {\frac{N_1}{N_2} + \frac{N_2}{N_1} + 1}(N_1+N_2)\\
        ={}& \frac{\frac{(N_1+N_2)^2}{N_1N_2} -\frac{N_2}{N_2}-\frac{N_1}{N_1}}
            {\frac{(N_1+N_2)^2}{N_1N_2}-1}(N_1+N_2)\\
        ={}& \frac{\Gamma_1\Gamma_2(N_1+N_2)-\Gamma_1N_2-\Gamma_2N_1}
            {\Gamma_1\Gamma_2-1}.
\end{align}

\subsubsection{Case 3 (Everything Else)}

We now consider all remaining antenna configurations: all configurations where $\max\{N_1,N_2\}< M_1+M_2$ and $M_1<N_1+N_2$. For all such configurations, we claim the following.
\begin{claim}\label{cl:case3}
    If $\max\{N_1,N_2\} < M_1+M_2$ and $M_1<N_1+N_2$ then 
    \begin{align}
    \Gamma_i = \min\left\{ \frac{M_1+M_2}{N_{i^\prime}}, \frac{Q_{i1}N_{i^\prime}+Q_{i2}(N_1+N_2)}{N_{i\prime}(Q_{i2}+N_{i^\prime})}\right\},
    \end{align}
    and 
    \begin{align}
     N_1+N_2 > \Gamma_iN_{i^\prime}.
    \end{align}
\end{claim}
\begin{IEEEproof}
Since, $N_{i^\prime}< M_1+M_2$, we observe from (\ref{eq:rankratio}) that $\Gamma_i>1$. 
Also, when $M_1<N_1+N_2$, then $Q_{i1}<N_1+N_2$, and
\begin{align}
\frac{Q_{i1}N_{i^\prime}+Q_{i2}(N_1+N_2)}{N_{i\prime}(Q_{i2}+N_{i\prime})}
    <{}&\frac{(N_{i^\prime}+Q_{i2})(N_1+N_2)}{N_{i\prime}(Q_{i2}+N_{i\prime})}
    ={}\frac{N_1+N_2}{N_{i\prime}}.
\end{align}
Applying these oberservations to the definition in (\ref{eq:rankratio}), we arrive at the claim.
\end{IEEEproof}
Using Claim~\ref{cl:case3}, we evaluate (\ref{eq:chanperrnd}) and (\ref{eq:decode_2}), and find that
\begin{align}
    S_i %={}& \frac{M_2\xi_i}{\Gamma_i N_{i^\prime}-M_1}\nonumber\\
        %={}& \frac{M_2\xi_i}{\min\{M_2,\frac{Q_{i1}N_{i^\prime}+Q_{i2}(N_1+N_2)}{Q_{i2}+N_{i^\prime}}-M_1\}}\nonumber\\
        ={}& \xi_i\max\left\{1,\frac{M_2(Q_{i2}+N_{i^\prime})}{(Q_{i1}-M_1)N_{i^\prime}+Q_{i2}(N_1+N_2-M_1)}\right\}.\label{eq:ach1a}
\end{align}
and 
\begin{align}
    \Lambda_i ={}& S_iM_1+\xi_iM_2.\label{eq:ach1b}
\end{align}

We first verify that all symbols sent per round are indeed decodable at Receiver~$i$. Recall that during Phase~$i$ delayed CSIT was used such that, for each round, the image of Transmitter~2 symbols, $\vec{\mathbf{u}}_{i2}[k]$, at Receiver~$i^\prime$ almost surely has dimension $\xi_i\min\{M_2,N_{i^\prime}\}$. Due to the continuously distributed channel matrices, the image of Transmitter~1 symbols, $\vec{\mathbf{u}}_{i1}[k]$, at Receiver~$i^\prime$ almost surely has dimension $S_i\min\{M_1,N_{i^\prime}\}$.
Therefore, the number of equations in terms of Transmitter~1 symbols, $\vec{\mathbf{u}}_{i1}[k]$, that Receiver~$i^\prime$ can reconstruct is almost surely  %$S_iN_{i^\prime}-\xi_i\min\{M_2,N_{i^\prime}\}$. 
\begin{align}
\min\{S_iM_1,S_iN_{i^\prime}-\xi_i\min\{M_2,N_{i^\prime}\}\},
\end{align}
and the number of equations in terms of Transmitter~2 symbols, $\vec{\mathbf{u}}_{i2}[k]$, that Receiver~$i^\prime$ can reconstruct is almost surely  %$S_i[N_{i^\prime}-M_1]_+$. 
\begin{align}
\min\{\xi_i\min\{M_2,N_{i^\prime}\},S_i[N_{i^\prime}-M_1]_+\}.
\end{align}
We define the combined total number of equations that Receiver~$i^\prime$ can create in terms of variables from only one transmitter as 
\begin{align}
\overline{\lambda}_i ={}& \min\{S_iM_1,S_iN_{i^\prime}-\xi_i\min\{M_2,N_{i^\prime}\}\} + 
\min\{\xi_i\min\{M_2,N_{i^\prime}\},S_i[N_{i^\prime}-M_1]_+\}\nonumber\\
% ={}&\min\{S_iM_1+\xi_i\min\{M_2,N_{i^\prime}\},S_iN_{i^\prime},S_iN_{i^\prime}-\xi_i\min\{M_2,N_{i^\prime}\}+S_i[N_{i^\prime}-M_1]_+\}\nonumber\\
={}&\min\{
    \overbrace{\Lambda_i}^{\theta_1},
    \overbrace{S_iM_1+\xi_iN_{i^\prime}}^{\theta_2},
    \overbrace{S_iN_{i^\prime}}^{\theta_3},
    \overbrace{S_iN_{i^\prime}-\xi_i\min\{M_2,N_{i^\prime}\}+S_i[N_{i^\prime}-M_1]_+}^{\theta_4}\}.
\end{align}
If $\overline{\lambda}_i\geq\Lambda_i-S_iN_i$, then enough equations can be buffered and sent to Receiver~$i$ in Phase~3 to allow Receiver~$i$ to decode all symbols in the round. Consequently, satisfying the decoding condition for each round, (\ref{eq:decode_1}), requires
\begin{align}
    \theta_\ell \geq \Lambda_i-S_iN_i,\quad\text{ for }\ell=1,2,3,4.\label{eq:ach2}
\end{align}
For $\theta_1$, we see that this is clearly true. For $\theta_2$, we find
\begin{align}
 S_iM_1+\xi_iN_{i^\prime} 
    ={}& S_iM_1-S_iN_i+S_iN_i+\xi_iN_{i^\prime}\\
    \geq{}& S_iM_1-S_iN_i+\xi_i(N_i+N_{i^\prime})\\
    >{}& S_iM_1-S_iN_i+\xi_iM_1\\
    \geq{}& S_iM_1-S_iN_i+\xi_iM_2\\
    ={}& \Lambda_i-S_iN_i.
\end{align}
For $\theta_3$, we find
\begin{align}
    S_iN_{i^\prime} ={}& S_iN_{i^\prime} + S_iN_{i} - \xi_iM_2 - S_iN_i + \xi_iM_2 \\
        \stackrel{(a)}{=}{}& S_i(N_1+N_2) - S_i\min\left\{M_2,\frac{Q_{i1}N_{i^\prime}+Q_{i2}\left(N_{i^\prime} + N_i\right)}{N_{i^\prime}(Q_{i2} + N_{i^\prime})}-M_1\right\} 
        - S_iN_i + \xi_iM_2 \\
        \stackrel{(b)}{=}{}& S_iM_1 + S_i(N_1+N_2 - \Gamma_iN_{i^\prime})- S_iN_i + \xi_iM_2 \\
        \stackrel{(c)}{=}{}& S_iM_1 - S_iN_i + \xi_iM_2 \\
    ={}& \Lambda_i-S_iN_i,
\end{align}
where in (a) we applied (\ref{eq:ach1a}) and in (b) and (c) we used Claim~\ref{cl:case3}.
Finally for $\theta_4$, we find
\begin{align}
    S_iN_{i^\prime} &-\xi_i\min\{M_2,N_{i^\prime}\} + S_i[N_{i^\prime}-M_1]_+\nonumber\\
%         ={} S_i(N_{i^\prime} +[N_{i^\prime}-M_1]_+)-\xi_i\min\{M_2,N_{i^\prime}\}\\
        ={}& S_i(N_{i^\prime} +Q_{i1}-M_1)-\xi_i\min\{M_2,N_{i^\prime}\}\\
        ={}& S_i(M_1-N_i)+\xi_iM_2+S_i((N_1+N_{2}-M_1) +(Q_{i1}-M_1))-\xi_i\min\{2M_2,M_2+N_{i^\prime}\}\\
        \stackrel{(a)}{=}{}& \Lambda_i-S_iN_i\nonumber\\
            &+\xi_i\max\{((N_1+N_{2}-M_1) +(Q_{i1}-M_1)),
                \frac{M_2(Q_{i2}+N_{i^\prime})((N_1+N_{2}-M_1) +(Q_{i1}-M_1))}{(Q_{i1}-M_1)N_{i^\prime}+Q_{i2}(N_1+N_2-M_1)}\}
                \nonumber\\
            &-\xi_i\min\{2M_2,M_2+N_{i^\prime}\}\\
        ={}& \Lambda_i-S_iN_i\nonumber\\
            &+\xi_i\max\Bigg\{
            (N_1+N_{2}-(M_1+M_2)) +(Q_{i1}-(M_1+M_2)),\nonumber\\
            &\qquad\qquad\quad (N_i-M_1-M_2) +(Q_{i1}-M_1),\nonumber\\
            &\qquad\qquad\quad \overbrace{M_2\frac{(Q_{i2}-N_{i^\prime})(Q_{i1}-(N_1+N_{2})) }{(Q_{i1}-M_1)N_{i^\prime}+Q_{i2}(N_1+N_2-M_1)}}^{\Omega_1},   
                \nonumber\\
            &\qquad\qquad\quad \overbrace{\frac{(M_2Q_{i2}-N_{i^\prime}^2)(Q_{i1}-M_1)+(M_2-Q_{i2})(N_1+N_2-M_1)N_{i^\prime})}
            {(Q_{i1}-M_1)N_{i^\prime}+Q_{i2}(N_1+N_2-M_1)}}^{\Omega_2}\Bigg\},\label{eq:ach3}
\end{align}
where in (a) we used the definition (\ref{eq:chanperrnd}). We point out two cases: If $M_2<N_{i^\prime}$, then $Q_{i2}=N_{i^\prime}$ and $\Omega_1=0$. On the other hand, if $M_2\geq N_{i^\prime}$, then $Q_{i2}=M_2\geq N_{i^\prime}$ and $\Omega_2\geq 0$. Hence, the $\max$ term in (\ref{eq:ach3}) is nonnegative, and (\ref{eq:ach2}) and (\ref{eq:decode_1}) are satisfied.
We therefore express the number of equations buffered during each round as exactly
\begin{align}
    \lambda_i={}& S_i(M_1-N_i)+\xi_iM_2.
\end{align}
Choosing the number of rounds for Phase~1 and 2, $\kappa_1$ and $\kappa_2$, such that they satisfy (\ref{eq:balance}), we have
\begin{align}
    \kappa_1 
    ={}& \kappa_2\frac{N_1}{\lambda_1}\frac{\lambda_2}{N_2},%\\
%    ={}& \kappa_2\frac{N_1}{N_2}\frac{S_2(M_1-N_2)+\xi_2M_2}{S_1(M_1-N_1)+\xi_1M_2}\\
\end{align}
and evaluating (\ref{eq:achsumdof1}), we find
\begin{align}
    \underline{\DoF} 
        ={}& \frac{\kappa_1\Lambda_1+\kappa_2\Lambda_2}
            {\kappa_1S_1 + \kappa_2S_2 + \frac{\kappa_1\lambda_1}{\min\{M_1+M_2,N_1\}}}\\
        ={}& \frac{\frac{N_1}{N_2}\frac{\Lambda_1(\Lambda_2-S_2N_2)}{\Lambda_1-S_1N_1} + \Lambda_2}
            {\frac{N_1}{N_2}\frac{\Lambda_2-S_2N_2}{\Lambda_1-S_1N_1}S_1 + S_2 + \frac{\Lambda_2-S_2N_2}{N_2}}\\
%         ={}& \frac{\Lambda_1N_1(\Lambda_2-S_2N_2)+\Lambda_2N_2(\Lambda_1-S_1N_1)}
%             {S_1N_1(\Lambda_2-S_2N_2)+S_2N_2(\Lambda_1-S_1N_1)+(\Lambda_1-S_1N_1)(\Lambda_2-S_2N_2)}\\
%         ={}& \frac{\frac{\Lambda_1N_1(\Lambda_2-S_2N_2)}{S_1N_1S_2N_2}+\frac{\Lambda_2N_2(\Lambda_1-S_1N_1)}{S_1N_1S_2N_2}}
%             {\frac{\Lambda_1\Lambda_2-S_1N_1S_2N_2}{S_1N_1S_2N_2}}\\
        ={}& \frac{ \frac{\Lambda_1}{S_1N_2} \frac{\Lambda_2}{S_2N_1}(N_1+N_2) - \frac{\Lambda_1}{S_1} -\frac{\Lambda_1}{S_2} }
            {\frac{\Lambda_1}{S_1N_2} \frac{\Lambda_2}{S_2N_1}-1}\\
        \stackrel{(a)}={}& \frac{\Gamma_1\Gamma_2(N_1+N_2)-\Gamma_1N_2-\Gamma_2N_1}
            {\Gamma_1\Gamma_2-1},
\end{align}
where in (a) we noted from (\ref{eq:ach1a}) and (\ref{eq:ach1b}) that $\Gamma_i = \frac{\Lambda_i}{S_iN_{i^\prime}}$. 

\subsection{Symmetric Antenna Configurations}
\label{sec:sym}
We take this opportunity to focus on symmetric antenna configurations ($M_1=M_2=M$ and $N_1=N_2=N$) and to compare our transmission strategy with the previous best scheme of~\cite{GAK:isit2012}. Table~\ref{tab:DoF} displays the achieved sum DoF, $\underline{\DoF}$, for five symmetric antenna configuration regimes as well as the rank-ratio  (\ref{eq:rankratio}). The first three rows describe our proposed strategy, whereas the bottom two rows describe the schemes presented in~\cite{GAK:isit2012}. In addition to the achieved DoF, we also list the ratio between the number of symbols sent by Transmitter~2 versus the number sent by Transmitter~1 for each receiver. Because both transmitters have the same number of antennas, their roles within the scheme are interchangeable, and we therefore have defined this \emph{symbol ratio} such that it is less than or equal to 1.

\begin{table}[ht]\centering
\begin{tabular}{|c|c|c|c|c|c|}
\hline
Antenna Configuration Regime & $0<\frac{N}{M}\leq\frac{1}{2}$ & $\bm{\frac{1}{2}<\frac{N}{M}\leq 1}$ & $1<\frac{N}{M}\leq\frac{4}{3}$ & $\frac{4}{3}<\frac{N}{M}\leq 2$ & $2<\frac{N}{M}$ \\
\hline
\hline
Maximum rank-ratio ($\Gamma_i$) & $2$ & $\bm{\frac{3M}{M+N}}$ & $\frac{3}{2}$ & $\frac{2M}{N}$ & $1$ \\
\hline
Symbol ratio ($\frac{\xi_iM_2}{S_iM_1}$) & $0$ & $\bm{\frac{2N-M}{M+N}}$ & $\frac{3N-2M}{2M}$ & $1$ & $1$ \\
\hline
$\underline{\DoF}$ from (\ref{eq:sumDoF}) & $\frac{4}{3}N$ & $\bm{\frac{6M}{4M+N}N}$ & $\frac{6}{5}N$ & $\frac{4M}{2M+N}N$ & $2M$ \\
\hline
\hline
Symbol ratio from \cite{GAK:isit2012} & $0$ & $\bm{\frac{2N-M}{2M}}$ & $\frac{3N-2M}{2M}$ & $1$ & $1$ \\
\hline
$\underline{\DoF}$ from \cite{GAK:isit2012} & $\frac{4}{3}N$ & $\bm{\frac{2M+4N}{4N+M}N}$ & $\frac{6}{5}N$ & $\frac{4M}{2M+N}N$ & $2M$ \\
\hline
\end{tabular}\vspace{0.25cm}
\caption{Comparison between our scheme and existing scheme for symmetric MIMO XC with delayed CSIT.}\label{tab:DoF}
\end{table}

For three regimes ($\frac{N}{M}\leq\frac{1}{2}$, $\frac{4}{3}\leq\frac{N}{M}\leq 2$, and $\frac{N}{M}\geq 2$) the scheme proposed in \cite{GAK:isit2012} was already known to be DoF-optimal, even without the assumption of linear transmission strategies. Therefore it is not surprising that in these regimes the achieved DoF and symbol ratio of the strategies are the same.

What remained unknown in~\cite{GAK:isit2012} was the sum DoF when $\frac{1}{2}<\frac{N}{M}<\frac{4}{3}$. For symmetric antenna configurations where $1<\frac{N}{M}<\frac{4}{3}$, our converse confirms that the previously proposed scheme is at least linear sum DoF-optimal. Moreover, it can be shown that the previous approach and our approach are equivalent for such antenna configurations (e.g., the ratio between the number of symbols sent from each transmitter is the same). 

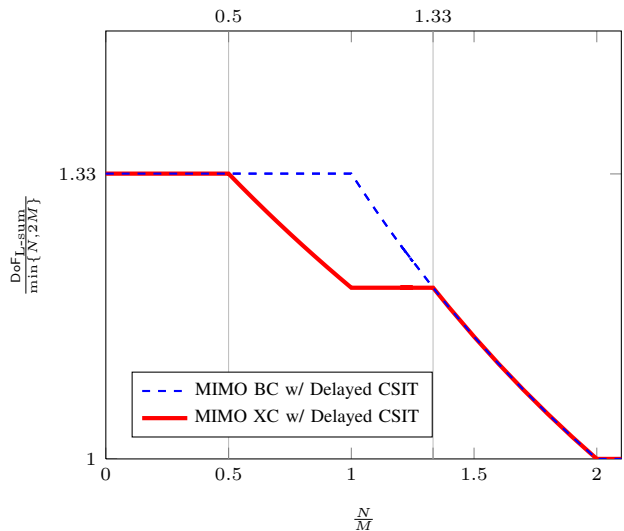
\begin{figure}[ht]
\centering
\begin{tikzpicture}[font=\scriptsize,
every axis/.style={
    xmax=2.1,%
    xmin=0,%
    ymax=1.5,%
    ymin=1,%
    ytick={1,1.333,...,2}}
]
\begin{axis}[
xlabel={$\frac{N}{M}$},
ylabel={$\frac{\LDoF}{\min\{N,2M\}}$},
every axis y label/.style=
    {at={(ticklabel cs:0.5)},rotate=90,anchor=near ticklabel},
extra x ticks={0.5,1.333},  % Add an extra tick at position x=1
extra x tick style={    % Set styles that only apply to the extra tick
            xticklabel pos=right,   % Put the label on the right ( = top) side of the plot
            %xticklabels={}, % Set the label text
            xmajorgrids=true            % Draw grid line
},
legend style={at={(0.05,0.05)},anchor=south west}
]
    \addplot[thick,dashed,color=blue] plot file {BDCSIT.data};
    \addlegendentry{MIMO BC w/ Delayed CSIT};
    \addplot[ultra thick,color=red] plot file {XDCSIT.data};
    \addlegendentry{MIMO XC w/ Delayed CSIT};
    \addplot[thick,dashed,color=blue] plot file {BDCSIT.data};
% %     \addplot[thick,color=red] plot file {XC.data};
% %     \addlegendentry{XC w/ Full CSIT};
     \end{axis}
\end{tikzpicture}
\caption{Linear sum DoF for symmetric MIMO XC (i.e. $M_1=M_2=M$ and $N_1=N_2=N$) as a function of receiver-transmitter antenna ratio ($\frac{N}{M}$). We also show the dum DoF (linear or otherwise) of the associated MIMO BC with $2M$ transmit antennas. We normalize the linear sum DoF curves in this plot by $\min\{N,2M\}$.}\label{fig:normDOF}
\end{figure}

On the other hand, in antenna configurations where $\frac{1}{2}<\frac{N}{M}\leq 1$, our strategy outperforms the previous one. The increased sum DoF is due to using delayed CSIT during Phases~1 and 2. The resulting linear sum DoF for all symmetric antenna configurations is shown in Figure~\ref{fig:normDOF}.

\section{Loss of Linear Sum DoF due to Distributed Transmitters}
\label{sec:distloss}

As a direct consequence of our main result, we now determine which antenna configurations for the MIMO XC result in a linear sum DoF equivalent to a MIMO BC that results from combining the two transmitters. These antenna configurations achieve the same sum DoF despite having messages that originate at two different transmitters and transmission schemes that are distributed. Such configurations are significant because these networks exhibit \emph{no distributed transmitter loss} from a sum DoF perspective.
Conversely, we also identify configurations that exhibit loss, which suggests that in these antenna configurations, the primary bottleneck to higher (linear) sum DoF is the distribution of communication tasks between the two transmitters. Determining whether a particular antenna configuration falls within the former or latter category is of fundamental importance in the architecture of a wireless network deployment.

Recall that proving the rank-ratio inequality of Lemma~\ref{lem:rankratioUB} (which formed the core of the converse) required proving two bounds: one based on cooperation, (\ref{eq:rankratio_UB1}), and one focused on normalized rank difference with distributed transmissions, (\ref{eq:rankratio_UB2}). The necessity of (\ref{eq:rankratio_UB1}) suggested that, for some antenna configurations, the MIMO XC with delayed CSIT may achieve the same rank ratio (and consequently same sum DoF) as a MIMO BC with $M_1+M_2$ transmit antennas. 
This is indeed the case, and can be easily verified in Figure~\ref{fig:normDOF}, where we see that for symmetric antenna configurations with $\frac{N}{M}\leq\frac{1}{2}$ or $\frac{N}{M}\geq\frac{4}{3}$, the sum DoF of the MIMO XC and the analogous MIMO BC are the same. 

On the other hand, the need for the bound (\ref{eq:rankratio_UB2}) suggests that there exist antenna configurations where the linear sum DoF of the MIMO XC is strictly less than that of the MIMO BC (e.g., symmetric antennas configurations where $\frac{1}{2} < \frac{N}{M} < \frac{4}{3}$). Such antenna configurations exhibit a distributed transmitter loss in linear sum DoF and in this section, we analyze the maximum rank-ratios expression (\ref{eq:rankratio}) to establish which antenna configurations exhibit such a loss.

In the regime where the right hand side (RHS) of (\ref{eq:rankratio_UB1}) is tighter than that of (\ref{eq:rankratio_UB2}) for both $\Gamma_1$ and $\Gamma_2$, there is no rank-ratio loss (and thus no DoF loss) due to distributed transmitters. 
Therefore, a necessary condition for distributed transmitter loss is that (\ref{eq:rankratio_UB2}) must be tighter than (\ref{eq:rankratio_UB1}) in evaluating either $\Gamma_1$ or $\Gamma_2$. We first determine when this is the case for one, $\Gamma_i$.

\begin{figure}[]
\centering
\begin{tikzpicture}[scale=2.25,font=\scriptsize]
    \fill[fill=red!50] (1,0) -- (1,1.5) -- (2,1) -- (2,0) -- cycle;
    \fill[fill=red!50] (1,0) -- (1,2) -- (0,2) -- (0,1) -- cycle;

    \draw[thick,black] (1,0) -- (1,1.5);
    \draw[thick,dotted,black] (1,1.5) -- (2,1) -- (2,0);

    \draw[thick,black,dotted] (0,1) -- (1,0) -- (1,1);

    \draw[thick,black,dotted] (1,1) -- (1,2) -- (0,2);

    \draw[thick,-latex] (0,0) -- (2.5,0) node[below]{$\frac{N_{i^\prime}}{M}$};
    \draw[thick,-latex] (0,0) -- (0,2.5) node[left]{$\frac{N_i}{M}$};
    \draw (0,0) node[anchor=north east]{(0,0)};
    \draw (1,0) --(1,-0.1) node[below]{1};
    \draw (2,0) --(2,-0.1) node[below]{2};
    \draw (0,1) --(-0.1,1) node[left]{1};
    \draw (0,2) --(-0.1,2) node[left]{2};
    \draw[dashed,thin] (0,0) -- (2.5,2.5);
    \draw (1.5,0.5) node{I};
    \draw (0.5,1.33) node{III};
    \draw[-stealth,shorten >=2pt] (0.175,0.5) node[fill=white,inner sep=1pt] {($\frac{1}{2}$,$\frac{1}{2}$)} -- (0.5,0.5);
    \draw[-stealth,shorten >=2pt] (1.5,2) node[fill=white] {($\frac{4}{3}$,$\frac{4}{3}$)} -- (1.333,1.333);
\end{tikzpicture}
\caption{Regime where Bound~2 is active for $\Gamma_i$, when $M_1=M_2=M$. Note that when $M_1=M_2=M$, Regime~II defines an empty set.}
\label{fig:MeqCases}
\end{figure}
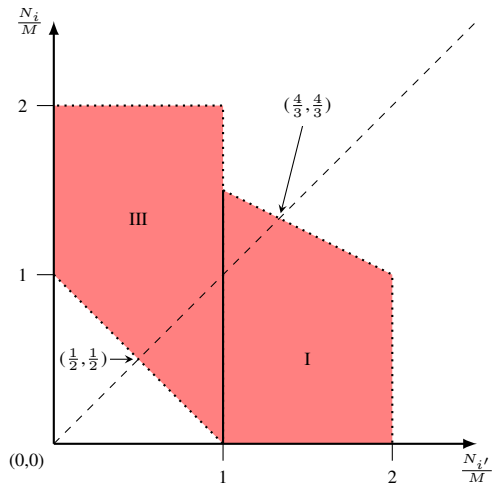

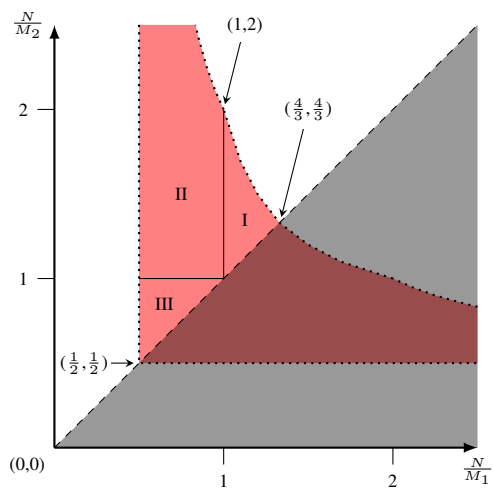
\begin{figure}[]
\centering
\begin{tikzpicture}[scale=2.25,font=\scriptsize]
    \draw[thick,-latex] (0,0) -- (2.5,0) node[below]{$\frac{N}{M_1}$};
    \draw[thick,-latex] (0,0) -- (0,2.5) node[left]{$\frac{N}{M_2}$};
    \draw (0,0) node[anchor=north east]{(0,0)};
    \draw (1,0) --(1,-0.1) node[below]{1};
    \draw (2,0) --(2,-0.1) node[below]{2};
    \draw (0,1) --(-0.1,1) node[left]{1};
    \draw (0,2) --(-0.1,2) node[left]{2};
    \fill[red!50] (0.5,2.5) -- (0.5,0.5) -- (2.5,0.5) -- (2.5,0.8333) 
        -- (2.4286,0.85)
        -- (2.25,0.9)
        -- (2.1111,0.95) 
        -- (2,1) -- (1.6923,1.1) -- (1.5,1.2) -- (1.3684,1.3) -- (1.33,1.33) -- (1.3,1.3684) -- (1.2,1.5) -- (1.1,1.6923)
        -- (1,2)
        -- (0.95,2.1111) 
        -- (0.9,2.25)
        -- (0.85,2.4286)
        -- (0.8333,2.5) -- cycle;
    \draw[black,thick,dotted] (0.5,2.5) -- (0.5,0.5) -- (2.5,0.5);
    \draw[black,thick,dotted] (1,2)
        -- (0.95,2.1111) 
        -- (0.9,2.25)
        -- (0.85,2.4286)
        -- (0.8333,2.5);
    \draw[black] (1,1)
        -- (0.5,1);
    \draw[black,thick,dotted] (2.5,0.8333) 
        -- (2.4286,0.85)
        -- (2.25,0.9)
        -- (2.1111,0.95) 
        -- (2,1);
    \draw[black,thick,dotted] (2,1) -- (1.6923,1.1) -- (1.5,1.2) -- (1.3684,1.3) -- (1.33,1.33) -- (1.3,1.3684) -- (1.2,1.5) -- (1.1,1.6923) -- (1,2);
    \draw[black] (1,2) -- (1,1);
    \draw[dashed,thin] (0,0) -- (2.5,2.5);
    \draw (1.125,1.35) node{I};
    \draw (0.75,1.5) node{II};
    \draw (0.65,0.85) node{III};
    \fill[black,opacity=0.4] (0,0) -- (2.5,2.5) -- (2.5,0) -- cycle;
    \draw[-stealth,shorten >=2pt] (1.125,2.5) node[fill=white] {(1,2)} -- (1,2);
    \draw[-stealth,shorten >=2pt] (0.175,0.5) node[fill=white,inner sep=1pt] {($\frac{1}{2}$,$\frac{1}{2}$)} -- (0.5,0.5);
    \draw[-stealth,shorten >=2pt] (1.5,2) node[fill=white] {($\frac{4}{3}$,$\frac{4}{3}$)} -- (1.333,1.333);
\end{tikzpicture}
\caption{Regime where Bound~2 is active, when $N_1=N_2=N$. Though we have assumed $M_1\geq M_2$, the black shaded region consists of configurations where $M_2>M_1$, and we have depicted the region when Bound~2 is active assuming a relabeling of transmitter indices. Due to symmetry (i.e. if $N_1=N_2$, then $\Gamma_1 = \Gamma_2$) the colored antenna configurations also the set $\mathcal{E}$ where there exists a loss of LDOF due to distributed transmitters.}
\label{fig:NeqCases}
\end{figure}

Through direct comparison of the RHS of (\ref{eq:rankratio_UB1}) and (\ref{eq:rankratio_UB2}), we find the following conditions for when (\ref{eq:rankratio_UB2}) is tighter than (\ref{eq:rankratio_UB1}):
\begin{align}
N_{i^\prime} <{}& M_1+M_2,\label{eq:dofloss1}\\
\frac{Q_{i1}N_{i^\prime}+Q_{i2}\left(N_{i^\prime} + N_i\right)}{(Q_{i2} + N_{i^\prime})} <{}& M_1+M_2,\label{eq:dofloss2}\\
\frac{Q_{i1}N_{i^\prime}+Q_{i2}\left(N_{i^\prime} + N_i\right)}{(Q_{i2} + N_{i^\prime})} <{}& N_1+N_2.\label{eq:dofloss3}
\end{align}
Since $M_1\geq M_2$ by assumption, we have three possible cases of values for $Q_{i1}$ and $Q_{i2}$: 
\begin{enumerate}
\item $(Q_{i1},Q_{i2}) = (N_{i^\prime},N_{i^\prime})$ which occurs when $M_2\leq M_1\leq N_{i^\prime}$,
\item $(Q_{i1},Q_{i2}) = (M_1,N_{i^\prime})$ which occurs when $M_2\leq N_{i^\prime} < M_1$,
\item $(Q_{i1},Q_{i2}) = (M_1,M_2)$ which occurs when $N_{i^\prime} < M_2\leq M_1$.
\end{enumerate}
For each such case, we evaluate the conditions in (\ref{eq:dofloss1})--(\ref{eq:dofloss3}), and summarize the resulting antenna configuration regimes (with redundancies removed) in Table~\ref{tab:boundregimes}. Figures~\ref{fig:MeqCases} and~\ref{fig:NeqCases} depict the boundaries defined by the antenna configuration regimes listed in Table~\ref{tab:boundregimes}, when either transmitters have the same number of antennas ($M_1=M_2$) or receivers have the same number of antennas ($N_1=N_2$) respectively.
\begin{table}[ht]\centering
\begin{tabular}{|c|c|}
\hline
\multirow{2}{*}{Regime I} & $M_2\leq M_1\leq N_{i^\prime} < M_1+M_2$\\
 & $N_i+\frac{N_{i^\prime}}{2} < M_1+M_2$\\
\hline
\multirow{2}{*}{Regime II} & $M_2 \leq N_{i^\prime} < M_1$\\
 & $M_1 < N_1+N_2 < M_1+2M_2$\\
\hline
\multirow{3}{*}{Regime III} & $N_{i^\prime} < M_2 \leq M_1$\\
 & $N_i < M_1+M_2$\\
 & $M_1 < N_1+N_2$\\
\hline
\end{tabular}
\caption{Regimes where the RHS  of (\ref{eq:rankratio_UB2}) is strictly less than the RHS of (\ref{eq:rankratio_UB1}). Recall that WLOG we assume $M_1\geq M_2$.}\label{tab:boundregimes}
\end{table}

\begin{figure}[]
\centering
\begin{tikzpicture}[scale=2.25,font=\scriptsize]
    \filldraw[thick,draw=black,fill=red!50] (0,1) -- (0,2) -- (1,2) -- (1.333,1.333) -- (2,1) -- (2,0) -- (1,0) -- cycle;
    \draw [dotted,thin] (0,1) -- (0,2) -- (1,2) -- (1,1.5) -- (2,1) -- (2,0) -- (1,0) -- cycle;
    \draw [dashed,thin] (0,1) -- (0,2) -- (1,2) -- (1.5,1) -- (2,1) -- (2,0) -- (1,0) -- cycle;
    \draw[thick,-latex] (0,0) -- (2.5,0) node[below]{$\frac{N_1}{M}$};
    \draw[thick,-latex] (0,0) -- (0,2.5) node[left]{$\frac{N_2}{M}$};
    \draw (0,0) node[anchor=north east]{(0,0)};
    \draw (1,0) --(1,-0.1) node[below]{1};
    \draw (2,0) --(2,-0.1) node[below]{2};
    \draw (0,1) --(-0.1,1) node[left]{1};
    \draw (0,2) --(-0.1,2) node[left]{2};
    \draw[dashed,thin] (0,0) -- (2.5,2.5);
    \draw[-stealth,shorten >=2pt] (0.175,0.5) node[fill=white,inner sep=1pt] {($\frac{1}{2}$,$\frac{1}{2}$)} -- (0.5,0.5);
    \draw[-stealth,shorten >=2pt] (1.5,2) node[fill=white] {($\frac{4}{3}$,$\frac{4}{3}$)} -- (1.333,1.333);
\end{tikzpicture}
\caption{Antenna configurations where the linear sum DoF of the MIMO XC is strictly less than that of the associated MIMO BC, when $M_1=M_2 = M$. The dashed and dotted lines indicate the boundaries of the region depicted in Figure~\ref{fig:MeqCases} and its reflection across the 45-degree ray, respectively.}\label{fig:boundregimes}
\end{figure}
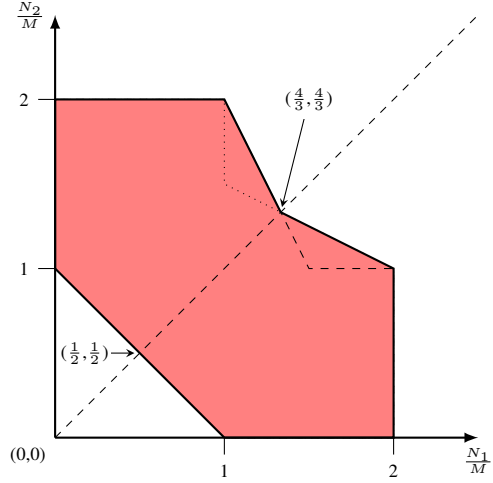

% Notice that the regime where Bound~2 is tighter than Bound~1, is not symmetric (e.g., Figure~\ref{fig:MeqCases}). However, 
Because an antenna configuration exhibits distributed transmitter loss in linear sum DoF if Bound~2 is tighter for \emph{either} $\Gamma_1$ or $\Gamma_2$, we observe that the entire set of antenna configrations exhibiting loss can be computed in the following way. 
Let $\mathcal{E}_1$ and $\mathcal{E}_2$ denote the sets of antenna configurations satisfying all conditions for any regime in Table~\ref{tab:boundregimes} when $i=1$ and $i=2$, respectively.
The set of antenna configurations exhibiting distributed transmitter loss in linear sum DoF is $\mathcal{E}=\mathcal{E}_1\cup\mathcal{E}_2$.
As an example, Figure~\ref{fig:boundregimes} depicts $\mathcal{E}$ when $M_1=M_2=M$. Note that it is computed from taking the union of the shaded region of Figure~\ref{fig:MeqCases} and its reflection across the 45-degree ray.
% Notice that, in all of our examples, the points $(\frac{1}{2},\frac{1}{2})$ and $(\frac{4}{3},\frac{4}{3})$ lie on the boundary of the depicted region which is consistent. Also generalized from the symmetric case, is the conclusion that there is no loss in LDOF due to distributed transmitters when the total number of transmit antennas greatly outnumber receive antennas or vice versa. 
% Finally, Recall that in the symmetric antenna configuration case (Figure~\ref{fig:normDOF}) the regime where such a loss occurred was $\frac{1}{2}< \frac{N}{M}<\frac{4}{3}$. 
% 

\section{Proof of Theorem~\ref{thm:DoFreg}}
\label{sec:dofreg}

In this section, we first prove the converse for Theorem~\ref{thm:DoFreg}. We then demonstrate that either using our scheme or simpler versions of it, the corner points of the linear DoF region are achievable for five regimes which span all symmetric antenna configurations.

\subsection{Converse}

The converse consists of four types of outer bounds each stated as a separate inequality in (\ref{eq:dofreg1})--(\ref{eq:dofreg4}). Inequality (\ref{eq:dofreg1}) provides four outer bounds, and is by definition of the DoF region as a set of non-negative tuples. Inequality (\ref{eq:dofreg2}) provides four outer bounds, and is a direct result of considering Transmitter~$j$ and the two receivers in isolation as a two-user MIMO BC with delayed CSIT (i.e., Theorem~1 of~\cite{VV:isit2011}). Two outer bounds equivalent to (\ref{eq:dofreg3}) were proven in equations (\ref{eq:WSM1}) and (\ref{eq:WSM2}) as part of the general linear sum DoF converse. Therefore, we need only prove inequality (\ref{eq:dofreg4}), which itself provides four outer bounds.

We prove (\ref{eq:dofreg4}) by proving two statements. The first, supplies an outer bound for linear DoF regions for general antenna configurations. The second statement applies only to symmetric antenna configurations. To arrive  at (\ref{eq:dofreg4}), we now combine the resulting inequalities, and evaluate it for symmetric antenna configurations taking whichever is dominant given values of $N$ and $M$. 

The linear DoF outer bound for general antenna configurations is given in the following lemma:
\begin{lemma} 
Any DoF tuple, $(d_{11},d_{12},d_{21},d_{22})$, that is achievable according to Definition~\ref{def:linDoF} must satisfy
\begin{align}
d_{i1} + d_{i2} 
+ \frac{N_i}{\min\{Q_{i^\prime j},N_1+N_2\}}d_{i^\prime j} 
+ \frac{[N_i-M_{j}]_+}{M_{j^\prime}}d_{i^\prime j^\prime} 
\leq{}& \min\{N_i,M_1+M_2\}.\label{eq:dofregbnd1}
\end{align}
\label{lem:regbnd1}
\end{lemma}
\begin{IEEEproof}
First, consider the case where $N_i\leq M_j$. In this case we provide each receiver with a genie aided signal fo $\vec{\mathbf{u}}_{i^\prime j^\prime}$ and observe
\begin{align}
T\big(d_{i1} + d_{i2} &+ \frac{N_i}{\min\{Q_{i^\prime j},N_1+N_2\}}d_{i^\prime j} \big)\nonumber\\
\stackrel{(a)}{\leq}{} &
    \rank[\mathbf{G}_{i1}^T\mathbf{V}_{i1}^T \quad \mathbf{G}_{i2}^T\mathbf{V}_{i2}^T 
    		\quad \mathbf{G}_{ij}^T\mathbf{V}_{i^\prime j}^T] 
    - \rank[\mathbf{G}_{ij}^T\mathbf{V}_{i^\prime j}^T] \nonumber\\
    &+ \frac{N_i}{\min\{Q_{i^\prime j},N_1+N_2\}}\left(
        \rank[\mathbf{G}_{i^\prime 1}^T\mathbf{V}_{i1}^T \quad \mathbf{G}_{i^\prime 2}^T\mathbf{V}_{i2}^T \quad \mathbf{G}_{i^\prime j}^T\mathbf{V}_{i^\prime j}^T] 
        - \rank[\mathbf{G}_{i^\prime 1}^T\mathbf{V}_{i1}^T \quad \mathbf{G}_{i^\prime 2}^T\mathbf{V}_{i2}^T] 
    \right)\nonumber\\
\stackrel{(b)}{\leq}{}& 
    \rank[\mathbf{G}_{i1}^T\mathbf{V}_{i1} \quad \mathbf{G}_{i2}^T\mathbf{V}_{i2}^T \quad \mathbf{G}_{ij}^T\mathbf{V}_{i^\prime j}^T] 
    - \frac{N_i}{\min\{Q_{i^\prime j},N_1+N_2\}}\rank[\mathbf{G}_{i^\prime j}^T\mathbf{V}_{i^\prime j}^T] \nonumber\\
    &+ \frac{N_i}{\min\{Q_{i^\prime j},N_1+N_2\}}\left(
        \rank[\mathbf{G}_{i^\prime 1}^T\mathbf{V}_{i1}^T \quad \mathbf{G}_{i^\prime 2}^T\mathbf{V}_{i2}^T \quad \mathbf{G}_{i^\prime j}^T\mathbf{V}_{i^\prime j}^T] 
        - \rank[\mathbf{G}_{i^\prime 1}^T\mathbf{V}_{i1}^T \quad \mathbf{G}_{i^\prime 2}^T\mathbf{V}_{i2}^T] 
    \right)\nonumber\\
\stackrel{(c)}{\leq}{}& 
    \rank[\mathbf{G}_{i1}^T\mathbf{V}_{i1}^T \quad \mathbf{G}_{i2}^T\mathbf{V}_{i2}^T \quad \mathbf{G}_{ij}^T\mathbf{V}_{i^\prime j}^T] \nonumber\\
    &+ \frac{N_i}{\min\{Q_{i^\prime j},N_1+N_2\}}\left(
        \rank[\mathbf{G}_{i^\prime 1}^T\mathbf{V}_{i1}^T \quad \mathbf{G}_{i^\prime 2}^T\mathbf{V}_{i2}^T \quad \mathbf{G}_{i^\prime j}^T\mathbf{V}_{i^\prime j}^T] 
        -  \rank[\mathbf{G}_{i^\prime 1}^T\mathbf{V}_{i1} \quad \mathbf{G}_{i^\prime 2}^T\mathbf{V}_{i2}^T \quad \mathbf{G}_{i^\prime j}^T\mathbf{V}_{i^\prime j}^T] 
    \right)\nonumber\\
\stackrel{}{\leq}{}& 
    \rank[\mathbf{G}_{i1}^T\mathbf{V}_{i1}^T \quad \mathbf{G}_{i2}^T\mathbf{V}_{i2}^T \quad \mathbf{G}_{ij}^T\mathbf{V}_{i^\prime j}^T] \nonumber\\
\stackrel{}{\leq}{}& T\min\{N_i,M_1+M_2\}.
\end{align}
In step (a), we applied the decodability condition (\ref{eq:decode2}), in step (b) we used Lemma~\ref{lem:BCratio}, and in step (c) we used submodularity of rank. Now, consider the case where $N_i > M_j$. In this case $\frac{N_i}{\min\{Q_{i^\prime j},N_1+N_2\}}=1$ and we have
\begin{align}
T\big(d_{i1} + d_{i2} &+ d_{i^\prime j} +\frac{[N_i-M_{j}]_+}{M_{j^\prime}}d_{i^\prime j^\prime} \big)\nonumber\\
\stackrel{(d)}{\leq}{} &
    \rank[\mathbf{G}_{i1}^T\mathbf{V}_{i1}^T \quad \mathbf{G}_{i2}^T\mathbf{V}_{i2}^T 
    		\quad \mathbf{G}_{ij}^T\mathbf{V}_{i^\prime j}^T \quad \mathbf{G}_{ij^\prime}^T\mathbf{V}_{i^\prime j^\prime}^T] 
    - \rank[\mathbf{G}_{ij}^T\mathbf{V}_{i^\prime j}^T \quad \mathbf{G}_{ij^\prime}^T\mathbf{V}_{i^\prime j^\prime}^T] \nonumber\\
    &+\rank[\mathbf{V}_{i^\prime j}^T] + \frac{[N_i-M_{j}]_+}{M_{j^\prime}}\rank[\mathbf{V}_{i^\prime j^\prime}^T] \nonumber\\
\stackrel{(e)}{\leq}{}& 
    \rank[\mathbf{G}_{i1}^T\mathbf{V}_{i1}^T \quad \mathbf{G}_{i2}^T\mathbf{V}_{i2}^T 
    		\quad \mathbf{G}_{ij}^T\mathbf{V}_{i^\prime j}^T \quad \mathbf{G}_{ij^\prime}^T\mathbf{V}_{i^\prime j^\prime}^T] \nonumber\\
\stackrel{}{\leq}{}& T\min\{N_i,M_1+M_2\}.\label{eq:dofregex2}
\end{align}
In step (d), we again applied the condition (\ref{eq:decode2}) and for step (e) we observe through expansion of rank terms into a telescoping series:
\begin{align}
\rank[\mathbf{G}_{ij}^T\mathbf{V}_{i^\prime j}^T \quad \mathbf{G}_{ij^\prime}^T\mathbf{V}_{i^\prime j^\prime}^T]
={}& \sum_{t=1}^T 
		\rank[\mathbf{G}_{ij}^t\mathbf{V}_{i^\prime j}^t \quad \mathbf{G}_{ij^\prime}^t\mathbf{V}_{i^\prime j^\prime}^t] -
		\rank[\mathbf{G}_{ij}^{t-1}\mathbf{V}_{i^\prime j}^{t-1} \quad \mathbf{G}_{ij^\prime}^{t-1}\mathbf{V}_{i^\prime j^\prime}^{t-1}]
		\nonumber\\
\leq{}& \sum_{t=1}^T 
		\rank[\mathbf{V}_{i^\prime j}^t] - \rank[\mathbf{V}_{i^\prime j}^{t-1}]
		+\frac{N_i - M_j}{M_{j^\prime}} \left(\rank[\mathbf{V}_{i^\prime j^\prime}^t] - \rank[\mathbf{V}_{i^\prime j^\prime}^{t-1}]\right)
		,
\end{align}
which implies that $\rank[\mathbf{V}_{i^\prime j}^T] + \frac{[N_i-M_{j}]_+}{M_{j^\prime}}\rank[\mathbf{V}_{i^\prime j^\prime}^T] - \rank[\mathbf{G}_{ij}^T\mathbf{V}_{i^\prime j}^T \quad \mathbf{G}_{ij^\prime}^T\mathbf{V}_{i^\prime j^\prime}^T] \leq 0$, thus justifying step (e) in inequality (\ref{eq:dofregex2}).
\end{IEEEproof}

We now claim the following statement, which provides a linear DoF outer bound only for symmetric antenna configurations:
\begin{claim} 
For the MIMO XC with delayed CSIT and symmetric antenna configurations where $M<N$, any DoF tuple, $(d_{11},d_{12},d_{21},d_{22})$, that is achievable according to Definition~\ref{def:linDoF} must satisfy:
\begin{align}
d_{i1} + d_{i2} 
+ d_{i^\prime j} 
+ \frac{N-M}{\Gamma N-M}d_{i^\prime j^\prime} 
\leq{}& \min\{N, 2M\}.\label{eq:dofregbnd2}
\end{align}\label{cl:dofregbnd2}
\end{claim}
\begin{IEEEproof}
To prove the claim we begin by applying the decoding condition (\ref{eq:decode2}) twice and then claiming that some of the terms when combined may be upper bounded by zero:
\begin{align}
	T\big(d_{i1} + d_{i2} &+ d_{i^\prime j} +\frac{N-M}{\Gamma N-M}d_{i^\prime j^\prime} \big)\nonumber\\
\stackrel{}{\leq}{} &
    \rank[\mathbf{G}_{i1}^T\mathbf{V}_{i1}^T \quad \mathbf{G}_{i2}^T\mathbf{V}_{i2}^T 
    		\quad \mathbf{G}_{ij}^T\mathbf{V}_{i^\prime j}^T \quad \mathbf{G}_{ij^\prime}^T\mathbf{V}_{i^\prime j^\prime}^T] 
    - \rank[\mathbf{G}_{ij}^T\mathbf{V}_{i^\prime j}^T \quad \mathbf{G}_{ij^\prime}^T\mathbf{V}_{i^\prime j^\prime}^T] \nonumber\\
    &+\rank[\mathbf{V}_{i^\prime j}^T] + \frac{N-M}{\Gamma N-M}\rank[\mathbf{V}_{i^\prime j^\prime}^T],\nonumber\\
\stackrel{}{\leq}{} &
    \rank[\mathbf{G}_{i1}^T\mathbf{V}_{i1}^T \quad \mathbf{G}_{i2}^T\mathbf{V}_{i2}^T 
    		\quad \mathbf{G}_{ij}^T\mathbf{V}_{i^\prime j}^T \quad \mathbf{G}_{ij^\prime}^T\mathbf{V}_{i^\prime j^\prime}^T] 
    - \rank[\mathbf{G}_{ij}^T\mathbf{V}_{i^\prime j}^T \quad \mathbf{G}_{ij^\prime}^T\mathbf{V}_{i^\prime j^\prime}^T] \nonumber\\
    &+\rank[\mathbf{V}_{i^\prime j}^T] 
    	+ \frac{N-M}{\Gamma N-M}
    	\left(
    	\rank[\mathbf{G}_{i^\prime j}^T\mathbf{V}_{i^\prime j}^T \quad \mathbf{G}_{i^\prime j^\prime}^T\mathbf{V}_{i^\prime j^\prime}^T]     	-\rank[\mathbf{V}_{i^\prime j}^T]
    \right)\nonumber\\
\stackrel{(a)}{\leq}{} &
    \rank[\mathbf{G}_{i1}^T\mathbf{V}_{i1}^T \quad \mathbf{G}_{i2}^T\mathbf{V}_{i2}^T 
    		\quad \mathbf{G}_{ij}^T\mathbf{V}_{i^\prime j}^T \quad \mathbf{G}_{ij^\prime}^T\mathbf{V}_{i^\prime j^\prime}^T] \nonumber\\
\stackrel{}{\leq}{} &
    T\min\{N,2M\} .
\end{align}
To prove (a), we will now demonstrate
\begin{align}
    \rank[\mathbf{G}_{i^\prime j}^T\mathbf{V}_{i^\prime j}^T \quad \mathbf{G}_{i^\prime j^\prime}^T\mathbf{V}_{i^\prime j^\prime}^T]     	-\rank[\mathbf{V}_{i^\prime j}^T]
\leq{}&
\frac{\Gamma N-M}{N-M}\left(\rank[\mathbf{G}_{ij}^T\mathbf{V}_{i^\prime j}^T \quad \mathbf{G}_{ij^\prime}^T\mathbf{V}_{i^\prime j^\prime}^T] -\rank[\mathbf{V}_{i^\prime j^\prime}^T]\right).
\end{align}
Notice that this is a bound on the ratio between the dimensions of subspaces of only $\mathbf{V}_{i^\prime j^\prime}^T$ recoverable at Receiver~$i^\prime$ versus Receiver~$i$. Intuitively, because $M<N$ and transmitters only have delayed CSIT, to maximize this ratio, one would expect to hide as much of the subspace as possible within that of transmissions from Transmitter $j$ at Receiver~$i$.

Consider the expansion in time of the rank difference term
\begin{align}
\rank[\mathbf{G}_{ij}^T\mathbf{V}_{i^\prime j}^T \quad \mathbf{G}_{ij^\prime}^T\mathbf{V}_{i^\prime j^\prime}^T] -\rank[\mathbf{V}_{i^\prime j^\prime}^T]
	={}&
	\sum_{t=1}^T 
		\Delta_i[t] - \delta_{i^\prime j}[t],
\end{align}
where
\begin{align*}
	\Delta_{i}[t] \triangleq & \rank[\mathbf{G}_{ij}^t\mathbf{V}_{i^\prime j}^t \quad \mathbf{G}_{ij^\prime}^t\mathbf{V}_{i^\prime j^\prime}^t] -
		\rank[\mathbf{G}_{ij}^{t-1}\mathbf{V}_{i^\prime j}^{t-1} \quad \mathbf{G}_{ij^\prime}^{t-1}\mathbf{V}_{i^\prime j^\prime}^{t-1}]\\
	\delta_{i^\prime j}[t]\triangleq & \rank[\mathbf{V}_{i^\prime j}^t] - \rank[\mathbf{V}_{i^\prime j}^{t-1}].
\end{align*}
We also define $\mathcal{L}[t] \triangleq \rowspan [\mathbf{G}_{ij}^{t-1}\mathbf{V}_{i^\prime j}^{t-1} \quad \mathbf{G}_{ij^\prime}^{t-1}\mathbf{V}_{i^\prime j^\prime}^{t-1}]$.
Notice that $\Delta_i[t] < N$ implies almost surely that $\mathbf{V}_{i^\prime j^\prime}[t] \in \mathcal{L}[t+1]$ (i.e., all of the relevant subspace transmitted at time $t$ will be recoverable, undesired, at Receiver~$i$). Therefore, time instances $t\in\mathcal{T}$,
where 
$\mathcal{T}\triangleq\{t:\quad\Delta_i[t]=N\}$, provide almost surely the only opportunities to ``hide'' a subspace of $\mathbf{V}_{i^\prime j^\prime}[t]$ the signal from Receiver~$i$.

Using this, we now have
\begin{align}
\rank[\mathbf{G}_{i^\prime j}^T\mathbf{V}_{i^\prime j}^T \quad \mathbf{G}_{i^\prime j^\prime}^T\mathbf{V}_{i^\prime j^\prime}^T]     	-\rank[\mathbf{V}_{i^\prime j}^T]
={}& \sum_{t=1}^T \Delta_{i^\prime}[t] - \delta_{i^\prime j}[t]\nonumber\\
\stackrel{\stackrel{(a)}{a.s.}}{\leq}{}& \sum_{t\in\mathcal{T}} \Gamma\Delta_{i}[t] - \delta_{i^\prime j}[t] + \sum_{t\notin\mathcal{T}} \Delta_{i}[t] - \delta_{i^\prime j}[t]\nonumber\\
\stackrel{(b)}{\leq}{}& \sum_{t\in\mathcal{T}} \frac{\Gamma N - \delta_{i^\prime j}[t]}{N- \delta_{i^\prime j}[t]}\left(\Delta_{i}[t] - \delta_{i^\prime j}[t]\right) + \sum_{t\notin\mathcal{T}} \Delta_{i}[t] - \delta_{i^\prime j}[t]\nonumber\\
\stackrel{(c)}{\leq}{}& \sum_{t\in\mathcal{T}} \frac{\Gamma N - M}{N- M}\left(\Delta_{i}[t] - \delta_{i^\prime j}[t]\right) + \sum_{t\notin\mathcal{T}} \Delta_{i}[t] - \delta_{i^\prime j}[t]\nonumber\\
\stackrel{(d)}{\leq}{}& \frac{\Gamma N - M}{N- M}\left(\sum_{t\in\mathcal{T}} \Delta_{i}[t] - \delta_{i^\prime j}[t] + \sum_{t\notin\mathcal{T}} \Delta_{i}[t] - \delta_{i^\prime j}[t]\right)\nonumber\\
={}&\frac{\Gamma N-M}{N-M}\left(\rank[\mathbf{G}_{ij}^T\mathbf{V}_{i^\prime j}^T \quad \mathbf{G}_{ij^\prime}^T\mathbf{V}_{i^\prime j^\prime}^T] -\rank[\mathbf{V}_{i^\prime j^\prime}^T]\right),
\end{align}
as desired. In step (a) we noted that a rank ratio greater than one can almost surely occur only if there exists subspaces of $\mathbf{V}_{i^\prime j^\prime}[t]$ that are not recoverable by Receiver $i$, in step (b) we observe that, by definition, $t\in\mathcal{T}$ implies $\Delta_i[t]=N$, in (c) we note that $\delta_{i^\prime j}[t]\leq M$, and in (d) we note that for $N>M$, the factor $\frac{\Gamma N - M}{N- M}\geq 1$.
\end{IEEEproof}

To arrive  at (\ref{eq:dofreg4}), we now apply (\ref{eq:dofregbnd1}) for $N\leq M$ and the tighter inequality out of (\ref{eq:dofregbnd1}) and (\ref{eq:dofregbnd2}) for $N>M$. 

\subsection{Achievability}

We now address achievability of the region given by (\ref{eq:dofreg1})--(\ref{eq:dofreg4}). We do so by considering five symmetric antenna configuration regimes. Notably, the five regimes are those identified in Section~\ref{sec:sym} and listed in Table~\ref{tab:DoF}. For each regime, we evaluate bounds (\ref{eq:dofreg1})--(\ref{eq:dofreg4}), often resulting in simpler expressions for the region. We then determine the resulting corner points of the linear DoF region and demonstrate that either using the scheme described in Section~\ref{sec:scheme} or simpler transmission strategy, all corner points are achievable. Since the corner points are achievable, through time division the regions are achievable.

\subsubsection*{Regime 1 ($\frac{N}{M}\leq\frac{1}{2}$)}
In this regime, we evaluate and see that $Q=M$, $\Gamma=2$ and noting that for these values (\ref{eq:dofreg3}) is a stricter condition than either (\ref{eq:dofreg2}) and (\ref{eq:dofreg4}), the region may be expressed simply as the set of tuples $(d_{11},d_{12},d_{21},d_{22})$ satisfying
\begin{align}
d_{ij} \geq{}& 0,\label{eq:dofreg_1_1}\\
%d_{ij} + \frac{1}{2}d_{i^\prime j} \leq{}& N,\label{eq:dofreg_1_2}\\
d_{i1} + d_{i2} + \frac{1}{2}(d_{i^\prime 1}+d_{i^\prime 2}) \leq{}& N.\label{eq:dofreg_1_3}
%d_{i1} + d_{i2} + \frac{1}{2}d_{i^\prime j} \leq{}& N.\label{eq:dofreg_1_4}
\end{align}
The region has eight corner points we we list in two groups:
\begin{itemize}
\item $(0,0,0,N)$, $(0,0,N,0)$, $(0,N,0,0)$, $(N,0,0,0)$
\item $(0,\frac{2N}{3},0,\frac{2N}{3})$, $(0,\frac{2N}{3},\frac{2N}{3},0)$, $(\frac{2N}{3},0,0,\frac{2N}{3})$, $(\frac{2N}{3},0,\frac{2N}{3},0)$
\end{itemize}
The first four may be achieved by selecting a single Transmitter-Receiver pair to communicate. For the second set of four, we point out that our scheme, as defined in Section~\ref{sec:scheme} achieves the tuple
\begin{align}
    (d_{11},d_{12},d_{21},d_{22}) = \left(\frac{2N}{3},0,\frac{2N}{3},0\right),
\end{align}
and did so by keeping Transmitter~2 silent during both Phases~1 and 2. However, through reassigning the roles of Transmitters in Phases~1 and 2 (i.e., reindexing transmitter nodes for each phase), we may achieve the remaining corner points.

\subsubsection*{Regime 2 ($\frac{1}{2}<\frac{N}{M}\leq 1$)}
In this regime, we evaluate and see that $Q=M$, $\Gamma=\frac{3M}{M+N}$. Using these, we note that (\ref{eq:dofreg4}) is a stricter condition than (\ref{eq:dofreg2}), and the region may be expressed simply as the set of tuples $(d_{11},d_{12},d_{21},d_{22})$ satisfying 
\begin{align}
d_{ij} \geq{}& 0,\label{eq:dofreg_2_1}\\
%d_{ij} + \frac{N}{M}d_{i^\prime j} \leq{}& N,\label{eq:dofreg_2_2}\\
d_{i1} + d_{i2} + \frac{M+N}{3M}(d_{i^\prime 1}+d_{i^\prime 2}) \leq{}& N,\label{eq:dofreg_2_3}\\
d_{i1} + d_{i2} + \frac{N}{M}d_{i^\prime j} \leq{}& N.\label{eq:dofreg_2_4}
\end{align}
The region has 12 corner points listed here in three groups:
\begin{itemize}
\item $(0,0,0,N)$, $(0,0,N,0)$, $(0,N,0,0)$, $(N,0,0,0)$ 
\item $(0,\frac{NM}{N+M},0,\frac{NM}{N+M})$, $(0,\frac{NM}{N+M},\frac{NM}{N+M},0)$, $(\frac{NM}{N+M},0,0,\frac{NM}{N+M})$, $(\frac{NM}{N+M},0,\frac{NM}{N+M},0)$
\item $(\frac{M(N+M)}{N+4M},\frac{M(2N-M)}{N+4M},\frac{M(N+M)}{N+4M},\frac{M(2N-M)}{N+4M})$, 
    $(\frac{M(2N-M)}{N+4M},\frac{M(N+M)}{N+4M},\frac{M(N+M)}{N+4M},\frac{M(2N-M)}{N+4M})$, \\
    $(\frac{M(N+M)}{N+4M},\frac{M(2N-M)}{N+4M},\frac{M(2N-M)}{N+4M},\frac{M(N+M)}{N+4M})$, 
    $(\frac{M(2N-M)}{N+4M},\frac{M(N+M)}{N+4M},\frac{M(2N-M)}{N+4M},\frac{M(N+M)}{N+4M})$
\end{itemize}

The first four may be achieved by selecting a single Transmitter-Receiver pair to communicate. 

With the second set of four, we note that, as in Regime~1, when we force a Transmitter~2 to be silent during Phases~1 and 2 of our scheme (i.e., allowing Transmitter~1 to act as a MIMO BC with delayed CSIT), we achieve the tuple
\begin{align}
    (d_{11},d_{12},d_{21},d_{22}) = \left(\frac{NM}{N+M},0,\frac{NM}{N+M},0\right).
\end{align}
Again, through reassigning the roles of Transmitters in Phases~1 and 2 (i.e., specifying who should be silent) we may achieve the remaining corner points of the set.

For the last set, one may verify from Section~\ref{sec:sym} that the linear sum DoF optimal tuple achieved by our scheme in this regime is
\begin{align}
    (d_{11},d_{12},d_{21},d_{22}) = \left(\frac{M(N+M)}{N+4M},\frac{M(2N-M)}{N+4M},\frac{M(N+M)}{N+4M},\frac{M(2N-M)}{N+4M}\right),
\end{align}
which is the first tuple in the third set of corner points. The other tuples within the third set of DoF tuples may be achieved by reassigning the roles of
transmitters in Phases~1 and 2.

\subsubsection*{Regime 3 ($1<\frac{N}{M}\leq \frac{4}{3}$)}
In this regime, we evaluate and see that $Q=N$, $\Gamma=\frac{3}{2}$. For these values, (\ref{eq:dofreg1})--(\ref{eq:dofreg4}) yields
\begin{align}
d_{ij} \geq{}& 0,\label{eq:dofreg_3_1}\\
d_{ij} + d_{i^\prime j} \leq{}& M,\label{eq:dofreg_3_2}\\
d_{i1} + d_{i2} + \frac{2}{3}(d_{i^\prime 1}+d_{i^\prime 2}) \leq{}& N,\label{eq:dofreg_3_3}\\
d_{i1} + d_{i2} + d_{i^\prime j} + \frac{2N-2M}{3N-2M}d_{i^\prime j^\prime} \leq{}& N.\label{eq:dofreg_3_4}
\end{align}

The region has twelve corner points listed here in three groups:
\begin{itemize}
\item 
    $(M,0,0,0)$,
    $(0,M,0,0)$, 
    $(0,0,M,0)$,
    $(0,0,0,M)$, 
\item 
    $(M,N-M,0,0)$,
    $(M,0,0,N-M)$
    $(N-M,M,0,0)$, 
    $(0,M,N-M,0)$, \\
    $(0,N-M,M,0)$,
    $(0,0,M,N-M)$, 
    $(N-M,0,0,M)$, 
    $(0,0,N-M,M)$, 
\item 
$(\frac{2M}{5},\frac{3N-2M}{5},\frac{2M}{5},\frac{3N-2M}{5})$,
$(\frac{2M}{5},\frac{3N-2M}{5},\frac{3N-2M}{5},\frac{2M}{5})$,\\
$(\frac{3N-2M}{5},\frac{2M}{5},\frac{2M}{5},\frac{3N-2M}{5})$,
$(\frac{3N-2M}{5},\frac{2M}{5},\frac{3N-2M}{5},\frac{2M}{5})$,
\end{itemize}

The first set of four may be achieved by activating a single transmitter-receiver pair and using a single user code.

The second set of eight corner points may be achieved by selecting one transmitter to communicate $M$ symbols to one receiver, and allowing the other transmitter to communicate $N-M$ symbols to either receiver. The receivers decode both messages, and the result is a multiple access-like scenario.

The DoF tuple achieved by our linear sum DoF optimal scheme is
\begin{align}
    (d_{11},d_{12},d_{21},d_{22}) = \left(\frac{2M}{5},\frac{3N-2M}{5},\frac{2M}{5},\frac{3N-2M}{5}\right),
\end{align}
which is the first tuple in the final set of four corner points. Again, the other tuples within the third set of DoF tuples may be achieved by reassigning the roles of transmitters in Phases~1 and 2.

\subsubsection*{Regime 4 ($\frac{4}{3}<\frac{N}{M}\leq 2$)}
In this regime, we evaluate and see that $Q=N$, $\Gamma=\frac{2M}{N}$ and for these values, (\ref{eq:dofreg1})--(\ref{eq:dofreg4}) yields
\begin{align}
d_{ij} \geq{}& 0,\label{eq:dofreg_4_1}\\
d_{ij} + d_{i^\prime j} \leq{}& M,\label{eq:dofreg_4_2}\\
d_{i1} + d_{i2} + \frac{N}{2M}(d_{i^\prime 1}+d_{i^\prime 2}) \leq{}& N,\label{eq:dofreg_4_3}\\
d_{i1} + d_{i2} + d_{i^\prime j} + \frac{N-M}{M}d_{i^\prime j^\prime} \leq{}& N.\label{eq:dofreg_4_4}
\end{align}

The region has thirteen corner points listed here in three groups:
\begin{itemize}
\item 
    $(M,0,0,0)$,
    $(0,M,0,0)$, 
    $(0,0,M,0)$,
    $(0,0,0,M)$, 
\item 
    $(M,N-M,0,0)$,
    $(M,0,0,N-M)$
    $(N-M,M,0,0)$, 
    $(0,M,N-M,0)$, \\
    $(0,N-M,M,0)$,
    $(0,0,M,N-M)$, 
    $(N-M,0,0,M)$, 
    $(0,0,N-M,M)$, 
\item $(\frac{NM}{N+2M},\frac{NM}{N+2M},\frac{NM}{N+2M},\frac{NM}{N+2M})$
\end{itemize}

The first set of four may be achieved by activating a single transmitter-receiver pair and using a single user code.

The second set of eight corner points may be achieved by selecting one transmitter to communicate $M$ symbols to one receiver, and allowing the other transmitter to communicate $N-M$ symbols to either receiver. The receivers decode both messages, and the result is a multiple access-like scenario.

The DoF tuple achieved by our linear sum DoF optimal scheme is
\begin{align}
    (d_{11},d_{12},d_{21},d_{22}) = \left(\frac{NM}{N+2M},\frac{NM}{N+2M},\frac{NM}{N+2M},\frac{NM}{N+2M}\right),
\end{align}
which is the final corner point.

\subsubsection*{Regime 5 ($2<\frac{N}{M}$)}
In this regime, we evaluate and see that $Q=N$, $\Gamma=1$. For these values we note that (\ref{eq:dofreg2}) implies both (\ref{eq:dofreg3}) and (\ref{eq:dofreg4}), thus the region is given by tuples tuples $(d_{11},d_{12},d_{21},d_{22})$ satisfying 
\begin{align}
d_{ij} \geq{}& 0,\label{eq:dofreg_5_1}\\
d_{ij} + d_{i^\prime j} \leq{}& M.\label{eq:dofreg_5_2}
%d_{i1} + d_{i2} + d_{i^\prime 1}+d_{i^\prime 2} \leq{}& 2M,\label{eq:dofreg_5_3}\\
%d_{i1} + d_{i2} + d_{i^\prime j} + \frac{N-M}{M}d_{i^\prime j^\prime} \leq{}& 2M\label{eq:dofreg_5_4}
\end{align}
We see that the region has four corner points:
$(0,0,M,M)$, $(0,M,M,0)$, $(M,0,0,M)$, and $(M,M,0,0)$. For each corner point we select only one receiver for each transmitter (possibly the same receiver for both transmitters). Transmitters transmit simultaneously and receivers have sufficient number of antennas to separate signals from both transmitters and linearly decode. Notice that in this regime no CSIT is needed.

\section{Summary}
\label{sec:end}
We provided new results on the linear degrees of freedom of MIMO XC with delayed CSIT. First, we established the linear sum degrees of freedom of the MIMO XC with delayed CSIT, for all antenna configurations at nodes. Second, we characterized for symmetric antenna configurations the linear DoF region for the MIMO XC with delayed CSIT. The notion of maximum rank-ratio was critical in developing new converses and a linear sum DoF-optimal transmission strategy. Additionally, we used the analysis of the maximum rank-ratio to identify the set of antenna configurations where linear sum DoF of the MIMO XC with delayed CSIT is strictly less than in an analogous MIMO broadcast channel, i.e., antenna configurations which exhibit ``distributed transmitter loss''. 

\bibliographystyle{IEEEtran}
\bibliography{references}

\appendix
%%%%%%%%%% %%%%%%%%%% 
\subsection{Proof of Lemma~\ref{lem:BCratio}}
\label{append:BCrankratio}
\begin{secproof}
We prove the lemma for the case of $i=1$. A similar analysis yields the analogous result for $i=2$. Also for notational simplicity we use $\mathbf{V}^T=\mathbf{V}_i^T$ since we are focused only on a single precoding matrix. We proceed by analyzing a genie-aided channel where the genie provides the output of Receiver~2 to Receiver~1. This genie at best increases the rank of Receiver~1's output, and therefore maximum rank-ratio of the genie-enhanced channel upper bounds the rank-ratio of the original channel. The resulting channel is a physically degraded MIMO BC with $M$ transmit antennas, $\underline{N}_1 = N_1+N_2$ receive antennas at the genie-aided Receiver~1, and $N_2$ antennas at Receiver~2. We denote the genie aided channel to Receiver~1 as
\begin{align}
    \underline{G}_1[t]\triangleq{}&
        \begin{bmatrix} G_1[t]\\G_2[t]\end{bmatrix},
\end{align}
and adhere to conventions stated in Section~\ref{sec:main} for $\underline{\mathbf{G}}_1[t]$ and $\underline{\mathbf{G}}_1^t$.

The following claim will be used in the proof:
\begin{claim}\label{cl:BClemma}
For any linear coding strategies $f^{(T)}$ with corresponding precoding matrix $\mathbf{V}^T$, if
\begin{align}
\rank[\mathbf{G}_2^t\mathbf{V}^t]& - \rank[\mathbf{G}_2^{t-1}\mathbf{V}^{t-1}] < N_2,
\end{align}
then
\begin{align}
     \rank[\underline{\mathbf{G}}_1^t\mathbf{V}^t] - \rank[\underline{\mathbf{G}}_1^{t-1}\mathbf{V}^{t-1}] \stackrel{a.s.}{\leq} \rank[\mathbf{G}_2^t\mathbf{V}^t] - \rank[\mathbf{G}_2^{t-1}\mathbf{V}^{t-1}].
\end{align}
\end{claim}
\begin{IEEEproof}
Let sets $\mathcal{A}_t$ and $\mathcal{B}_t$ be defined as
\begin{align}
\mathcal{A}_t \triangleq \{\mathcal{G}^T:\quad &\rank[{G}_2^t{V}^t] - \rank[{G}_2^{t-1}{V}^{t-1}] <N_2\}, \\
\mathcal{B}_t \triangleq \{\mathcal{G}^T : \quad &\exists {W}[t] \in\mathbb{C}^{M\times M},
        \ s.t.\ \rowspan{W}[t] {V}[t]\subseteq 
        \rowspan[{G}_2^{t-1}{V}^{t-1}], \quad \rank{W}[t] > M-N_2 \}.
\end{align}
We first claim the following, which is stated more generally and proven in Appendix~\ref{app:pAUBc} for the MIMO XC;
note that if one of the transmitters in the MIMO XC has zero antennas, then the resulting network is equivalent to a MIMO BC. 
\begin{claim*}
$\Pr(\bm{\mathcal{G}}^T\in\mathcal{A}_t\cap\mathcal{B}_t^c) = 0$.
\end{claim*} 
The claim implies that satisfying the if condition of Claim~\ref{cl:BClemma} almost surely requires $\bm{\mathcal{G}}^T\in\mathcal{B}_t$. 

Assuming $\Pr(\bm{\mathcal{G}}^T\in\mathcal{A}_t\cap\mathcal{B}_t^c) = 0$, we now show that 
\begin{align}
\bm{\mathcal{G}}^T\in\mathcal{B}_t \quad \Rightarrow \quad \rank[\underline{\mathbf{G}}_1^t\mathbf{V}^t] - \rank[\underline{\mathbf{G}}_1^{t-1}\mathbf{V}^{t-1}] 
\stackrel{a.s.}{\leq}{}& \rank[\mathbf{G}_2^t\mathbf{V}^t] - \rank[\mathbf{G}_2^{t-1}\mathbf{V}^{t-1}]. \label{eq:cl:BClemma:last}
\end{align}

For a given channel realization $\mathcal{G}^T$, let spaces $\mathcal{J}$ and $\mathcal{L}$ be defined as
\begin{align}
    \mathcal{J} \triangleq{}& \rowspan [\underline{G}_1^{t-1}V^{t-1}],\\
    \mathcal{L} \triangleq{}& \rowspan [{G}_2^{t-1}V^{t-1}],
\end{align}
and note that rows of ${G}_2^{t-1}V^{t-1}$ are also rows of $\underline{G}_1^{t-1}V^{t-1}$, and therefore $\mathcal{L} \subseteq \mathcal{J}$.

Now, let $\mathbf{W}[t]\in\mathbb{C}^{M\times M}$ be a matrix satisfying the condition in the definition of $\mathcal{B}_t$, and define $\mathbf{W}^c[t]\in\mathbb{C}^{M\times M}$ and adapted channel matrices $\mathbf{\widetilde{G}}_1[t]$ and $\mathbf{\widetilde{G}}_2[t]$ such that they satisfy 
\begin{align}
    \rowspan\mathbf{W}[t]\mathbf{V}[t] \subseteq {}& \ker(\Proj_{\mathcal{L}^c}),\\
    \rowspan\mathbf{W}^c[t]\mathbf{V}[t] \subseteq {}& \im(\Proj_{\mathcal{L}^c}),\\
    \rowspan\mathbf{W}[t]\mathbf{V}[t] \cup \rowspan \mathbf{W}^c[t]\mathbf{V}[t] = & \rowspan \mathbf{V}[t],\\
%    \rank (\mathbf{W}[t] + \mathbf{W}^c[t])={}& M,\\
    \mathbf{\widetilde{G}}_1[t] ={}& \mathbf{G}_1[t](\mathbf{W}[t] + \mathbf{W}^c[t])^{-1},\\
    \mathbf{\widetilde{G}}_2[t] ={}& \mathbf{G}_2[t](\mathbf{W}[t] + \mathbf{W}^c[t])^{-1}.
\end{align}
Since $\rank\mathbf{W}[t]\mathbf{V}[t]> M-N_2$, by rank-nullity theorem~(\cite{Roman_book} page 63) $\rank\mathbf{W}^c[t]\mathbf{V}[t] < N_2$. 

Finally we establish (\ref{eq:cl:BClemma:last}):
\begin{align}
    \rank[\underline{\mathbf{G}}_1^t\mathbf{V}^t] 
            - \rank[\underline{\mathbf{G}}_1^{t-1}\mathbf{V}^{t-1}]
        \stackrel{}{=}{}&\rank\Proj_{\mathcal{J}^c} \begin{bmatrix}\mathbf{G}_1[t] \\ \mathbf{G}_2[t]\end{bmatrix}\mathbf{V}[t]\label{eq:BCclaim1}\\
        \stackrel{(a)}{\leq}{}&\rank\Proj_{\mathcal{L}^c} \begin{bmatrix}\mathbf{G}_1[t] \\ \mathbf{G}_2[t]\end{bmatrix}\mathbf{V}[t]\label{eq:BCclaim1}\\
        \stackrel{}{=}{}&\rank\Proj_{\mathcal{L}^c} 
            \begin{bmatrix}\mathbf{\widetilde{G}}_1[t] \\ \mathbf{\widetilde{G}}_2[t]\end{bmatrix}
            (\mathbf{W}[t]\mathbf{V}[t] + \mathbf{W}^c[t]\mathbf{V}[t])\\
        \stackrel{(b)}{=}{}&\rank\Proj_{\mathcal{L}^c} 
            \begin{bmatrix}\mathbf{\widetilde{G}}_1[t] \\ \mathbf{\widetilde{G}}_2[t]\end{bmatrix}
            \mathbf{W}^c[t]\mathbf{V}[t]\label{eq:BCclaim2}\\
        \stackrel{\stackrel{(c)}{a.s.}}{=}&\rank\mathbf{W}^c[t]\mathbf{V}[t]\label{eq:BCclaim3}\\
        \stackrel{\stackrel{(d)}{a.s.}}{=}&\rank\Proj_{\mathcal{L}^c} \mathbf{\widetilde{G}}_2[t]
            \mathbf{W}^c[t]\mathbf{V}[t]\\
        \stackrel{}{=}{}&\rank\Proj_{\mathcal{L}^c} \mathbf{\widetilde{G}}_2[t]
            (\mathbf{W}[t]\mathbf{V}[t] + \mathbf{W}^c[t]\mathbf{V}[t])\\
        \stackrel{}{=}{}&\rank\Proj_{\mathcal{L}^c} \mathbf{G}_2[t]\mathbf{V}[t]\label{eq:BCclaim4}\\
        ={}&\rank[\mathbf{G}_2^t\mathbf{V}^t] - \rank[\mathbf{G}_2^{t-1}\mathbf{V}^{t-1}]
\end{align}
In step (a) we note that $\mathcal{L} \subseteq \mathcal{J}$ implies $\mathcal{J}^c \subseteq \mathcal{L}^c$ and that projection onto a larger subspace can only increase rank.
In step (b) we note that by definition $\rowspan\mathbf{W}[t]\mathbf{V}[t]\in\ker\left(\Proj_{\mathcal{L}^c}\right)$. In steps (c) and (d) we observe that, 1) by definition, $\rowspan\mathbf{W}^c[t]\mathbf{V}[t]\in\im\left(\Proj_{\mathcal{L}^c}\right)$, and 2) because the adapted channel matrices are continuously distributed and thus almost surely full rank, $\begin{bmatrix}\mathbf{\widetilde{G}}_1[t] \\ \mathbf{\widetilde{G}}_2[t]\end{bmatrix}$ and $\mathbf{\widetilde{G}}_2[t]$ are almost surely full rank.
\end{IEEEproof}

We now use Claim~\ref{cl:BClemma} to complete the proof of Lemma~\ref{lem:BCratio}.  For the special case of $M\leq N_2$, regardless of the linear encoding chosen at the transmitter, the if condition of Claim~\ref{cl:BClemma} holds, and thus when $M\leq N_2$, we have \footnote{An alternative proof for $M\leq N_2$ may also be constructed from the argument that both receivers in the genie-aided MIMO BC can almost surely recover the full span of transmitted linear equations.}
\begin{align}
    \rank[\underline{\mathbf{G}}_1^T\mathbf{V}^T] 
    = \rank\begin{bmatrix} \mathbf{G}_1^T\mathbf{V}^T \\ \mathbf{G}_2^T\mathbf{V}^T\end{bmatrix} 
    \stackrel{a.s.}{=} \rank[\mathbf{G}_2^T\mathbf{V}^T].\label{eq:BCratioproof1}
\end{align}

When $N_2 < M$, we define $\mathcal{T}_0 = \{t: \quad \rank[{G}_2^t{V}^t] - \rank[{G}_2^{t-1}{V}^{t-1}] <N_2\}$ and apply Claim~\ref{cl:BClemma} in the following way:
\begin{align}
    \rank[\underline{\mathbf{G}}_1^T\mathbf{V}^T]& - \rank[\mathbf{G}_2^T\mathbf{V}^T]\nonumber\\
        \leq{}& \sum_{t=1}^T \rank[\underline{\mathbf{G}}_1^t\mathbf{V}^t] - \rank[\underline{\mathbf{G}}_1^{t-1}\mathbf{V}^{t-1}] - \rank[\mathbf{G}_2^t\mathbf{V}^t] + \rank[\mathbf{G}_2^{t-1}\mathbf{V}^{t-1}]\\
        ={}& \sum_{t\in\mathcal{T}_0} \rank[\underline{\mathbf{G}}_1^t\mathbf{V}^t] - \rank[\underline{\mathbf{G}}_1^{t-1}\mathbf{V}^{t-1}] - \rank[\mathbf{G}_2^t\mathbf{V}^t] + \rank[\mathbf{G}_2^{t-1}\mathbf{V}^{t-1}]\nonumber\\
        &+ \sum_{t\notin\mathcal{T}_0} \rank[\underline{\mathbf{G}}_1^t\mathbf{V}^t] - \rank[\underline{\mathbf{G}}_1^{t-1}\mathbf{V}^{t-1}] - \rank[\mathbf{G}_2^t\mathbf{V}^t] + \rank[\mathbf{G}_2^{t-1}\mathbf{V}^{t-1}]\\
        \stackrel{\stackrel{\text{Claim~\ref{cl:BClemma}}}{a.s.}}{\leq}{}&\sum_{t\notin\mathcal{T}_0} \rank[\underline{\mathbf{G}}_1^t\mathbf{V}^t] - \rank[\underline{\mathbf{G}}_1^{t-1}\mathbf{V}^{t-1}] - \rank[\mathbf{G}_2^t\mathbf{V}^t] + \rank[\mathbf{G}_2^{t-1}\mathbf{V}^{t-1}]\\
        \stackrel{(a)}{\leq}{}& \sum_{t\notin\mathcal{T}_0}\frac{\min\{M,N_1+N_2\}-N_2}{N_2}\left(\rank[\mathbf{G}_2^t\mathbf{V}^t] - \rank[\mathbf{G}_2^{t-1}\mathbf{V}^{t-1}]\right)\\
        \leq{}& \sum_{t=1}^T\frac{\min\{M,N_1+N_2\}-N_2}{N_2}\left(\rank[\mathbf{G}_2^t\mathbf{V}^t] - \rank[\mathbf{G}_2^{t-1}\mathbf{V}^{t-1}]\right)\\
        ={}& \frac{\min\{M,N_1+N_2\}-N_2}{N_2}\rank[\mathbf{G}_2^T\mathbf{V}^T].
\end{align}
In step (a) we note that for every $t \notin \mathcal{T}_0$, $\rank[\mathbf{G}_2^t\mathbf{V}^t] - \rank[\mathbf{G}_2^{t-1}\mathbf{V}^{t-1}]=N_2$, and the maximum rank increase for the term $\rank[\overline{\mathbf{G}}_1^t\mathbf{V}^t] - \rank[\overline{\mathbf{G}}_1^{t-1}\mathbf{V}^{t-1}]$ is $\min\{M,N_1+N_2\}$. Rearranging terms, we have that when $N_2 < M$
\begin{align}
\rank[\overline{\mathbf{G}}_1^T\mathbf{V}^T] \leq \frac{\min\{M,N_1+N_2\}}{N_2}\rank[\mathbf{G}_2^T\mathbf{V}^T].\label{eq:BCratioproof2}
\end{align}
Noting that $\rank[\mathbf{G}_1^T\mathbf{V}^T] \leq \rank[\overline{\mathbf{G}}_1^T\mathbf{V}^T]$, and combining (\ref{eq:BCratioproof1}) and (\ref{eq:BCratioproof2}) we arrive at the desired statement.
\end{secproof}
%%%%%%%%%% %%%%%%%%%% 
\subsection{Proof of Lemma~\ref{lem:diff}}
\label{append:normdiff}
\begin{secproof}

We prove the lemma for the case of $i=1$. A similar analysis yields the analogous result for $i=2$. Also, for notational simplicity we use $\mathbf{V}_{j}^T=\mathbf{V}_{ij}^T$ since only precoding matrices favoring a single receiver are considered. Before proving Lemma~\ref{lem:diff}, we define the following notions used in the proof.
\begin{definition}
For any linear coding strategy $\{f_1^{(T)},f_2^{(T)}\}$ and a realization $\mathcal{G}^T$ of the random variable $\bm{\mathcal{G}}^T$, let the set ${\mathcal{T}}_{\{f_1^{(T)},f_2^{(T)}\}}({\mathcal{G}}^T)$ be defined as
\begin{align}
    {\mathcal{T}}_{\{f_1^{(T)},f_2^{(T)}\}}({\mathcal{G}}^T) \triangleq \{t: \rank[{G}_{21}^{t}{V}_1^{t}\quad {G}_{22}^{t}{V}_2^{t}] - \rank[{G}_{21}^{t-1}{V}_1^{t-1}\quad {G}_{22}^{t-1}{V}_2^{t-1}] < N_2\}.
\end{align}
Accordingly, $\bm{\mathcal{T}}_{\{f_1^{(T)},f_2^{(T)}\}}$ is a random subset of $\{1,2,\ldots,T\}$ and is a function of the strategy $\{f_1^{(T)},f_2^{(T)}\}$ and random channels $\bm{\mathcal{G}}^n$. For the remainder of the analysis we a single fixed strategy and use the shorthand, $\bm{\mathcal{T}}$.
%If $t_1, t_2,\ldots t_{|\mathcal{T}|}$ are the elements of $\mathcal{T}$, and $t_i<t_j$ for $i<j$, 
We also use the notation $\mathbf{V}_j^{\tau_k}$ and $\mathbf{G}_{ij}^{\tau_k}$ to refer to submatrices of $\mathbf{V}_j^T$ and $\mathbf{G}_{ij}^T$ respectively, representing the concatenation of precoding matrices and channel matrices of only time instances $\tau_1, \tau_2,\ldots \tau_{k} \in \mathcal{T}$, where $\tau_{k-1} < \tau_{k}$ and $k\leq |\mathcal{T}|$.  
\end{definition}

\begin{definition}
For a linear coding strategy $\{f_1^{(T)},f_2^{(T)}\}$ and a realization $\mathcal{G}^T$ of the random variable $\bm{\mathcal{G}}^T$, let the random variables
$\mathbf{r}_1$ and $\mathbf{r}_2$ be defined in the following manner. Let
\begin{align}
    r_i(\mathcal{G}^T) \triangleq{}& \dimspan \mathcal{E}_i(\mathcal{G}^T),
\end{align}
where
\begin{align}
    \mathcal{E}_1(\mathcal{G}^T) \triangleq{}& \{\vec{s}\in\mathbb{C}^{1\times m_1(T)}: \exists\vec{\ell}_{N_2T\times 1}
        \text{ s.t. }[\vec{s} \quad \vec{0}_{1\times m_2(T)}] = \vec{\ell}^\top[G_{21}^{T}V_1^{T} \quad G_{22}^{T}V_2^{T}]\},\\
    \mathcal{E}_2(\mathcal{G}^T) \triangleq{}& \{\vec{s}\in\mathbb{C}^{1\times m_2(T)}: \exists\vec{\ell}_{N_2T\times 1}
        \text{ s.t. }[\vec{0}_{1\times m_1(T)} \quad \vec{s}] = \vec{\ell}^\top[G_{21}^{T}V_1^{T} \quad G_{22}^{T}V_2^{T}]\}.
\end{align}
We denote as $\mathbf{r}_1$ and $\mathbf{r}_2$ the resulting random variables. Each $\mathbf{r}_j$ may be interpreted as the number of linearly independent equations that Receiver~2 can recover from its received signal, which only involve symbols from Transmitter~$j$.
\end{definition}

Using the definitions, we now state the steps to prove Lemma~\ref{lem:diff}: 
\begin{align}
\frac{\rank[\mathbf{G}_{11}^n \mathbf{V}_1^n \quad \mathbf{G}_{12}^n \mathbf{V}_2^n]}{N_1} 
    &-\frac{\rank[\mathbf{G}_{21}^n \mathbf{V}_1^n \quad \mathbf{G}_{22}^n \mathbf{V}_2^n]}{N_2} \nonumber\\
        \stackrel{\stackrel{(a)}{a.s.}}{\leq}&\frac{\rank[\mathbf{G}_{11}^\mathcal{T} \mathbf{V}_1^\mathcal{T} 
                            \quad \mathbf{G}_{12}^\mathcal{T} \mathbf{V}_2^\mathcal{T}]}{N_1}\label{eq:diffstep1}\\
    \stackrel{(b)}{\leq}{}&\frac{\rank[\mathbf{G}_{11}^\mathcal{T} \mathbf{V}_1^\mathcal{T}] 
                             +\rank[\mathbf{G}_{12}^\mathcal{T} \mathbf{V}_2^\mathcal{T}]}{N_1}\label{eq:diffproof1}\\
    \stackrel{(c)}{\leq}{}&\frac{\rank[\mathbf{V}_1^\mathcal{T}] 
                             +\rank[\mathbf{V}_2^\mathcal{T}]}{N_1}\label{eq:diffproof2}\\
    \stackrel{\stackrel{(d)}{a.s.}}{\leq}&\frac{\mathbf{r}_1 + \mathbf{r}_2}{N_1}\label{eq:diffstep2}\\
    \stackrel{\stackrel{(e)}{a.s.}}{\leq}&
        \frac{\rank[\mathbf{G}_{21}^n \mathbf{V}_1^n \quad \mathbf{G}_{22}^n \mathbf{V}_2^n] - \rank[\mathbf{G}_{22}^n \mathbf{V}_{2}^n ]}{N_1}
        + \frac{\rank[\mathbf{G}_{21}^n \mathbf{V}_1^n \quad \mathbf{G}_{22}^n \mathbf{V}_2^n] - \rank[\mathbf{G}_{21}^n \mathbf{V}_{1}^n ]}{N_1}.\label{eq:diffstep3}
\end{align}
Step (a) is proven in Appendix~\ref{app:step1}. 
Step (b) results from submodularity of the rank operation and step (c) observes that for any two matrices $A$ and $B$ where $AB$ is defined, $\rank AB \leq \rank B$. Step (d) is proven in Appendix~\ref{app:step2}, and step (e) was proven in part~C of Appendix~A in~\cite{LAS:ARXIV2013}.
\end{secproof}

%\noindent\textbf{Remark: } We point out that steps (a), (d), and (e) are similar to those highlighted in Lemma~3 of \cite{LAS:ARXIV2013}. Indeed, as we mentioned, the proof for step (e) is exactly the one given in~\cite{LAS:ARXIV2013}. This is because our definition of $\mathbf{r}_1$ and $\mathbf{r}_2$ is identical to the one given in~\cite{LAS:ARXIV2013}. Notice, however, that our definition for $\bm{\mathcal{T}}$ is different and that steps (a) and (d) require new proofs to accomodate the existence of multiple antennas.

\subsubsection{Proof of $\frac{\rank[\mathbf{G}_{11}^n \mathbf{V}_1^n \quad \mathbf{G}_{12}^n \mathbf{V}_2^n]}{N_1} -\frac{\rank[\mathbf{G}_{21}^n \mathbf{V}_1^n \quad \mathbf{G}_{22}^n \mathbf{V}_2^n]}{N_2} \stackrel{a.s.}{\leq} \frac{\rank[\mathbf{G}_{11}^\mathcal{T} \mathbf{V}_1^\mathcal{T} \quad \mathbf{G}_{22}^\mathcal{T} \mathbf{V}_2^\mathcal{T}]}{N_1}$}\hfill\\
\label{app:step1}
\begin{secproof}
To prove the statement, we require the following claim.
\begin{claim}\label{cl:incr}
For a realization, $\mathcal{G}^T$, of random channel $\bm{\mathcal{G}}^T$, if $t = \tau_k$ and $\tau_k\in\mathcal{T}$,
\begin{align}
    \rank[G_{11}^tV_1^t \quad G_{12}^tV_2^t ] - \rank[G_{11}^{t-1}V_1^{t-1} \quad G_{12}^{t-1}V_2^{t-1} ]
        \stackrel{}{\leq}{}& \rank[G_{11}^{\tau_k}V_1^{\tau_k} \quad G_{12}^{\tau_k}V_2^{\tau_k} ] 
            - \rank[G_{11}^{\tau_{k-1}}V_1^{\tau_{k-1}} \quad G_{12}^{\tau_{k-1}}V_2^{\tau_{k-1}} ].\label{eq:cl:incr}
\end{align}
\end{claim}
\begin{IEEEproof}
We first point out that, by definition,
\begin{align}
    \begin{bmatrix}
        G_{11}^tV_1^t & G_{12}^tV_2^t 
    \end{bmatrix}
    ={}& 
        \begin{bmatrix}
            G_{11}^{t-1}V_1^{t-1} & G_{12}^{t-1}V_2^{t-1}\\G_{11}[t]V_1[t] & G_{12}[t]V_2[t]
        \end{bmatrix},\\
    \begin{bmatrix}
        G_{11}^{\tau_k}V_1^{\tau_k} & G_{12}^{\tau_k}V_2^{\tau_k} 
    \end{bmatrix}
    ={}& 
        \begin{bmatrix}
        G_{11}^{\tau_{k-1}}V_1^{\tau_{k-1}} & G_{12}^{\tau_{k-1}}V_2^{\tau_{k-1}}\\G_{11}[t]V_1[t] & G_{12}[t]V_2[t]
        \end{bmatrix}.
\end{align}
Note that $\rank[G_{11}^tV_1^t \quad G_{12}^tV_2^t ] - \rank[G_{11}^{t-1}V_1^{t-1} \quad G_{12}^{t-1}V_2^{t-1} ] = L$ implies
\begin{align}
 \exists s_1,\ldots,s_L \in\mathbb{C}^{1\times (M_1+M_2)}\text{ s.t. }{}
    & s_\ell \in \rowspan[G_{11}[t]V_1[t] \quad G_{12}[t]V_2[t]], \\
    & s_\ell \notin \rowspan[G_{11}^{t-1}V_1^{t-1} \quad G_{12}^{t-1}V_2^{t-1}],\\
    & s_\ell \notin \spn(\{s_1,\ldots,s_L\}\setminus s_\ell),\\
    & \forall \ell\in\{1,\ldots,L\}.
\end{align}
However, because $[G_{11}^{\tau_{k-1}}V_1^{\tau_{k-1}} \quad G_{12}^{\tau_{k-1}}V_2^{\tau_{k-1}}]$ is a submatrix of $[G_{11}^{t-1}V_1^{t-1} \quad G_{12}^{t-1}V_2^{t-1}]$, 
\begin{align}
\rowspan[G_{11}^{\tau_{k-1}}V_1^{\tau_{k-1}} \quad G_{12}^{\tau_{k-1}}V_2^{\tau_{k-1}}] \subseteq \rowspan[G_{11}^{t-1}V_1^{t-1} \quad G_{12}^{t-1}V_2^{t-1}],
\end{align}
and therefore 
\begin{align}
s_\ell \notin \rowspan[G_{11}^{t-1}V_1^{t-1} \quad G_{12}^{t-1}V_2^{t-1}] \forall \ell\in\{1,\ldots,L\} \Rightarrow
s_\ell \notin \rowspan[G_{11}^{\tau_{k-1}}V_1^{\tau_{k-1}} \quad G_{12}^{\tau_{k-1}}V_2^{\tau_{k-1}}].
\end{align}
Therefore every linearly independent vector that increases the rank difference 
\begin{align*}
\rank[G_{11}^tV_1^t \quad G_{12}^tV_2^t ] - \rank[G_{11}^{t-1}V_1^{t-1} \quad G_{12}^{t-1}V_2^{t-1} ],
\end{align*}
must also increase 
\begin{equation*}
 \rank[G_{11}^{\tau_k}V_1^{\tau_k} \quad G_{12}^{\tau_k}V_2^{\tau_k} ] 
 - \rank[G_{11}^{\tau_{k-1}}V_1^{\tau_{k-1}} \quad G_{12}^{\tau_{k-1}}V_2^{\tau_{k-1}} ]. \IEEEQEDhereeqn
\end{equation*}
\end{IEEEproof}

Using Claim~\ref{cl:incr}, we note that for any realization, $\mathcal{G}^T$, of random channel $\bm{\mathcal{G}}^T$
\begin{align}
    &\frac{\rank[G_{11}^T V_1^T \quad G_{12}^T V_2^T ]}{N_1} - \frac{\rank[G_{21}^T V_1^T \quad G_{22}^T V_2^T ]}{N_2} \nonumber\\
        &\ \ \stackrel{(a)}{=}{} \sum_{t=1}^T \frac{\rank[G_{11}^tV_1^t \quad G_{12}^tV_2^t ] 
            - \rank[G_{11}^{t-1}V_1^{t-1} \quad G_{12}^{t-1}V_2^{t-1} ]}{N_1}  
            - \frac{\rank[G_{21}^tV_1^t \quad G_{22}^tV_2^t ] 
            - \rank[G_{21}^{t-1}V_1^{t-1} \quad G_{22}^{t-1}V_2^{t-1} ]}{N_2}\\
        &\ \ \stackrel{(b)}{\leq}{} \sum_{t\in\mathcal{T}} \frac{\rank[G_{11}^tV_1^t \quad G_{12}^tV_2^t ] 
            - \rank[G_{11}^{t-1}V_1^{t-1} \quad G_{12}^{t-1}V_2^{t-1} ]}{N_1}  
            - \frac{\rank[G_{21}^tV_1^t \quad G_{22}^tV_2^t ] 
            - \rank[G_{21}^{t-1}V_1^{t-1} \quad G_{22}^{t-1}V_2^{t-1} ]}{N_2}\\
        &\ \ \stackrel{(c)}{\leq}{} \sum_{t\in\mathcal{T}} \frac{\rank[G_{11}^tV_1^t \quad G_{12}^tV_2^t ] 
            - \rank[G_{11}^{t-1}V_1^{t-1} \quad G_{12}^{t-1}V_2^{t-1} ]}{N_1}\\
        &\stackrel{\text{Claim~\ref{cl:incr}}}{\leq}\sum_{k=1}^{|\mathcal{T}|} \frac{\rank[G_{11}^{\tau_k}V_1^{\tau_k} \quad G_{12}^{\tau_k}V_2^{\tau_k} ] 
            - \rank[G_{11}^{\tau_{k-1}}V_1^{\tau_{k-1}} \quad G_{12}^{\tau_{k-1}}V_2^{\tau_{k-1}} ]}{N_1}.
\end{align}
In step (a) we expanded both terms into two telescoping series, and grouped terms based on time index. 
In (b) we observe that $\frac{\rank[G_{11}^tV_1^t \quad G_{12}^tV_2^t ] - \rank[G_{11}^{t-1}V_1^{t-1} \quad G_{12}^{t-1}V_2^{t-1} ]}{N_1}  \leq 1$, and that any time instances which evaluate to a positive difference require $\frac{\rank[G_{21}^tV_1^t \quad G_{22}^tV_2^t ] - \rank[G_{21}^{t-1}V_1^{t-1} \quad G_{22}^{t-1}V_2^{t-1} ]}{N_2}<1$, which are, by definition, time instances in $\mathcal{T}$. 
In step (c), we removed the negative term.
\end{secproof}

\subsubsection{Proof of $\rank[\mathbf{V}_j^\mathcal{T}] \stackrel{a.s.}{\leq} \mathbf{r}_j$}\hfill\\
\label{app:step2}
\begin{secproof}
We first define the following sets:
\begin{align}
\mathcal{A}_t \triangleq \{\mathcal{G}^T : \quad &
\rank[{G}_{21}^{t}{V}_1^{t}\quad {G}_{22}^{t}{V}_2^{t}] 
    - \rank[{G}_{21}^{t-1}{V}_1^{t-1}\quad {G}_{22}^{t-1}{V}_2^{t-1}] < N_2\}\label{eq:Asetdef}\\
\mathcal{B}_t \triangleq \{\mathcal{G}^T : \quad &\exists {W}_1[t] \in\mathbb{C}^{M_1\times M_1}, {W}_2[t] \in\mathbb{C}^{M_2\times M_2},\nonumber\\
        & \rowspan[{W}_1[t] {V}_1[t]\quad {0}]\subseteq 
        \rowspan[{G}_{21}^{t-1}{V}_1^{t-1}\quad {G}_{22}^{t-1}{V}_2^{t-1}],\nonumber\\
        & \rowspan[{0}\quad {W}_2[t] {V}_2[t]]\subseteq
        \rowspan[{G}_{21}^{t-1}{V}_1^{t-1}\quad {G}_{22}^{t-1}{V}_2^{t-1}],\nonumber\\
        & \rank{W}_1[t]+\rank{W}_2[t]> M_1+M_2-N_2\},\label{eq:Bsetdef}
\end{align}
and note that $\bm{\mathcal{G}}^T\in\mathcal{A}_t$ is equivalent to the statement $t\in\bm{\mathcal{T}}$. 
We first claim the following, which is restated more generally and proven in Appendix~\ref{app:pAUBc}. 
\begin{claim*}
$\Pr(\bm{\mathcal{G}}^T\in\mathcal{A}_t\cap\mathcal{B}_t^c) = 0$.
\end{claim*} 
Notice that the claim implies that $t\in\bm{\mathcal{T}}$ almost surely requires $\bm{\mathcal{G}}^T\in\mathcal{B}_t$. 

Assuming $\Pr(\bm{\mathcal{G}}^T\in\mathcal{A}_t\cap\mathcal{B}_t^c) = 0$, we now show that $\rank[\mathbf{V}_1^\mathcal{T}] \stackrel{a.s.}{\leq} \mathbf{r}_1$. An analogous proof holds for $\rank[\mathbf{V}_2^\mathcal{T}] \stackrel{a.s.}{\leq} \mathbf{r}_2$. Consider a channel realization $\mathcal{G}^T$, such that for $t\in\mathcal{T}$, and corresponding precoding matrices ${V}_1[t]$ and ${V}_2[t]$. 
Let $\widetilde{W}_1[t]$ and $\widetilde{W}_2[t]$ be any two matrices satisfying the conditions stated in the definition of $\mathcal{B}_t$ in (\ref{eq:Bsetdef}). Additionally, let $\widetilde{W}_1^c[t]$ and $\widetilde{W}_2^c[t]$ be $M_1\times M_1$ and $M_2\times M_2$ matrices respectively, and assume that $\widetilde{W}_j[t]$ and  $\widetilde{W}_j^c[t]$ satisfy for $j = 1,2$,
\begin{align}
    \rowspan \left[\widetilde{W}_j^c[t] V_j[t]\right] ={}& \rowspan\big[V_j[t]\big]\setminus\rowspan\left[\widetilde{W}_j[t] V_j[t]\right],\label{eq:Ucompdef}
\end{align}
where $\setminus$ is the set difference operator, i.e., for sets $\mathcal{S}_1$ and $\mathcal{S}_2$
\begin{align}
    \mathcal{S}_1\setminus\mathcal{S}_2 = \left\{s\in\mathcal{S}_1 | s\notin \mathcal{S}_2\right\}.
\end{align}

Consider
\begin{align}
\rowspan&\left[V_1^\mathcal{T}\right] = \bigcup_{t\in\mathcal{T}}\rowspan[V_1[t]]\\
    ={}& \bigcup_{ t\in\mathcal{T}} \rowspan \left[\widetilde{W}_1[t]V_1[t]\right] 
        \cup \rowspan\left[\widetilde{W}_1^c[t]V_1[t]\right]\\
    ={}& \left(\bigcup_{ t\in\mathcal{T}} \rowspan\left[\widetilde{W}_1[t]V_1[t]\right]\right) 
        \cup \left(\bigcup_{ t\in\mathcal{T}} \rowspan\left[\widetilde{W}_1^c[t]V_1[t]\right]\right).\label{eq:rjproof0}
\end{align}
Notice first that
\begin{align}    
    \rowspan\left[\widetilde{W}_1[t]V_1[t]\right] 
        \subseteq{}&
            \bigcup_{
            \substack{A\in\mathbb{C}^{M_1\times m_1(T)}, s.t.\\ 
            \rowspan\left[A \quad 0\right] \subseteq \rowspan\left[G_{21}^{t-1}V_1^{t-1} \quad G_{22}^{t-1}V_2^{t-1}\right]}}
            \rowspan\left[A\right]\\
        \subseteq{}&
            \bigcup_{
            \substack{A\in\mathbb{C}^{M_1\times m_1(T)}, s.t.\\ 
            \rowspan\left[A \quad 0\right] \subseteq \rowspan\left[G_{21}^{T}V_1^{T} \quad G_{22}^{T}V_2^{T}\right]}}
            \rowspan\left[A\right]\\
    ={}& \left\{\vec{s}\in\mathbb{C}^{1\times m_1(T)}: \exists\vec{\ell}_{N_2T\times 1}
        \text{ s.t. }[\vec{s} \quad \vec{0}_{1\times m_2(T)}] = \vec{\ell}^\top \left[G_{21}^{T}V_1^{T} \quad G_{22}^{T}V_2^{T}\right]\right\},\\
     ={}& \mathcal{E}_1(\mathcal{G}^T).
\end{align}

We also observe from the definition of $\mathcal{B}_t$ in (\ref{eq:Bsetdef}), $\rank[\widetilde{W}_1[t]]+\rank[\widetilde{W}_2[t]]> M_1+M_2-N_2$, which coupled with the definition given in (\ref{eq:Ucompdef}) implies
\begin{align}
\rank\left[\widetilde{W}_1^c[t]V_1[t]\right]+\rank\left[\widetilde{W}_2^c[t]V_2[t]\right]
    \leq{}& \rank\left[\widetilde{W}_1^c[t]\right]+\rank\left[\widetilde{W}_2^c[t]\right]\\
    \leq{}& M_1-\rank\left[\widetilde{W}_1[t]\right]+M_2-\rank\left[\widetilde{W}_2[t]\right]\\
    <{}& N_2.
\end{align}
In other words, if $\widetilde{W}_1[t]V_1[t]$ and $\widetilde{W}_1[t]V_1[t]$ are already known by Receiver~2, then the combined residual rank of precoders at both transmitters is less than the number of antennas at Receiver~2. Noting that the random channel matrices $\mathbf{G}_{2j}[t]$ almost surely have rank $\min\{M_j,N_2\}$, we find
\begin{align}
    \rowspan \left[\mathbf{\widetilde{W}}_1^c[t] \mathbf{V}_1[t] \quad 0\right]
        \stackrel{a.s.}{\subseteq}{}& \rowspan \left[\mathbf{G}_{21}[t] \mathbf{V}_1[t] \quad \mathbf{G}_{22}[t] \mathbf{V}_2[t] \right] \nonumber\\
        &\cup \rowspan \left[\mathbf{\widetilde{W}}_1[t] \mathbf{V}_1[t] \quad 0 \right] \cup 
            \rowspan \left[0\quad \mathbf{\widetilde{W}}_2[t] \mathbf{V}_2[t] \right].\label{eq:rjproof}
\end{align}

Using (\ref{eq:rjproof}), we find
\begin{align}
    \rowspan& \left[\mathbf{V}_1[t] \quad 0\right] ={} \rowspan \left[\mathbf{\widetilde{W}}_1[t] \mathbf{V}_1[t] \quad 0\right] \cup \rowspan \left[\mathbf{\widetilde{W}}_1^c[t] \mathbf{V}_1[t] \quad 0\right]\\
        \stackrel{a.s.}{\subseteq}{}& \rowspan \left[\mathbf{G}_{21}^{t-1} \mathbf{V}_1^{t-1} \quad \mathbf{G}_{22}^{t-1} \mathbf{V}_2^{t-1} \right] \nonumber\\
            &\cup 
            \left(\rowspan \left[\mathbf{G}_{21}[t] \mathbf{V}_1[t] \quad \mathbf{G}_{22}[t] \mathbf{V}_2[t] \right] \cup 
            \rowspan \left[\mathbf{\widetilde{W}}_1[t] \mathbf{V}_1[t] \quad 0\right] \cup 
            \rowspan \left[0\quad \mathbf{\widetilde{W}}_2[t] \mathbf{V}_2[t]\right] \right)\\
        ={}& \rowspan \left[\mathbf{G}_{21}^{t} \mathbf{V}_1^{t} \quad \mathbf{G}_{22}^{t} \mathbf{V}_2^{t} \right],
\end{align}
and may now state
\begin{align}    
\rowspan \left[\mathbf{\widetilde{W}}_1^c[t] \mathbf{V}_1[t] \quad 0 \right]
        \stackrel{a.s.}{\subseteq}{}&
            \bigcup_{
            \substack{A\in\mathbb{C}^{M_1\times m_1(T)}, s.t.\\ 
            \rowspan\left[A \quad 0\right] \subseteq \rowspan\left[\mathbf{G}_{21}^{t}\mathbf{V}_1^{t} \quad \mathbf{G}_{22}^{t}\mathbf{V}_2^{t}\right]}}
            \rowspan[A]\\
        \subseteq{}&
            \bigcup_{
            \substack{A\in\mathbb{C}^{M_1\times m_1(T)}, s.t.\\ 
            \rowspan\left[A \quad 0\right] \subseteq \rowspan\left[\mathbf{G}_{21}^{T}\mathbf{V}_1^{T} \quad \mathbf{G}_{22}^{T}\mathbf{V}_2^{T}\right]}}
            \rowspan[A]\\
    ={}& \left\{\vec{s}\in\mathbb{C}^{1\times m_1(T)}: \exists\vec{\ell}_{N_2T\times 1}
        \text{ s.t. }\left[\vec{s} \quad \vec{0}_{1\times m_2(T)}\right] = \vec{\ell}^\top\left[\mathbf{G}_{21}^{T}\mathbf{V}_1^{T} \quad \mathbf{G}_{22}^{T}\mathbf{V}_2^{T}\right]\right\},\\
     ={}& \mathcal{E}_1(\bm{\mathcal{G}}^T).
\end{align}

Since $\rowspan [\mathbf{\widetilde{W}}_1[t] \mathbf{V}_1[t] \quad 0]$ and $\rowspan [\mathbf{\widetilde{W}}_1^c[t] \mathbf{V}_1[t] \quad 0]
$ are both almost surely subsets of $\mathcal{E}_1(\bm{\mathcal{G}}^T)$ for all $t\in\bm{\mathcal{T}}$, their union, and in fact the union over $t\in\bm{\mathcal{T}}$ expressed in (\ref{eq:rjproof0}) is also almost surely a subset of $\mathcal{E}_1(\bm{\mathcal{G}}^T)$. Noting that by definition $r_1 = \dimspan \mathcal{E}_1(\mathcal{G}^T)$, we arrive at the claim that $\rank [\mathbf{V}_1^\mathcal{T}] \stackrel{a.s.}{\leq} \mathbf{r}_1$.
\end{secproof}

%%%%%%%%%% %%%%%%%%%% 
\subsection{Formal Statement and Proof of $\Pr(\mathcal{G}^T\in\mathcal{A}_t\cap\mathcal{B}_t^c)=0$}
\label{app:pAUBc}

\begin{claim}
\label{cl:claim}
Consider a MIMO XC with delayed CSIT employing fixed linear coding strategies $f_1^{(T)}$ and $f_2^{(T)}$. Let sets $\mathcal{A}_t$ and $\mathcal{B}_t$ be defined as in (\ref{eq:Asetdef}) and (\ref{eq:Bsetdef}) respectively.
For all time slots $t$
\begin{align}
    \Pr(\bm{\mathcal{G}}^T\in\mathcal{A}_t\cap\mathcal{B}_t^c)=0.
    %|\bm{\mathcal{G}}^{t-1} = \mathcal{G}^{t-1}) = 0
\end{align}
\end{claim}

\begin{IEEEproof}
The following proof applies even if $M_2=0$, which describes a MIMO BC setting of Lemma~\ref{lem:BCratio}.
Notice first that when $N_2 \geq M_1+M_2$, the matrices ${W}_1[t] =\mathbf{0}$ and ${W}_2[t] =\mathbf{0}$ satisfy the condition for the set $\mathcal{B}_t$; thus if $N_2 \geq M_1+M_2$, then the set $\mathcal{B}_t^c$ is empty and $\Pr(\mathcal{G}^T\in\mathcal{A}_t\cap\mathcal{B}_t^c) = 0$. 

When $N_2<M_1+M_2$, we consider an arbitrary channel realization, ${\mathcal{G}}^{T}$, of the random channel, $\bm{\mathcal{G}}^{T}$, and resulting precoding matrices up to time $t$, ${V}_1^t$ and ${V}_2^t$, and suppose that ${\mathcal{G}}^T\in\mathcal{B}_t^c$. 
Let $\mathcal{L}\triangleq \rowspan[{G}_{21}^{t-1}{V}_1^{t-1} \quad {G}_{22}^{t-1}{V}_2^{t-1}]$. 
Since ${\mathcal{G}}^T\in\mathcal{B}_t^c$, 
for any $\widetilde{W}_1[t] \in\mathbb{C}^{M_1\times M_1}$ and $\widetilde{W}_2[t] \in\mathbb{C}^{M_2\times M_2}$ where $\rank\widetilde{W}_1[t]+ \rank\widetilde{W}_2[t] > M_1+M_2-N_2$ one of the following must hold:
\begin{align*}
    \rowspan[{W}_1[t] {V}_1[t]\quad {0}]\nsubseteq {}&
        \rowspan[{G}_{21}^{t-1}{V}_1^{t-1}\quad {G}_{22}^{t-1}{V}_2^{t-1}],\\
    \rowspan[{0}\quad {W}_2[t] {V}_2[t]]\nsubseteq{}&
        \rowspan[{G}_{21}^{t-1}{V}_1^{t-1}\quad {G}_{22}^{t-1}{V}_2^{t-1}],
\end{align*}
which implies either
\begin{align}
    \Proj_{\mathcal{L}^c}&[{W}_1[t]{V}_1[t] \quad 0 ] \neq \mathbf{0}
        \quad\text{or}\quad 
    \Proj_{\mathcal{L}^c}[0\quad {W}_2[t]{V}_2[t]] \neq \mathbf{0},\nonumber\\
        &\forall \widetilde{W}_1[t] \in\mathbb{C}^{M_1 \times M_1}, \widetilde{W}_2[t] \in\mathbb{C}^{M_2\times M_2}  \text{ s.t. } \rank[\widetilde{W}_1[t]]+\rank[\widetilde{W}_2[t]] > M_1+M_2-N_2.\label{eq:nonull}
\end{align}
In other words, the null space of the projection with respect to the spaces spanned by the rows of $[{V}_1[t] \quad 0]$ and $[0\quad {V}_2[t]]$ has dimension of at most $M_1+M_2-N_2$. %Conversely, the dimension of its image is strictly greater than $N_2$. 
Now consider two invertible basis transformation matrices, ${A}_1\in\mathbb{C}^{M_1\times M_1}$ and ${A}_2\in\mathbb{C}^{M_2\times M_2}$, that satisfy%applies the following structure to $\mathbf{V}[t]$:
\begin{align*}
    {A}_1{V}_1[t] ={}& \begin{bmatrix} {V}_1^o[t] \\ {V}_1^+[t] \end{bmatrix}\\
    \rowspan[{V}_1^o[t]\quad 0] \subseteq {}& \ker(\Proj_{\mathcal{L}^c}),
\end{align*}
and \begin{align*}
    {A}_2{V}_2[t] ={}& \begin{bmatrix} {V}_2^o[t] \\ {V}_2^+[t] \end{bmatrix}\\
    \rowspan[0 \quad {V}_2^o[t]] \subseteq {}& \ker(\Proj_{\mathcal{L}^c}).
\end{align*}
Let $s_1^o$, $s_1^+$, $s_2^o$, and $s_2^+$ denote the dimensions of $\rowspan[{V}_1^o[t]\quad 0]$, $\rowspan[{V}_1^+[t]\quad 0]$,  $\rowspan[0 \quad {V}_2^o[t]]$, and  $\rowspan[0 \quad {V}_2^+[t]]$, respectively. From (\ref{eq:nonull}), $s_1^o+s_2^o\leq M_1+M_2-N_2$, and therefore, by rank-nullity theorem~(\cite{Roman_book} page 63) $s_1^+ + s_2^+ = M_1+M_2-(s_1^o + s_2^o) \geq N_2$. 

We also define the adjusted random channel matrices $\mathbf{\widetilde{G}}_{21}[t] \triangleq \mathbf{G}_{21}[t]{A}_1^{-1}$ and $\mathbf{\widetilde{G}}_{22}[t] \triangleq \mathbf{G}_{22}[t]{A}_2^{-1}$ which, because $\mathbf{G}_{2j}[t]$ is drawn from a continuous distribution and $A_j^{-1}$ is full rank, is also continuously distributed. Within the realization ${\widetilde{G}}_{2j}[t]$ we identify two submatrices, ${\widetilde{G}}_{2j}^o[t]\in \mathbb{C}^{N_2\times s_j^o}$ and ${\widetilde{G}}_{2j}^+[t]\in\mathbb{C}^{N_2\times s_j^+}$, such that
\begin{align*}
    {\widetilde{G}}_{2j}[t] ={}& \begin{bmatrix} {\widetilde{G}}_{2j}^o[t] & {\widetilde{G}}_{2j}^+[t] \end{bmatrix}.
\end{align*}

Using these we see
\begin{align}
\Pr&(\bm{\mathcal{G}}^T\in\mathcal{A}_t\cap\mathcal{B}_t^c|\bm{\mathcal{G}}^{t-1}=\mathcal{G}^{t-1})\nonumber\\    
        ={}& \Pr(\bm{\mathcal{G}}^T\in\mathcal{A}_t|\bm{\mathcal{G}}^{t-1}=\mathcal{G}^{t-1},\bm{\mathcal{G}}^T\in\mathcal{B}_t^c)
            \Pr(\bm{\mathcal{G}}^T\in\mathcal{B}_t^c)\\
        ={}& \Pr(\rank(\Proj_{\mathcal{L}^c}
            [\mathbf{G}_{21}[t]{V}_1[t] \quad \mathbf{G}_{22}[t]{V}_2[t] ]) < N_2
            |\bm{\mathcal{G}}^{t-1}=\mathcal{G}^{t-1},\bm{\mathcal{G}}^T\in\mathcal{B}_t^c)\Pr(\bm{\mathcal{G}}^T\in\mathcal{B}_t^c)\\
        ={}& \Pr(\rank(\Proj_{\mathcal{L}^c}
            [\mathbf{\widetilde{G}}_{21}[t]{A}_1{V}_1[t]\quad \mathbf{\widetilde{G}}_{22}[t]{A}_2{V}_2[t]]) < N_2
            |\bm{\mathcal{G}}^{t-1}=\mathcal{G}^{t-1},\bm{\mathcal{G}}^T\in\mathcal{B}_t^c)\Pr(\bm{\mathcal{G}}^T\in\mathcal{B}_t^c)\\
%        \stackrel{}{=}{}& \Pr(\rank(\Proj_{\mathcal{L}^c}[\mathbf{\widetilde{G}}_2^o[t]\mathbf{V}^o[t]] 
%            +\Proj_{\mathcal{L}^c}[\mathbf{\widetilde{G}}_2^+[t]\mathbf{V}^+[t]]) < N_2
%            |\bm{\mathcal{G}}^{t-1}=\mathcal{G}^{t-1},\bm{\mathcal{G}}^T\in\mathcal{B}_t^c)\Pr(\bm{\mathcal{G}}^T\in\mathcal{B}_t^c)\\
        \stackrel{(a)}{=}{}& \Pr(\rank(\Proj_{\mathcal{L}^c}\mathbf{\widetilde{G}}_{21}^+[t]{V}_1^+[t]] + \Proj_{\mathcal{L}^c}\mathbf{\widetilde{G}}_{22}^+[t]{V}_2^+[t]]) < N_2
            |\bm{\mathcal{G}}^{t-1}=\mathcal{G}^{t-1},\bm{\mathcal{G}}^T\in\mathcal{B}_t^c)\Pr(\bm{\mathcal{G}}^T\in\mathcal{B}_t^c)\\
        \stackrel{(b)}{=}{}& \Pr(\rank\mathbf{\widetilde{G}}_{21}^++\rank\mathbf{\widetilde{G}}_{22}^+[t]<N_2
            |\bm{\mathcal{G}}^{t-1}=\mathcal{G}^{t-1},\bm{\mathcal{G}}^T\in\mathcal{B}_t^c)\Pr(\bm{\mathcal{G}}^T\in\mathcal{B}_t^c)\\
        \stackrel{a.s.}{=}& 0.
\end{align}
In step (a), we observed that any linear transformation of $\mathbf{V}_j^o[t]$ still lies within the nullspace of the projection.
In step (b), we noted that $s_1^++s_2^+\geq N_2$ and from (\ref{eq:nonull}) the preimage of non-zero values of the projection operator with respect to the transmitters' combined rowspaces has dimension $N_2$. This implies the condition $\rank\left[\Proj_{\mathcal{L}^c}\mathbf{\tilde{G}}_2^+[t]\mathbf{V}^+[t]\right] < N_2$ is satisfied only if the linear transformation $\mathbf{\tilde{G}}_2^+$ has rank less than $N_2$. In the final step, we observe that the probability of a continuously distributed channel matrix being rank deficient is zero. 
This proves that $\Pr(\bm{\mathcal{G}}^T\in\mathcal{A}_t\cap\mathcal{B}_t^c) = 0$. 
\end{IEEEproof}
%%%%%%%%%% %%%%%%%%%% 
\subsection{Phase 1 and 2 Linear Encoding Strategy when $\Gamma_iN_{i^\prime}\leq M_1$}\label{app:othercase}
For all antenna configurations where $\Gamma_i N_{i^\prime} \leq M_1$, our strategy ignores Transmitter~2 and simply employs a strategy given for the MIMO BC with delayed CSIT originally described in~\cite{VV:2012}. To keep consistent with existing notation, we say that the number of channel uses per round is $S_i=1$, and the parameter for Transmitter~2 is assigned the value $\xi_i=0$. 

During each round of transmission, Transmitter~1 broadcasts $\min\{M_1,N_1+N_2\}$ symbols per round, each on one of $\min\{M_1,N_1+N_2\}$ different antennas. Delayed CSIT is not used during the phase for these antenna configurations, simply because Transmitter~2 plays no role in the phase.
\end{document}